# Cluster Computing White Paper

Status – Final Release

Version 2.0

Date – 28th December 2000

Editor - Mark Baker, University of Portsmouth, UK

Contents and Contributing Authors:

1. *An Introduction to PC Clusters for High Performance Computing*, Thomas Sterling California Institute of Technology and NASA Jet Propulsion Laboratory, USA
2. *Network Technologies*, Amy Apon, University of Arkansas, USA, and Mark Baker, University of Portsmouth, UK.
3. *Operating Systems*, Steve Chapin, Syracuse University, USA and Joachim Worringen, RWTH Aachen, University of Technology, Germany
4. *Single System Image (SSI)*, Rajkumar Buyya, Monash University, Australia, Toni Cortes, Universitat Politecnica de Catalunya, Spain and Hai Jin, University of Southern California, USA
5. *Middleware*, Mark Baker, University of Portsmouth, UK, and Amy Apon, University of Arkansas, USA.
6. *Systems Administration*, Anthony Skjellum, MPI Software Technology, Inc. and Mississippi State University, USA, Rossen Dimitrov and Srihari Angulari, MPI Software Technology, Inc., USA, David Lifka and George Coulouris, Cornell Theory Center, USA, *Putchong Uthayopas, Kasetsart University, Bangkok, Thailand,* Stephen Scott, Oak Ridge National Laboratory, USA, Rasit Eskicioglu, University of Manitoba, Canada
7. *Parallel I/O*, Erich Schikuta, University of Vienna, Austria and Helmut Wanek, University of Vienna, Austria
8. *High Availability*, Ira Pramanick, Sun Microsystems, USA
9. *Numerical Libraries and Tools for Scalable Parallel Cluster Computing*, Jack Dongarra, University of Tennessee and ORNL, USA, Shirley Moore, University of Tennessee, USA, and Anne Trefethen, Numerical Algorithms Group Ltd, UK
10. *Applications*, David Bader, New Mexico, USA and Robert Pennington, NCSA, USA
11. *Embedded/Real-Time Systems*, Daniel Katz, Jet Propulsion Laboratory, California Institute of Technology, Pasadena, CA, USA and Jeremy Kepner, MIT Lincoln Laboratory, Lexington, MA, USA
12. *Education*, Daniel Hyde, Bucknell University, USA and Barry Wilkinson, University of North Carolina at Charlotte, USA

# Preface

Cluster computing is not a new area of computing. It is, however, evident that there is a growing interest in its usage in all areas where applications have traditionally used parallel or distributed computing platforms. The mounting interest has been fuelled in part by the availability of powerful microprocessors and high-speed networks as off-the-shelf commodity components as well as in part by the rapidly maturing software components available to support high performance and high availability applications.

This rising interest in clusters led to the formation of an IEEE Computer Society Task Force on Cluster Computing (TFCC[1]) in early 1999. An objective of the TFCC was to act both as a magnet and a focal point for all cluster computing related activities. As such, an early activity that was deemed necessary was to produce a White Paper on cluster computing and its related technologies.

Generally a White Paper is looked upon as a statement of policy on a particular subject. The aim of this White Paper is to provide a relatively unbiased report on the existing, new and emerging technologies as well as the surrounding infrastructure deemed important to the cluster computing community. This White Paper is essentially a snapshot of cluster-related technologies and applications in year 2000.

This White Paper provides an authoritative review of all the hardware and software technologies that can be used to make up a cluster now or in the near future. These technologies range from the network level, through the operating system and middleware levels up to the application and tools level. The White Paper also tackles the increasingly important areas of High Availability and Embedded/Real Time applications, which are both considered crucial areas for future clusters.

The White Paper has been broken down into twelve chapters, each of which has been put together by academics and industrial researchers who are both experts in their fields and where willing to volunteer their time and effort to put together this White Paper.

On a personal note, I would like to thank all the contributing authors for finding the time to put the effort into their chapters and making the overall paper an excellent state-of-the-art review of clusters. In addition, I would like to thank the reviewers for their timely comments.

Mark Baker
University of Portsmouth, UK
December 2000

---

[1] TFCC – http://www.ieeetfcc.org/

# 1. An Introduction to PC Clusters for High Performance Computing

Thomas Sterling
California Institute of Technology and NASA Jet Propulsion Laboratory

## 1.1. High Performance Computing in Transition

The halcyon days of specialized supercomputers, built and acquired at any cost to deliver orders of magnitude greater floating point performance than contemporary mainframes on grand challenge (often defense related) applications are rapidly diminishing in to the past along with the Cold War to be replaced by a more pragmatic cost-constrained reality that today permeates the community of high performance computing. Once the stars and vanguard of advanced computing, in appearance at least, supercomputers have been eclipsed by an Internet-enabled Web-driven future of opportunity and challenge. High end computing had been perceived as having become just too hard, too expensive, and of too narrow interest to justify a self-sustained industry.

Yet, in spite of statements to the contrary by a number of authority figures, performance or *capability computing* is important and, in fact, may matter very much. In no uncertain way, the futures of many of the path-finding fields that will determine the evolution of civilization throughout this new century will themselves be determined in their contributions by the degree to which they exploit the possibilities of trans-Teraflops (even Petaflops) scale computation. Energy including controlled fusion, medical science and molecular biology including genetic engineering, space exploitation and cosmology, climate and Earth science, materials including semiconductors and composites, machine intelligence and robotics, financial modeling and commerce, and information communication and human-machine interface will all be governed in their rate of progress and degree of impact on human culture by the magnitude of computational performance they are able to engage. In fact many of the strategic problems that plague our planet and species may only be solved through our ability to harness the most powerful computing engines possible and the capabilities of those machines.

Unfortunately, the market value of such extreme performance systems appears to be, in the short term at least, well below that required to justify industry investment in the development of these specialty-class supercomputer architectures. Fortunately, this conflict between the requirements for high performance and the availability of resources needed to provide it is being addressed through an innovative synergy of some old ideas from the parallel computing community and some new low-cost technologies from the consumer digital electronics industry. The legacy concepts are the physical clustering of general-purpose hardware resources and the abstract message-passing model of distributed computing. The low cost computing capability is derived from the mass market COTS PC and local networking industries. Together, these basic capabilities and founding principles, derived for very different purposes, have emerged as the new domain of high performance: commodity cluster computing. Commodity clusters including clusters of workstations, PC clusters, and Beowulf-class systems have become the single most rapidly growing class of high performance computing systems.

## 1.2. The Emergence of Cluster Computing for HPC

The annual Supercomputing Conference provides a snapshot of the state, accomplishments, and directions of the field and industry of high performance computing. At SC99 in Portland, Oregon, it was a surprise to some and a lesson to all to find a broad range of industrial and research exhibits highlighting their production and application of commercial clustered computer systems. These ranged from the high end commercial clustering of SMP servers with high-speed proprietary networks and commercial software to the self-assembled Beowulf-class PC clusters using freely distributed open source Linux and tools. A year later at SC00 in Dallas, commodity clusters dominated the exhibitions with more clusters operational on the floor than all other forms of parallel/distributed computing systems combined. Thus the beginning of the new century has been marked by a major shift in the means by which much of high performance computing will be accomplished in the future. Cluster systems are exhibiting the greatest rate in growth of any class of parallel computer and may dominate high performance computing in the near future.

But why should this have come to be? In contrast, conventional wisdom easily exposes serious problems with the cluster approach. Parallel processing and parallel computer architecture is a field with decades of experience that clearly demonstrates the critical factors of interconnect latency and bandwidth, the value of shared memory, and the need for lightweight control software. Generally, clusters are known to be weak on all these points. Bandwidths and latencies both could differ by two orders of magnitude (or more) between tightly couple MPPs and PC clusters. The shared memory model is more closely related to how applications programmers consider their variable name space and such hardware support can provide more efficient mechanisms for such critical functions as global synchronization and automatic cache coherency. And custom node software agents can consume much less memory and respond far more quickly than full-scale standalone operating systems usually found on PC clusters. In fact, for some application classes these differences make clusters unsuitable. But experience over the last few years has shown that the space and requirements of applications are rich and varying. While some types of applications may be difficult to efficiently port to clusters, a much broader range of workloads can be adequately supported on such systems, perhaps with some initial effort in optimization. Where conventional application MPP codes do not work well on clusters, new algorithmic techniques that are latency tolerant have been devised in some cases to overcome the inherent deficiencies of clusters. As a consequence, on a per node basis in many instances applications are performed at approximately the same throughput as on an MPP for a fraction of the cost. Indeed, the price-performance advantage in many cases exceeds an order of magnitude. It is this factor of ten that is driving the cluster revolution in high performance computing.

## 1.3. A Definition

A "*commodity cluster*" is a local computing system comprising a set of independent computers and a network interconnecting them. A cluster is *local* in that all of its component subsystems are supervised within a single administrative domain, usually residing in a single room and managed as a single computer system. The constituent *computer nodes* are commercial-off-the-shelf (COTS), are capable of full independent operation as is, and are of a type ordinarily employed individually for standalone mainstream workloads and applications. The nodes may incorporate a single microprocessor or multiple microprocessors in a symmetric multiprocessor (SMP) configuration. The *interconnection network* employs COTS local area network (LAN) or systems area network (SAN) technology that may be a hierarchy of or multiple separate network structures. A cluster network is dedicated to the integration of the cluster compute nodes and is separate from the cluster's external (worldly) environment. A cluster may be employed in many modes including but not limited to: high capability or sustained performance on a single problem, high capacity or throughput on a job or process workload, high availability through redundancy of nodes, or high bandwidth through multiplicity of disks and disk access or I/O channels. A "*Beowulf-class system*" is a cluster with nodes that are personal computers (PC) or small symmetric multiprocessors (SMP) of PCs integrated by COTS local area networks (LAN) or system area networks (SAN), and hosting an open source Unix-like node operating system. An Windows-Beowulf system also exploits low cost mass market PC hardware but instead of hosting an open source Unix-like O/S, it runs the mass market widely distributed Microsoft Windows and NT operating systems. A "*Constellation*" differs from a commodity cluster in that the number of processors in its node SMPs exceeds the number of SMPs comprising the system and the integrating network interconnecting the SMP nodes may be of custom technology and design. Definitions such as these are useful in that they provide guidelines and help focus analysis. But they can also be overly constraining in that they inadvertently rule out some particular system that intuition dictates should be included in the set. Ultimately, common sense must prevail.

1.4. Benefits of Clusters

Clusters allow *trickle-up*; hardware and software technologies that were developed for broad application to mainstream commercial and consumer markets can also serve in the arena of high performance computing. It was this aspect of clusters that initially made them possible and triggered the first wave of activity in the field. Both network of workstations and Beowulf-class PC clusters were possible because they required no expensive or long-term development projects prior to their initial end use. Such early systems were far from perfect but they were usable. Even these inchoate cluster systems exhibited price-performance advantage with respect to contemporary supercomputers that approached a factor of 50 in special cases while delivering per node sustained performance for real-world applications often within a range of a factor of 3 and sometimes well within 50% of the more costly systems with the same number of processors. But the rapid rate of improvement in PC microprocessor performance and advances in local area networks have led to systems capable of tens or even hundreds of Gigaflops performance while retaining exceptional price-performance benefits as recognized by the 1997 and 1998 Gordon Bell Prizes for Price-Performance which were awarded to PC cluster systems.

Commodity clusters permit a flexibility of configuration not ordinarily encountered through conventional MPP systems. Number of nodes, memory capacity per node, number of processors per node, and interconnect topology are all parameters of system structure that may be specified in fine detail on a per system basis without incurring additional cost due to custom configurability. Further, system structure may easily be modified or augmented over time as need and opportunity dictate without the loss of prior investment. This expanded control over system structure not only benefits the end user but the system vendor as well, yielding a wide array of system capabilities and cost tradeoffs to better meet customer demands. Commodity clusters also permit rapid response to technology improvements. As new devices including processors, memory, disks, and networks become available, they are most likely to be integrated in to desktop or server nodes most quickly allowing clusters to be the first class of parallel systems to benefit from such advances. The same is true of benefits incurred through constantly improving price-performance trends in delivered technology. Commodity clusters are best able to track technology improvements and respond most rapidly to new component offerings.

1.5. Hardware Components

The key components comprising a commodity cluster are the nodes performing the computing and the dedicated interconnection network providing the data communication among the nodes. Significant advances for both have been accomplished over the last half-decade and future improvements are ensured for at least the next two to three years. While originally clusters were almost always assembled at the user site by staff local to the user organization, now increasingly clusters are being delivered by vendors as turnkey solutions to user specifications. Where once such systems were packaged in conventional "tower" cases, now manufacturers are providing slim, rack mounted, nodes to deliver more capacity per unit floor area. Although original clusters leveraged industry investment in the development of local area networks, today multiple manufacturers have developed and are marketing system area networks optimized to the needs of commodity clusters. Here is briefly summarized the basic elements that comprise both the node and network hardware and some of the principal offerings available today from major vendors.

1.5.1 Cluster Node Hardware

A node of a cluster provides the system computing and data storage capability. It is differentiated from nodes of fully integrated systems such as an MPP in that it is derived from fully operational standalone computing subsystems that are typically marketed as desktop or server systems. Thus, no additional hardware development cost is incurred by vendors to provide nodes for cluster configurations and they directly benefit from the cost advantage of the consumer mass market. The node integrates several key subsystems in a single unit. These are:

- *Processor* – the central processing unit today is a complex subsystem including the core processing element, two layers of cache, and external bus controllers. Both 32 bit and 64 bit processors are available with popular architectures including the Intel Pentium II/III family and the Compaq Alpha 21264. Others include the IBM PowerPC, the Sun Microsystems Super Sparc III, and the AMD K7 Athelon. Clock rates in excess of 1 GHz are becoming increasingly available along with peak performance greater than 1 Gflops. For some nodes, more than one processor is combined in SMP configurations providing full mutual cache coherency.
- *Memory* – Dynamic Random Access Memory (DRAM) has dominated the main memory business for two decades providing high density with moderate access times at reasonable cost. Node main memory capacities range from 64 Mbytes to greater than 1 Gbytes packaged as DIMMs or SIMMs. Individual memory chips may contain 64 Mbits and employ several possible interface protocols. SDRAM is the most widely used to provide high memory bandwidth.
- *Secondary Storage* – for both high capacity and nonvolatile storage, most nodes (but not all) will incorporate one or more forms of mass storage. Hard disks are the primary form of secondary storage with SCSI II and EIDE the two most popular classes. SCSI provides higher capacity and superior access bandwidth while EIDE provides exceptional cost advantage. Current generation disk drives can provide between 20 and 100 Gbytes of spinning storage with access times on the order of a few milliseconds. Optical CD-ROMs and CD-RW are becoming another source of storage, especially for low cost portability, code and data delivery, and backing store. A single disk may hold more than 600 Mbytes and cost less than $1 in small quantities. Writable disks are rapidly becoming typical, at least for master nodes. Floppy disks are still found on many nodes, especially for bootup, initial installs, and diagnostics, but with a capacity of only 1.4 Mbytes they are of diminishing value to system nodes. Zip drives holding up to 250 Mbytes and of a form factor only slightly larger than that of a floppy drive are another type of removable storage found on some nodes.
- *External Interface* - standard interface channels are provided to connect the compute node to external devices and networks. While most nodes still include venerable

EISA ports on the motherboards, these are slow and provided almost entirely for backward compatibility. The primary interface today is PCI to which the majority of control cards including network NICs are connected. PCI comes in four configurations employing 32 or 64 bit connections driven at 33 MHz or 66 MHz clock rates with a peak throughput of between approximately 1 and 4 Gbps. USB is rapidly replacing other interfaces to connect medium speed external devices but parallel printer ports and serial RS-232 ports will still be found on most nodes, again for backward compatibility. Keyboard and video monitor interfaces ordinarily are provided for local diagnostics and control although such functionality is increasingly being handled remotely over the internal system area network.

1.5.2 Cluster Network Hardware

Commodity clusters are made possible only because of the availability of adequate inter-node communication network technology. Interconnect networks enable data packets to be transferred between logical elements distributed among a set of separate processor nodes within a cluster through a combination of hardware and software support. Commodity clusters incorporate one or more dedicated networks to support message packet communication within the distributed system. This distinguishes it from ensembles of standalone systems loosely connected by shared local area networks (LAN) that are employed primarily as desktop and server systems. Such computing environments have been successfully employed to perform combined computations using available unused resources. These practices are referred to as "cycle harvesting" or "workstation farms" and share the intercommunication network with external systems and services, not directly related to the coordinated multi-node computation. In comparison, the commodity cluster's system area network (SAN) is committed to the support of such distributed computation on the cluster, employing separate external networks for interaction with environment services.

Parallel applications exhibit a wide range of communication behaviors and impose diverse requirements on the underlying communication framework. Some problems require the high bandwidth and low latency found only in the tightly coupled massively parallel processing systems (MPP) and may not be well suited to commodity clusters. Other application classes referred to as "embarrassingly parallel" not only perform effectively on commodity clusters, they may not even fully consume the networking resources provided. Many algorithms fall in between these two extremes. As network performance attributes improve, the range of problems that can be effectively handled also expands. The two primary characteristics establishing the operational properties of a network are the bandwidth measured in millions of bits per second (Mbps) and the latency measured in microseconds. Peak bandwidth is the maximum amount of information that can be transferred in unit time through a single channel. Bi-section bandwidth is the total peak bandwidth that can be passed across a system. Latency is the amount of time to pass a packet from one physical or logical element to another. But the actual measurement values observed for these parameters will vary widely even on a single system depending on such secondary effects as packet size, traffic contention, and software overhead.

Early PC clusters such as Beowulf-class systems employed existing LAN Ethernet technology the cost of which had improved to the point that low cost commodity clusters were feasible. However, their 10 Mbps peak bandwidth was barely adequate for other than the most loosely coupled applications and some systems ganged multiple Ethernet networks together in parallel through a software means called "channel bonding" that made the multiplicity of available channels at a node transparent to the application code while delivering significantly higher bandwidth to it than provided by a single channel. With the emergence of Fast Ethernet exhibiting 100 Mbps peak bandwidth and the availability of low cost hubs and moderate cost switches, commodity clusters became practical and were found to be useful to an increasing range of applications for which latency tolerant algorithms could be devised. Hubs provide a shared communications backplane that provides connectivity between any two nodes at a time. While relatively simple and inexpensive, hubs limit the amount of communication within a cluster to one transaction at a time. The more expensive switches permit simultaneous transactions between disjoint pairs of nodes, thus greatly increasing the potential system throughput and reducing network contention.

While local area network technology provided an incremental path to the realization of low cost commodity clusters, the opportunity for the development of networks optimized for this domain was recognized. Among the most widely used is Myrinet with its custom network control processor that provides peak bandwidth in excess of 1 Gbps at latencies on the order of 20 microseconds. While more expensive per port than Fast Ethernet, its costs are comparable to that of the recent Gigabit Ethernet (1 Gbps peak bandwidth) even as it provides superior latency characteristics. Another early system area network is SCI (scalable coherent interface) that was originally designed to support distributed shared memory. Delivering several Gbps bandwidth, SCI has found service primarily in the European community. Most recently, an industrial consortium has developed a new class of network capable of moving data between application processes without requiring the usual intervening copying of the data to the node operating systems. This "zero copy" scheme is employed by the VIA (Virtual Interface Architecture) network family yielding dramatic reductions in latency. One commercial example is cLAN which provides bandwidth on the order of a Gbps with latencies well below 10 microseconds. Finally, a new industry standard is emerging, Infiniband, that in two years promises to reduce the latency even further approaching one microsecond while delivering peak bandwidth on the order of 10 Gbps. Infiniband goes further than VIA by removing the I/O interface as a contributing factor in the communication latency by directly connecting to the processor's memory channel interface. With these advances, the network infrastructure of the commodity cluster is becoming less dominant as the operational bottleneck. Increasingly, it is the software tools that are limiting the applicability of commodity clusters as a result of hardware advances that have seen improvements of three orders of magnitude in interconnect bandwidth within a decade.

## 1.6. Software Components

If low cost consumer grade hardware has catalyzed the proliferation of commodity clusters for both technical and commercial high performance computing, it is the software that has both enabled its utility and restrained its usability. While the rapid advances in hardware capability have propelled commodity clusters to the forefront of next generation systems, equally important has been the evolving capability and maturity of the support software systems and tools. The result is a total system environment that is converging on previous generation supercomputers and MPPs. And like these predecessors, commodity clusters present opportunity for future research and advanced development in programming tools and resource management software to enhance their applicability, availability, scalability, and usability. However, unlike these earlier system classes, the wide and rapidly growing distribution of clusters is fostering unprecedented work in the field even as it is providing a convergent architecture family in which the applications community can retain confidence. Therefore, software components, so necessary for the success of any type of parallel system, is improving at an accelerated rate and quickly approaching the stable and sophisticated levels of functionality that will establish clusters as the long-term solution to high-end computation.

The software components that comprise the environment of a commodity cluster may be described in two major categories: programming tools and resource management system software. Programming tools provide languages, libraries, and performance and correctness debuggers to construct parallel application programs. Resource management software relates to initial installation, administration, and scheduling and allocation of both hardware and software components as applied to user workloads. A brief summary of these critical components follows.

1.6.1 Application Programming Environments

Harnessing of parallelism within application programs to achieve orders of magnitude gain in delivered performance has been a challenge confronting high end computing since the 1970's if not before. Vector supercomputers, SIMD array processors, and MPP multiprocessors have exploited varying forms of algorithmic parallelism through a combination hardware and software mechanisms. The results have been mixed. The highest degrees of performance yet achieved have been through parallel computation. But in many instances, the efficiencies observed have been low and the difficulties in their accomplishment have been high. Commodity clusters, because of their superior price-performance advantage for a wide range of problems, are now undergoing severe pressure to address the same problems confronted by their predecessors. While multiple programming models have been pursued, one paradigm has emerged as the predominant form, at least for the short term. The "communicating sequential processes" model more frequently referred to as the "message passing" model has evolved through many different implementations resulting in the MPI or Message Passing Interface community standard. MPI is now found on virtually every vendor multiprocessor including SMPs, DSMs, MPPs, and clusters. MPI is not a full language but an augmenting library that allows users of C and Fortran to access libraries for passing messages between concurrent processes on separate but interconnected processor nodes. A number of implementations of MPI are available from system vendors, ISVs, and research groups providing open source versions through free distributions.

Parallel programming to be effective requires more than a set of constructs. There needs to be the tools and environment to understand the operational behavior of a program to correct errors in the computation and to enhance performance. The status of such tools for clusters is in its infancy although significant effort by many teams has been made. One popular debugger, Totalview, has met part of the need and is the product of an ISV, benefiting from years of evolutionary development. A number of performance profilers have been developed taking many forms. One good example is the set of tools incorporated with the PVM distribution. Work continues but more research is required and no true community wide standard has emerged.

While the message-passing model in general and MPI in particular dominate parallel programming of commodity clusters, other models and strategies are possible and may prove superior in the long run. Data-parallel programming has been supported through HPF and BSP. Message driven computation has been supported by Split-C and Fast messages. Distributed shared memory programming was pursued by the Shrimp project and through the development of UPC. And for the business community relying on coarse-grained transaction processing, a number of alternative tools have been provided to support a conventional master-slave methodology. It is clear that today for cluster programming there is an effective and highly portable standard for building and executing parallel programming and that work on possible alternative techniques is underway by researcher groups that may yield improved and more effective models in the future.

1.6.2 Resource Management Software

While a couple of strong standards have been accepted by the community for programming commodity clusters, such can not be said for the software environments and tools required to manage the resources, both hardware and software, that execute the programs. However, in the last two to three years substantial progress has been made with number of software products available from research groups and ISVs that are beginning to satisfy some of the more important requirements. The areas of support required are diverse and reflect the many challenges to the management of cluster resources.

- *Installation and Configuration* – the challenge of "build it yourself supercomputers" lies first with the means of assembling, installing, and configuring the necessary hardware and software components comprising the complete system. The challenge of implementing and maintaining common software image across nodes, especially for systems comprising hundreds of nodes requires sophisticated, effective, and easy to use tools. Many partial solutions have been developed by various groups and vendors (e.g. Scyld) and are beginning to make possible quick and easy creation of very large systems with a minimum of effort.
- *Scheduling and Allocation* - the placement of applications on to the distributed resources of commodity clusters requires tools to allocate the software components to the nodes and to schedule the timing of their operation. In its most simple form, the programmer performs this task manually. However, for large systems, especially those shared among multiple users and possibly performing multiple programs at one time, more sophisticated means are required for a robust and disciplined management of computing resources. Allocation can be performed at different levels of task granularity including: jobs, transactions, processes, or threads. They may be scheduled statically such that once an assignment is made it is retained until the culmination of the task or dynamically permitting the system to automatically migrate, suspend, and reinitiate tasks for best system throughput by load balancing. Examples of available schedulers that incorporate some, but not all, of these capabilities include Condor, the Maui Scheduler, and the Cluster Controller.
- *System Administration* – the supervision of industrial grade systems requires many mundane but essential chores to be performed including the management of user accounts, job queues, security, backups, mass storage, log journaling, operator interface, user shells, and other housekeeping activities. Traditionally, high-end computing systems have lacked some or most of these essential support utilities typical of commercial server class mainstream systems. However, incremental progress in this area has been made with PBS as a leading example.
- *Monitoring and Diagnosis* – continued operation of a system comprising a large number of nodes requires low level tools by which the operator may monitor the state, operation, and health of all system elements. Such tools can be as simple as distributed versions of command line Unix-like *ps* and *top* commands to GUI visual depictions of the total parallel system updated in real time. While a number of such tools has been devised by different groups, no single tool has achieved dominance in community wide usage.
- *Distributed Secondary Storage* – almost all computations require access to secondary storage including both local and remote disk drives for support of file systems. Many commercial applications and even some technical computations are disk access intensive. While NFS has continued to serve in many commodity cluster configurations, its many limitations in both functionality and performance have resulted in a new generation of parallel file systems to be developed. No one of these fully satisfies the diversity of requirements seen across the wide range of applications. As a result, commercial turnkey applications often incorporate their own proprietary distributed file management software tailored to the specific access patterns of the given application. But for the broader cluster user base, a general-purpose parallel file system is required that is flexible, scalable, and efficient as well as standardized. Examples of parallel file systems that have been applied to the cluster environment include PPFS, PVFS, and GPFS.
- *Availability* – as the scale of commodity clusters increases, the MTBF will decrease and active measures must be taken to ensure continued operation and minimal down time. Surprisingly, many clusters operate for months at a time without any outage and are often taken off line just for software upgrades. But especially during early operation, failures of both hardware (through infant mortality) and software (incorrect installation and configuration) can cause disruptive failures. For some cluster systems dedicated to a single long running application, continuous execution times of weeks or even months may be required. In all these cases, software tools to enhance system robustness and maximize up-time are becoming increasingly important to the practical exploitation of clusters. Checkpoint and restart tools are crucial to the realization of long-run applications on imperfect systems. Rapid diagnostics, recursion testing, fault detection, isolation, hot spares, and maintenance record analysis require software tools and a common framework to keep systems operational longer, resume operation rapidly in response to failures, and recover partial computations for continued execution in the presence of faults.

## 1.7. This Special Issue

This special of the International Journal of High Performance Computing and Applications is dedicated to providing the most current and in-depth representation of the field of PC Cluster Computing, its enabling technologies, system software components, programming environments, and research directions. A number of experts have collaborated on this important survey and summary of the state of the field to provide a comprehensive reflection of the practical capabilities, their application, and future research directions. This issue comprises a set of separate but coordinated papers that provide both useful guidance and exploratory insight into the present and future of commodity cluster computing. The issue begins with the critical subject of enabling network technologies. It describes the many choices, their properties, and tradeoffs as well as future directions. Equally important is the local node operating system of which a number are being employed for clusters. These include commercial versions of Unix, open-source Unix-like systems such as Linux, and other commercial systems such as Microsoft Windows-2000. But other experimental systems are considered as well along with the many underlying issues that determine the effectiveness of node operating systems to support cluster computation. The fourth paper addresses the broader issues related to making a distributed ensemble of essentially disparate processing nodes into a single homogenous computing resource. This paper considers the issues, requirements, and methodologies for achieving single system image at multiple levels of abstractions necessary for ease of management and programming. Following that is the central issue of managing distributed input output including secondary storage and the challenges faced by the sometimes-conflicting requirements of generality, robustness, and performance. The challenge to achieving high availability for large and complex systems is considered in the next paper that examines the hardware and software means of improving system reliability. The succeeding next two chapters explore the realm of computation on commodity clusters from the perspective of shared numerical libraries and example applications. These papers provide quantitative evidence of the effectiveness and power of the cluster approach while exposing the challenges to extracting their best performance. This special issue concludes with a particularly interesting domain of application that while special purpose has extremely wide usage nonetheless. Embedded computing represents an installed base more than an order of magnitude greater than all desktop and server systems combined and the role of cluster computing in this arena is only now being explored for commercial, industrial, and defense related applications. This last paper provides an introduction to a range of embedded applications and the enabling hardware and software technologies that are making cluster computing available to support their growing performance and availability requirements. This special issue is offered as a contribution to a rapidly growing field that is likely to dominate high performance computing in the near future.

# 2. Network Technologies

Amy Apon, University of Arkansas, USA, and Mark Baker, University of Portsmouth, UK.

## 2.1 Introduction

A broad and growing range of possibilities is available to designers of a cluster when choosing an interconnection technology. As the price of network hardware in a cluster can vary from almost free to several thousands of dollars per computing node, the decision is not a minor one in determining the overall price of the cluster. Many very effective clusters have been built from inexpensive products that are typically found in local area networks. However, some recent network products specifically designed for cluster communication have a price that is comparable to the cost of a workstation. The choice of network technology depends upon a number of factors, including price, performance, and compatibility with other cluster hardware and system software as well as communication characteristics of applications that will use the cluster.

Performance of a network is generally measured in terms of latency and bandwidth. Latency is the time to send data from one computer to another, and includes the overhead for the software to construct the message as well as the time to transfer the bits from one computer to another. Bandwidth is the number of bits per second that can be transmitted over the interconnection hardware. Ideally, applications that are written for a cluster will have a minimum amount of communication. However, if an application sends a large number of small messages, then its performance will be impacted by the latency of the network, and if an application sends large messages, then its performance will be impacted by the bandwidth of the network. In general, applications perform best when the latency of the network is low and the bandwidth is high. Achieving low latency and high bandwidth requires efficient communication protocols that minimize communication software overhead, and fast hardware.

Compatibility of network hardware with other cluster hardware and system software is a major factor in the selection of network hardware. From the user's perspective, the network should be interoperable with the selected end node hardware and operating system, and should be capable of efficiently supporting the communication protocols that are necessary for the middleware or application. Section two of this article gives a brief history of cluster communication protocols, leading to the development of two important standards for cluster communication.

Section three of this paper gives an overview of many common cluster interconnection products. A brief description of each technology is given, along with a comparison of products on the basis of price, performance, and support for standard communication protocols. Finally, section four presents a short conclusion and future developments for cluster network technologies.

## 2.2 Communication Protocols

A communication protocol defines the rules and conventions that will be used by two or more computers on the network to exchange information. Communication protocols can be classified as:
- Connection-oriented or connectionless,
- Offering various levels of reliability, including fully guaranteed to arrive in order (reliable), or not guaranteed (unreliable),

- Not buffered (synchronous), or buffered (asynchronous), or
- By the number of intermediate data copies between buffers, which may be zero, one, or more.

Several protocols are used in clusters. Traditional network protocols that were originally designed for the Internet are used, and protocols that have been designed specifically for cluster communication are used. In addition, two new protocol standards have been specially designed for use in cluster computing.

2.2.1 Internet Protocols

The Internet Protocol (IP) is the *de facto* standard for networking worldwide. IP offers a best-effort (unreliable) messaging service between two computers that have an IP address. The Transmission Control Protocol (TCP) and the User Datagram Protocol (UDP) are both transport layer protocols built over the Internet Protocol. TCP (or TCP/IP) offers a reliable, connection-oriented service between two hosts on a network. UDP is an unreliable, connectionless transport layer service. TCP and UDP protocols and the *de facto* standard BSD sockets Application Programmer's Interface (API) to TCP and UDP were among the first messaging libraries used for cluster computing [1].

Traditionally, TCP and UDP protocols are typically implemented using one or more buffers in system memory and with the aid of operating system services. To send a message, a user application constructs the message in user memory, and then makes an operating system request to copy the message into a system buffer. A system interrupt is required for this operating system intervention before the message can be sent out by network hardware. When a message is received by network hardware, the network copies the message to system memory. An interrupt is used to copy the message from system memory to user memory and to notify the receiving process that the message has arrived.

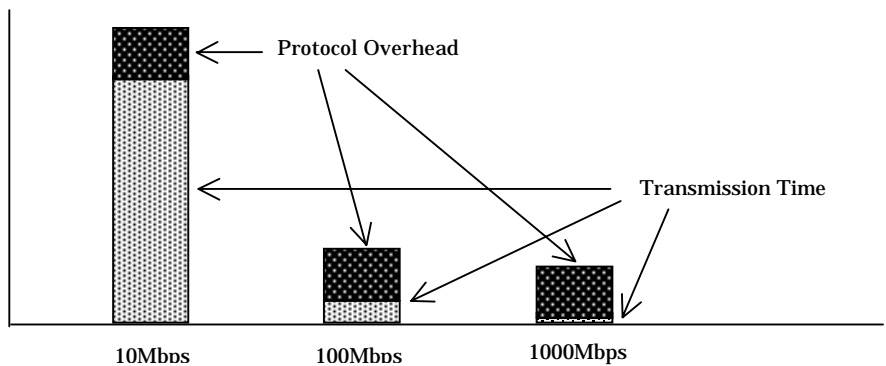

Figure 21. Traditional Protocol Overhead and Transmission Time

Operating system overhead and the overhead for copies to and from system memory are a significant portion of the total time to send a message. As network hardware became faster during the 1990's, the overhead of the communication protocols became significantly larger than the actual hardware transmission time for messages, as shown in Figure 2.1.

### 2.2.2 Low-latency Protocols

Several research projects during the 1990's led to the development low-latency protocols that avoid operating system intervention while at the same time providing user-level messaging services across high-speed networks. Low-latency protocols developed during the 1990's include Active Messages, Fast Messages, the VMMC (Virtual Memory-Mapped Communication) system, U-net, and Basic Interface for Parallelism (BIP), among others.

#### 2.2.2.1 Active Messages

Active Messages [2] is the enabling low-latency communications library for the Berkeley Network of Workstations (NOW) project [3]. Short messages in Active Messages are synchronous, and are based on the concept of a request-reply protocol. The sending user-level application constructs a message in user memory. To transfer the data, the receiving process allocates a receive buffer, also in user memory on the receiving side, and sends a request to the sender. The sender replies by copying the message from the user buffer on the sending side directly to the network. No buffering in system memory is performed. Network hardware transfers the message to the receiver, and then the message is transferred from the network to the receive buffer in user memory. This process requires that user virtual memory on both the sending and receiving side be pinned to an address in physical memory so that it will not be paged out during the network operation. However, once the pinned user memory buffers are established, no operating system intervention is required for a message to be sent. This protocol is also called a "zero-copy" protocol, since no copies from user memory to system memory are used.

Active Messages was later extended to Generic Active Messages (GAM) to support multiple concurrent parallel applications in a cluster. In GAM, a copy sometimes occurs to a buffer in system memory on the receiving side so that user buffers can be reused more efficiently. In this case, the protocol is referred to as a "one-copy" protocol.

#### 2.2.2.2 Fast Messages

Fast Messages was developed at the University of Illinois and is a protocol similar to Active Messages [4]. Fast Messages extends Active Messages by imposing stronger guarantees on the underlying communication. In particular, Fast Messages guarantees that all messages arrive reliably and in-order, even if the underlying network hardware does not. It does this in part by using flow control to ensure that a fast sender cannot overrun a slow receiver, thus causing messages to be lost. Flow control is implemented in Fast Messages with a credit system that manages pinned memory in the host computers. In general, flow control to prevent lost messages is a complication in all protocols that require the system to pin main memory in the host computers, since either a new buffer must be pinned or an already-pinned buffer must be emptied before each new message arrives.

#### 2.2.2.3 VMMC

The Virtual Memory-Mapped Communication [5] system, later extended to VMMC-2, is the enabling low-latency protocol for the Princeton SHRIMP project. One goal of VMMC is to view messaging as reads and writes into the user-level virtual memory system. VMMC works by mapping a page of user virtual memory to physical memory, and in addition makes a correspondence between pages on the sending and receiving sides. VMMC also uses specially designed hardware that allows the network interface to snoop writes to memory on the local host and have these writes automatically updated on the remote hosts memory. Various optimisations of these writes have been developed that help to minimize the total number of writes, network traffic, and overall application performance. VMMC is an example of a paradigm known as distributed shared memory (DSM). In DSM systems memory is physically distributed among the nodes in a system, but processes in an application may view "shared" memory locations as identical and perform reads and writes to the "shared" memory locations.

2.2.2.4 U-net

The U-net network interface architecture [6] was developed at Cornell University, and also provides zero-copy messaging where possible. U-net adds the concept of a virtual network interface for each connection in a user application. Just as an application has a virtual memory address space that is mapped to real physical memory on demand, each communication endpoint of the application is viewed as a virtual network interface mapped to a real set of network buffers and queues on demand. The advantage of this architecture is that once the mapping is defined, each active interface has direct access to the network without operating system intervention. The result is that communication can occur with very low latency.

2.2.2.5 BIP

BIP [7] (Basic Interface for Parallelism) is a low-latency protocol that was developed at the University of Lyon. BIP is designed as a low-level message layer over which a higher-level layer such as Message Passing Interface (MPI) [8] can be built. Programmers can use MPI over BIP for parallel application programming. The initial BIP interface consisted of both blocking and non-blocking calls. Later versions (BIP-SMP) provide multiplexing between the network and shared memory under a single API for use on clusters of symmetric multiprocessors. Currently BIP-SMP is only supported under the Linux operating system.

BIP achieves low latency and high bandwidth by using different protocols for various message sizes and by providing a zero or single memory copy of user data. To simply the design and keep the overheads low, BIP guarantees in-order delivery of messages, although some flow control issues for small messages are passed to higher software levels.

2.2.3 Standards for Cluster Communication

By 1997, research on low-latency protocols had progressed sufficiently for a new standard for low-latency messaging to be developed, the Virtual Interface Architecture (VIA).. During a similar period of time industrial researchers worked on standards for shared storage subsystems. The combination of the efforts of many researchers has resulted in the InfiniBand standard. The VIA and InfiniBand standards will likely dominate cluster network architectures in the coming years.

2.2.3.1 VIA

The Virtual Interface Architecture [9] is a communications standard that combines many of the best features of various academic projects. A consortium of academic and industrial partners, including Intel, Compaq, and Microsoft, developed the standard. Version 1.1 of VIA includes support for heterogeneous hardware and is available as of early 2001. Like U-net, VIA is based on the concept of a virtual network interface. Before a message can be sent in VIA, send and receive buffers must be allocated and pinned to physical memory locations.

No system calls are needed after the buffers and associated data structures are allocated. A send or receive operation in a user application consists of posting a descriptor to a queue. The application can choose to wait for a confirmation that the operation has completed, or can continue host processing while the message is being processed.

Several hardware vendors and some independent developers have developed VIA implementations for various network products [10][11]. VIA implementations can be classified as native or emulated. A native implementation of VIA off-loads a portion of the processing required to send and receive messages to special hardware on the network interface card. When a message arrives in a native VIA implementation, the network card performs at least a portion of the work required to copy the message into user memory. With an emulated VIA implementation, the host CPU performs the processing to send and receive messages. Although the host processor is used in both cases, an emulated implementation of VIA has less overhead than TCP/IP. However, the services provided by VIA are different than those provided by TCP/IP, since the communication may not be guaranteed to arrive reliably in VIA.

Although VIA can be used directly for application programming, the Virtual Interface Architecture is considered by many systems designers to be at too low a level for application programming. With VIA, the application must be responsible for allocating some portion of physical memory and using it effectively. It is expected that most operating system and middleware vendors will provide an interface to VIA that is suitable for application programming. For example, there is an implementation of the MPICH version of MPI [8] (Message Passing Interface), called MVICH [12] that runs over VIA. In fall, 2000, most database vendors offer a version of their product that runs over VIA, and other cluster software such as file systems that run over VIA are rapidly becoming available.

2.2.3.2 InfiniBand

A large consortium of industrial partners, including Compaq, Dell, Hewlett-Packard, IBM, Intel, Microsoft and Sun Microsystems, supports the InfiniBand standard [13]. The InfiniBand architecture replaces the shared bus that is standard for I/O on current computers with a high-speed serial, channel-based, message-passing, scalable, switched fabric. All systems and devices attach to the fabric through host channel adaptors (HCA) or target channel adaptors (TCA), as shown in Figure 2.2. A single InfiniBand link operates at 2.5Gbps point-to-point in a single direction. Data is sent in packets, and six types of transfer methods are available, including:

- Reliable and unreliable connections,
- Reliable and unreliable datagrams,
- Multicast connections, and
- Raw packets.

In addition, InfiniBand supports remote direct memory access (RDMA) read or write operations, which allows one processor to read or write the contents of memory at another processor, and also directly supports IPv6 [14] messaging for the Internet.

Since InfiniBand replaces the standard shared bus, new equipment is needed to introduce InfiniBand into the network, although this can be done incrementally. InfiniBand requires that new system software be implemented in the operating system of the host platforms, and that embedded logic be included within the enclosures and devices that attach to the fabric.

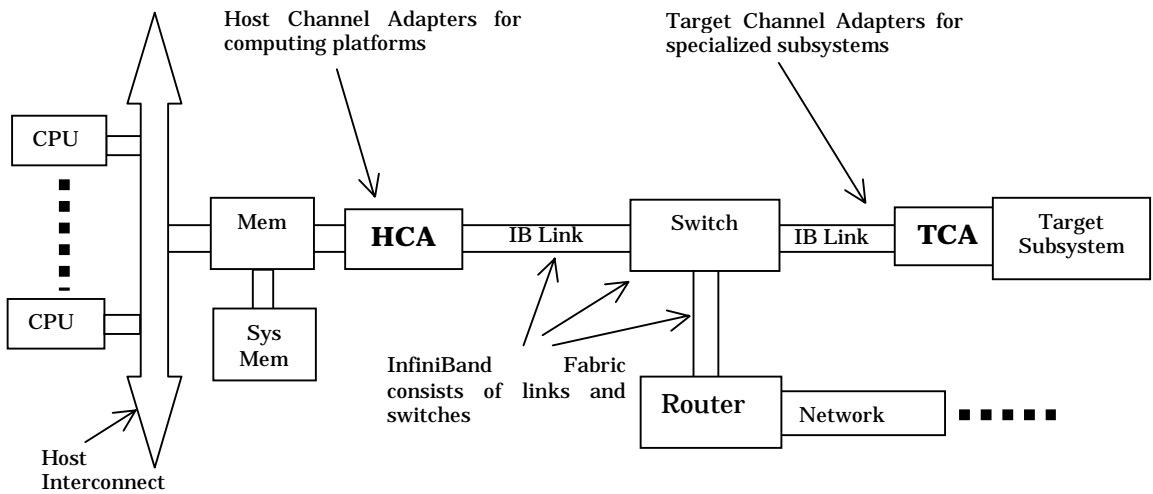

**Figure 2.2: The InfiniBand Architecture [13]**

2.3 Hardware Products

Network hardware products may be classified into four categories, depending on whether the internal connection is from the I/O bus or the memory bus, and depending on whether the communication between the computers is performed primarily using messages or using shared storage [15]. Table 1 illustrates the four types of interconnections.

|  | **Message Based** | **Shared Storage** |
|---|---|---|
| **I/O Attached** | Most common type, includes most high-speed networks; VIA, TCP/IP | Shared disk subsystems |
| **Memory Attached** | Usually implemented in software as optimizations of I/O attached message-based | Global shared memory, Distributed shared memory |

**Table 1: Categories of Cluster Interconnection Hardware**

Of the four interconnect categories, I/O attached message-based systems are by far the most common since this interface to the computer is well understood. The I/O bus provides, at least, a hardware interrupt that can inform the processor that data is waiting for it. How this interrupt is handled may vary, depending on the firmware or software stack that is used to receive the data.

I/O attached message-based systems includes all commonly-used wide-area and local-area network technologies, and includes several recent products that are specifically designed for cluster computing. I/O attached shared storage systems include computers that share a common disk sub-system. Memory attached systems are less common, since the memory bus of an individual computer generally has a design that is unique to that type of computer. However, many memory-attached systems are implemented in software or with memory-mapped I/O, such as Memory Channel [16]. Hybrid systems that combine the features of more than one category also exist, such as the Infiniband standard [13], which can be used to send data to a shared disk sub-system as well as to send messages to another computer. Similarly, SCI [17][18] may also be used for both shared memory and message passing. Networks that are designed specifically for cluster computing generally have additional hardware support, such as support for VIA, that lowers the latency in communication.

The topology of the network is generally determined by the selection of the specific network hardware. Some networks, such as Ethernet, only allow a configuration that can be viewed as a tree because of the hardware-implemented transparent routing algorithm. Other networks, such as Myrinet [19], allow a configuration with multiple redundant paths between pairs of nodes. In general, the primary factor with respect to topology that may affect the performance of the cluster is the maximum bandwidth that can be sustained between pairs of nodes in the cluster simultaneously. This is a function of both the topology and the capability of the switches in the network and depends on the specific network products that are chosen.

|  | Gigabit Ethernet | Giganet | Myrinet | Qsnet | SCI | ServerNet 2 |
|---|---|---|---|---|---|---|
| MPI Bandwidth (MBytes/s) | 35 – 50 | 105 | 140 | 208 | 80 | 65 |
| MPI Latency (µs) | 100 – 200 | 20-40 | ~18 | 5 | 6 | 20.2 |
| List price per port | $1.5K | $1.5K | $1.5K | $3.5K | $1.5K | $1.5K |
| Maximum #nodes | 1000's | 1000's | 1000's | 1000's | 1000's | 64K |
| VIA support | NT/Linux | NT/Linux | Over GM | None | Software | In hardware |
| MPI support | MPICH over MVIA, TCP | 3rd party | 3rd party | Quadrics/ Compaq | 3rd party | Compaq/ 3rd party |

**Table 2: Network Technology Comparison**

Table 2 compares several commonly used network products for high-performance clusters as of fall, 2000. Products are listed in alphabetic order across the table. Because new products are continually becoming available, this table only shows some of the most common, but is representative of the capabilities typically found for network products of this type. Note that the measured latencies and bandwidths shown here are the best published measurements available for MPI at the time of this article, but should only be used for a very rough comparison, as these numbers change depending on the configuration of the host computers and other factors. For example, the host CPU speed, bus speed, swap space, size and type of memory and cache will affect network performance. In addition, various parameters of the underlying communication protocol, such as whether or not a checksum is computed, can have an effect on network performance measurements.

2.3.1 Ethernet, Fast Ethernet, and Gigabit Ethernet

Ethernet has been the most widely used technology for local area networking. Standard Ethernet transmits at 10Mbps and has been successfully used in high-performance cluster computing [20]. However, this bandwidth is no longer sufficient for use in cluster computing since its bandwidth is not balanced compared to the computational power of the computers now available. Fast Ethernet provides 100Mbps bandwidth for very low cost, and Gigabit Ethernet [21] provides higher bandwidth for somewhat higher cost.

Both standard and many Fast Ethernet products are based on the concept of a collision domain. Within a collision domain, the communication channel is shared so that only one node can send at a time. If another node tries to send at the same time then a collision occurs, both messages are garbled, and both nodes retry their communication after a random waiting period. With a shared Ethernet or shared Fast Ethernet hub, all nodes share the same collision domain and only one node on the network can send at a time. With a switched hub, each node may be on a different collision domain, depending on the manufacturer's specifications. In this case, then two nodes can send at the same time as long as they are not both sending to the same receiver. In general, all switched networks, not just Ethernet networks, have this capability. For cluster computing, switches are preferred since the ability for more than one node to send at the same time can greatly improve overall performance.

All Gigabit Ethernet products, except for the lowest end hubs, are based on high-speed point-to-point switches in which each node is in its own separate collision domain. Gigabit Ethernet transmits signals at 1Gbps (hence the name Gigabit Ethernet) but because of the encoding required at high speeds it has a theoretical user data rate of 800Mbps, or about 100Mbytes/s. Because of the protocol overhead when using MPI, the measured user data rate of Gigabit Ethernet is approximately 50Mbytes/s with current generation computers. This rate will increase as CPU speeds continue to increase. Because of the physical nature of sending high-speed signals, Gigabit Ethernet products may have more restrictive distance and wiring requirements than Fast Ethernet products, but this is not generally a problem in a cluster environment.

Fast Ethernet products are affordable and provide adequate capacity and latency for many cluster-computing applications. Most new computers come with a built-in Fast Ethernet adapter, and small switches can be purchased for less than $30 [22]. Gigabit Ethernet products are more expensive than Fast Ethernet, but generally less expensive than other gigabit networking products.

One disadvantage of Gigabit Ethernet and Fast Ethernet products is that they are not capable of native support of the Virtual Interface Architecture (VIA) standard. However, Gigabit Ethernet and Fast Ethernet do have emulated VIA implementations that operate with lower latencies than TCP/IP [11].

2.3.2 Giganet (cLAN)

Giganet cLAN [23] was developed with the goal of supporting VIA in hardware, and was the first industry provider of a native hardware implementation of the VIA standard. Giganet began shipping the cLAN product suite for Windows NT in fall 1998 and for Linux in fall 1999. In 1999, Giganet products included PCI adapters that support 1.25Gbps high-speed cluster interconnection and 1024 virtual interfaces, switches with up to 30 ports, and a software tool for managing and testing of the cluster network. Measured bandwidth for Giganet is approximately 105Mbytes/s.

2.3.3 Myrinet

Myrinet [19] is a 1.28Gbps full duplex network supplied by Myricom. It was used in the Berkeley Network of Workstations (NOW) [3] cluster and many other academic clusters. Myrinet has a programmable on-board processor that allows for experimentation with communication protocols. Many protocols have been implemented and tested on Myrinet, including Active Messages, Generic Active Messages, Fast Messages, U-net, and BIP.

Myrinet uses switches that can be configured to provide redundant paths between nodes. The switches support cut-through routing, which allows messages to pass from end to end with minimal latency in the switches. Myrinet switches simplify network set up by automatically mapping the network configuration. Myrinet also provides "heartbeat" continuity monitoring on every link, for fault tolerance.

Flow control in done on each link in Myrinet using the concept of a slack buffer. As data is sent from one node to another, if the amount of data that is buffered in the receiving node exceeds a certain threshold, then a "stop" bit is sent to the sender to stall the transmission of more bits. As the amount of data in the buffer falls below another threshold, then a "go" bit is sent to the sender to start of the flow of bits again.

A low-latency protocol called GM is the preferred software for Myrinet, and MPI is supported over GM [8]. Efficient versions of TCP/IP and VIA are also available as a layer over GM. There are Myrinet drivers for every major processor and every major operating system, including Linux, Solaris, and Microsoft Windows. A version of the MPICH version of MPI running over BIP, called MPI-BIP [24], offers 7.6µs latency and 107Mbytes/s bandwidth across the Myrinet network, with 3.3µs latency and 150Mbytes/s bandwidth intra-node across shared memory. A version of MPI running over GM sustained a throughput of 140Mbytes/s. Sustained Myrinet bandwidth for TCP/IP and UDP has been measured at over 220Mbytes/s.

Myrinet prices are competitive with other high-performance cluster network products. In addition, by 2001, Myrinet switches with up to 32 ports are expected to be available, allowing Myrinet clusters to scale to 10,000's of nodes. In addition, its low latency, high bandwidth, and programmability make it competitive for cluster computing.

2.3.4 QsNet

QsNet [25] from Quadrics Supercomputers World Ltd. is a high bandwidth, low latency interconnection for commodity symmetric multiprocessing computers. QsNet is constructed from two QSW designed sub-systems:
- A network interface comprising one or more network adapters in each node;
- A high performance multi-rail data network that connects the nodes together.

The network interface is based on QSW's "Elan" ASIC. The Elan III integrates a dedicated I/O processor to offload messaging tasks from the main CPU, a 66Mhz 64-bit PCI interface, a QSW data link (a 400Mhz byte-wide, full duplex link), MMU, cache and local memory interface. The Elan performs three basic types of operation:
- Remote read and write;
- The direct transfer of data from a user virtual address space on one processor to another processor without requiring synchronization;
- Protocol handling.

The Elan has a programmable "Thread" processor that can generate network operations and execute code fragments to perform protocol handling without interrupting the main processor.

The data network is constructed from an 8-way cross-point switch component – the Elite III ASIC. Switches are connected in a fat tree. Two network products are available, a standalone 16-way network and a scalable switch chassis providing up to 128 ports.

QsNet provides parallel programming support via MPI, process shared memory, and TCP/IP. QsNet supports a true zero-copy (virtual-to-virtual memory) protocol, and has excellent performance. Although the current price for Qsnet is somewhat higher than for other high-performance cluster networks, it will likely drop as the product matures, making QsNet a very competitive cluster interconnection product.

### 2.3.5 ServerNet

ServerNet [26] has been a product from Tandem (now a part of Compaq) since 1995. ServerNet II offers direct support for VIA in hardware, 12-port non-blocking switches, and fat links between switches. Up to four physical links can be mapped into one logical link. Software support includes VIA drivers for NT and Linux. Although ServerNet II is a well-established project, it is only available from Compaq as packaged cluster solution, not as single components, which may limit its use in general-purpose clusters.

### 2.3.6 Scalable Coherent Interface (SCI)

SCI [17][18] was the first interconnection technology to be developed as a standard specifically for the purposes of cluster computing. To surpass the performance of bus-based systems, SCI is a point-to-point architecture with small packet sizes and split transactions while maintaining the impression of a bus-functionality from the upper layers. The IEEE 1596 standard for SCI was published in 1992 and specifies the physical layers of the network as well as higher layers for cache-coherent distributed shared memory across a network, which can optionally be cache-coherent. The standard allows serial fiber-optic links and parallel copper links on the physical layer. The bandwidth of current implementation of SCI links is located between 400 and 1000 Mbytes/s, and typical remote access latencies are little more than 2µs.

Since the lower SCI protocol layers are already meaningful for themselves, it has been possible for many commercial products to be developed using only these lower layers. Dolphin Interconnect Solutions, Inc. currently produces SCI adapter cards for IO-buses like the Sun SPARC SBus and PCI. This solution provides non-cache-coherent shared memory and is thus well suited for message passing [27][28] or efficient software-supported distributed shared memory [29][30].

At higher layers, SCI defines a distributed, pointer-based cache coherence scheme. This allows for caching of remote SCI memory: whenever data located in remote memory is modified, all cache lines on all nodes that, store this data are invalidated. Caching of remote SCI memory increases performance and allows for true, transparent shared memory programming. Examples for the implementation of cache-coherent SCI variants are the HP Convex, Sequent NUMA-Q (now IBM) and Data General AViiON NUMA systems. The IO-bus based solutions cannot implement the cache-coherence protocol since they have no access to the CPU bus to monitor memory accesses.

Next to multi-processor coupling, SCI is also used to implement I/O networks or transparently extend I/O buses like PCI: I/O address space from one bus is mapped into another one providing an arbitrary number of devices. Examples for this usage are the SGI/Cray GigaRing and Siemens' external I/O expansion for the RM600 enterprise servers [27].

Although SCI features a point-to-point architecture that makes the ring topology most natural, other (logical) topologies are possible as well using switches. Switches can be separate devices providing star – or switched-ringlet-topologies, or they can be integrated into the nodes allowing 2D- or even 3D-Torus topologies [27]. The cost of SCI components is in the range of other Gigabit-interconnects (less than $1000 per node). Switches are not needed for small clusters, but they improve performance and reliability if more than four nodes are to be used.

The Dolphin PCI-SCI adapters have support for most major operating systems (Windows, Solaris, Linux, Lynx and more), CPU architecture (x86, PPC, Alpha) and offer a unified user-level programming interface. The usual parallel programming models are available, either commercially [28], or as free software [29][30][32][33]. A native VIA interface is currently under development at [34].

SCI is an appealing cluster interconnection technology. Its performance is comparable or better than other Gigabit interconnects, in addition it enjoys an efficient user-level shared memory programming paradigm. Further information about SCI can be found in the IEEE standard and a recently published book [35].

2.3.7 ATM

ATM [36], or Asynchronous Transmission Mode, is an international network standard that grew out of the needs of the telecommunications industry. ATM is designed around the concept of cells, or small, fixed-sized packets. ATM is a switch-based technology that uses the idea of virtual circuits to deliver a connection-oriented service. ATM offers several classes of data delivery. Some classes are designed for voice and video communication, which require real-time guarantees on data delivery rates. Due to the recent consolidation of many telecommunications industry and Internet services, ATM is being increasingly used as a backbone technology in many organizations. While ATM products offer high bandwidth, in comparison to other similar technologies, it is relatively expensive.

2.3.8 Fibre Channel

Fibre Channel [37] is a set of standards that defines a high performance data transport connection technology that can transport many kinds of data at speeds up to 1Gbps, through copper wire or fiber optic cables. The Fibre Channel standard is defined in layers, beginning at the physical layer. Like the SCI standard, the standards layers are independent, enabling many companies to produce networking and cluster products based on the lower layers only. Fibre Channel is flexible, and mapping protocols have been defined which enable higher-level protocols, such as SCSI, IP, ATM, and HIPPI, to be used over Fibre Channel connections. Currently, the most frequent usage of Fibre Channel in clusters is for data storage devices. The high performance network capacity of Fibre Channel can be combined with SCSI and other I/O bridges to create a network of storage devices, separate from the primary cluster interconnect. A network of this type is sometimes called a Storage Area Network (also, SAN) and is an important subsystem of many commercial clusters. As Fibre Channel matures it may become more widely used as a general-purpose cluster interconnection technology.

2.3.9 HIPPI

HIPPI [38], or HIgh Performance Parallel Interface, is a gigabit network that was first designed for interconnecting high-performance parallel computers with network attached storage devices. When the first 800Mbps HIPPI standard was developed in the late 1980's, it was considered revolutionary. The standard has gone through several versions and now bears the name Gigabit System Network (GSN). GSN is the highest bandwidth and lowest latency interconnect standard, providing full duplex transmission rates of 6400Mbps (800 Mbytes/s) in each direction. GSN and HIPPI products are competitively priced when considering their high throughput. However, their cost prohibits their widespread usage, except in applications that require the highest performance.

2.3.10 Reflective Memory

Reflective memory [39][40] uses memory-mapped I/O to provide the illusion of a memory-attached interconnection. With reflective memory, nodes write to a local memory location on the network interface card, and the data is broadcast on the network so that it is available to all other nodes in the network. The broadcast of the data happens automatically, so that all nodes in the network have a shared view of the reflective memory. Reflective memory products have been available since the early 1990's and are available from several vendors [16][41][42]. Reflective memory is often used in a real-time fault-tolerant environment. In these environments, the replication of the data in the reflective memory gives a higher level of fault tolerance. Because each memory location has to be physically replicated on each node in the system, reflective memory products tend to be more expensive than other types of networks, and their cost grows as the size of the shared memory and the number of nodes in the system grows.

2.3.11 ATOLL

The Atoll [43], or Atomic Low Latency network, is one of the newest networks for cluster computing. Atoll has four independent network interfaces, an 8x8 crossbar switch and four link interfaces in a single chip. Message latency is as low as 4μs, and bandwidth between two nodes approaches 200 Mbytes/s. Atoll is available for Linux and Solaris operating systems and supports MPI over its own low-latency protocol. The price per node is expected to be around $700, with product release sometime in 2001.

2.4 Conclusions.

It is clear that all aspects of networks and their associated technologies are rapidly changing to meet the needs of current, new and emerging applications. Much progress has been made in the development of low-latency protocols and new standards that efficiently and effectively use network hardware components. Additionally, several types of network hardware that are specifically designed for cluster computing have been developed and made commercially available. Developers of network products are beginning to implement features such as the support of heartbeat monitoring for fault tolerance and specific protocols to deal with streaming, groups and object technologies, to fulfill the needs of specific applications.

The wide selection of available network hardware and software support makes the decision of selecting a cluster interconnection technology a difficult task. Indeed, the number of product choices appears to be increasing rather than decreasing, as some products such as ATOLL and QsNet have come onto the market in just the last couple of years. However, the competition among vendors will probably be good news for consumers who want a good product at a low cost. Network products for clusters are falling in price with respect to the price of host nodes. At the same time emerging products will increase in performance, compatibility with host nodes and operating systems, as well as having the ability to support a range of additional features. The increased competition will also be likely to force network vendors to provide more end-user support and better customer service for the products that they sell.

Multi-gigabit networks are used now with the highest-performance clusters, and will be in common usage for all types of clusters in the near future. There will be a standardisation on a small number of low-level and high-performance network protocols. These protocols will support the full range of services that are required of these new networks, including group, object, bulk, and streaming data transfer. In addition there will be more emphasis on the quality of service and reliability that a cluster network can provide.

## 2.5 Acknowledgements

We would like to thank Joachim Worringen, Salim Hariri, and Graham Fagg for their input to this paper.

# 3. Operating Systems

Steve Chapin, Syracuse University, USA and Joachim Worringen, RWTH Aachen, University of Technology, Germany

## 3.1. Introduction

Just as in a conventional desktop system, the operating system for a cluster lies at the heart of every node. Whether the user is opening files, sending messages, or starting additional processes, the operating system is omnipresent. While users may choose to use differing programming paradigms or middleware layers, the operating system is almost always the same for all users.

A generic design sketch of an operating system is given in Figure 3.1. It shows the major building blocks of a typical cluster node with the hardware resources located at the bottom, a monolithic kernel as the core of the OS, and system and user processes as well as some middleware extensions running on top of the kernel in the user space.

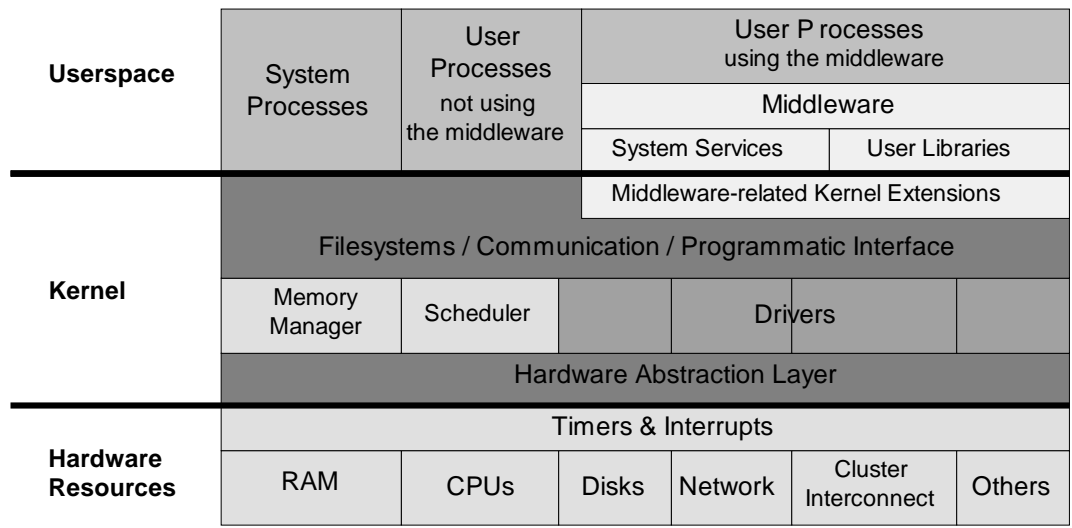

**Figure 3.1.** Generic OS design sketch for a typical cluster node

Some OS feature a distinct layer to abstract the hardware for the kernel. Porting such an OS to a different hardware requires only the need to adapt this abstraction layer. On top of this, the core functionalities of the kernel execute: memory manager, process and thread scheduler and device drivers to name the most important. Their services in turn are offered to the user processes by file systems, communication interfaces and a general programmatic interface to access kernel functionality in a secure and protected manner.

User (application) processes run in user space, along with system processes called daemons. Especially in clusters, the user processes often not only use the kernel functions, but also utilize additional functionality that is offered by middleware. This functionality is usually located in user libraries, often supported by additional system services (daemons). Some middleware extensions require extended kernel functionality, which is usually achieved by loading special drivers or modules into the kernel.

What then, is the role of the operating system in a cluster? The primary role is the same twofold task as in a desktop system: multiplex multiple user processes onto a single set of hardware components (resource management), and provide useful abstractions for high-level software (beautification). Some of these abstractions include protection boundaries, process/thread coordination and communication as well as device handling. Therefore, in the remainder of this section, we will examine the abstractions provided by current cluster operating systems, and explore current research issues for clusters.

## 3.2. Background and Overview

The ideal operating system would always help, and never hinder, the user. That is, it would help the user (which in this case is an application or middleware-designer) to configure the system for optimal program execution by supplying a consistent and well-targeted set of functions offering as many system resources as possible. After setting up the environment, it is desired to stay out of the user's way avoiding any time-consuming context switches or excessive set up of data structures for performance-sensitive applications. The most common example of this is in high-speed message passing, in which the operating system pins message buffers in DMA-able memory, and then allows the network interface card, possibly guided by user-level message calls, to transfer messages directly to RAM without operating system intervention. On the other hand, it might be desired that the operating system offer a rich functionality for security, fault-tolerance and communication facilities – which of course contrasts with the need for performance that is omnipresent.

Exactly what attributes a cluster operating system should possess is still an open question. Here is a small list of desirable features:
- *Manageability*: An absolute necessity is remote and intuitive system administration; this is often associated with a Single System Image (SSI) which can be realized on different levels, ranging from a high-level set of special scripts, perhaps controlled via Java-enabled graphical front-end, down to real state-sharing on the OS level.
- *Stability*: The most important characteristics are robustness against crashing processes, failure recovery by dynamic reconfiguration, and usability under heavy load.
- *Performance*: The performance critical parts of the OS, such as memory management, process and thread scheduler, file I/O and communication protocols should work in as efficiently as possible. The user and programmer should be able to transparently modify the relevant parameters to fine-tune the OS for his specific demands.
- *Extensibility*: The OS should allow the easy integration of cluster-specific extensions, which will most likely be related to the inter-node cooperation. This implies, at a minimum, user-loadable device drivers and profound documentation of interfaces in kernel- as well as in user-space to allow for the development of specific extensions. The best way to provide extensibility is probably the provision of the source code because it reveals all interfaces and allows for modification of existing functionality (instead of having to design replacement parts from scratch). A good example for this is the MOSIX system that is based on Linux (see chapter 4).
- *Scalability*: The scalability of a cluster is mainly influenced by the provision of the contained nodes, which is dominated by the performance characteristics of the interconnect. This includes the support of the OS to be able to use the potential performance of the interconnect by enabling low-overhead calls to access the interconnect (inter-node scalability). However, clusters are usually built with SMP nodes with an increasing number of CPUs contained in each node. The ability of the OS to benefit from these is determined by its intra-node scalability. The intra-node scalibility is dominated by the process and thread schedulers and by the degree of parallelism in the kernel that the OS allows. It also includes the resource limits that an OS inhibits, foremost the maximum size of usable address space.

- *Support*: Many intelligent and technically superior approaches in computing failed due to the lack of support in its various aspects: which tools, hardware drivers and middleware environments are available. This support depends mainly on the number of users of a certain system, which in the context of clusters is mainly influenced by the hardware costs (because usually dozens of nodes are to be installed). Additionally, support for interconnect hardware; availability of open interfaces or even open source; support or at least demand by the industry to fund and motivate research and development are important. All this leads to a user community that employs required middleware, environments and tools to, at least, enable cluster applications.
- *Heterogeneity*: Clusters provide a dynamic and evolving environment in that they can be extended or updated with standard hardware just as the user needs to or can afford. Therefore, a cluster environment does not necessarily consist of homogenous hardware. This requires that the same OS should run across multiple architectures. If such an OS dose not exist for the given hardware setup, different OS need to be used. To ensure the portability of applications, a set of standardized APIs (like [1]) should be supported by the different operating systems. The same is true for the required middleware layers, which are frequently used to enable a cluster for heterogeneous use.

It should be noted that experience shows that these goals may be mutually exclusive. For example, supplying a SSI at the operating system level, while a definite boon in terms of manageability, drastically inhibits scalability. Another example is the availability of the source code in conjunction with the possibility to extend (and thus modify) the operating system on this base. This property has a negative influence on the stability and manageability of the system: over time, many variants of the operating system will develop, and the different extensions may conflict when there is no single supplier.

## 3.3. Technological Scope

In this sub-section, we will touch on some of the aspects of operating systems for clusters that frequently lead to discussions among the developers and users in this area. Some of these aspects are covered in more detail in other sections of this paper.

### 3.3.1 Functionality: OS versus Middleware

There has been little work on operating systems specifically for clusters. Much of the work one might consider as affecting the operating system, e.g. the GLUnix [2] work at U.C. Berkeley, is actually middleware. There are good reasons for this. First, operating systems are complex, and it is not always clear how a proposed change will affect the rest of the system. Isolating these changes in middleware can make good sense. Second, the applicability of the new functionality added by such a middleware layer is usually not limited to a single operating system, but can be ported to other operating systems as well.

For example, support for Distributed Shared Memory (DSM) arises from shared address spaces, which in a conventional kernel would be inside the operating system. However, in distributed computing, DSM software such as SVMlib [3] is most often implemented in middleware, running on diverse operating systems such as Windows NT and Solaris. The advantage here is that a single middleware layer can provide services to multiple operating systems, without requiring access or changes to OS source code.

### 3.3.2 Single System Image (SSI)

Regarding the frequently desired feature of SSI, at least two variants of it should be distinguished: SSI for system administration or job scheduling purposes and SSI on a system-call level. The first is usually achieved by middleware, running daemons or services on each node delivering the required information to the administration tool or job scheduler. The latter would have to offer features like transparent use of devices located on remote nodes or using distributed storage facilities as one single standard file system. These features require extensions to current single-node operating systems.

The goal of these extensions is the cluster-wide transparent sharing of resources. Next to the sharing of I/O space via traditional network file systems or other, more sophisticated means of real parallel I/O, the sharing of RAM (Distributed Shared Memory, DSM) is an area in which a lot of research has been done. However, the usual approach to DSM is meant as a parallel programming paradigm to use shared memory between the processes of a parallel application distributed across a number of nodes. Therefore, it is mostly realized via user-level libraries [3] and not as a service offered by the OS. Integration of DSM into the OS has rarely been done [4], [6]. The reason for this is that the performance for general-purpose use (requiring strict sequential consistency) is often to low. Using relaxed consistency models improves performance, but requires that special care is taken by the user of the DSM system that prohibits offering it as a standard service by the OS. Next to using memory situated on remote nodes for DSM, some experiments [7] have been done to use it as OS-managed remote memory: current interconnect technologies such as SCI or Myrinet offer lower latencies and higher bandwidth of inter-node communication than what can be achieved between the primary storage level (RAM) and the secondary storage level (hard disk) in intra-node communication. This leads to the idea to use the memory of remote nodes instead of the local hard disk for purposes like swapping or paging. This is a promising approach, which is, however, limited by the fact that it requires permanent memory related load-imbalances inside the cluster that is not desired in environments of dedicated compute clusters.

An operating-system-level SSI implies detailed state sharing across all nodes of the cluster, and to this point, OS researchers and practitioners have been unable to scale this to clusters of significant size (more than a hundred nodes) using commodity interconnects. That does not mean that an OS-level SSI is a bad thing; for the vast majority of clusters, which have less than 32 nodes, an operating-system-level single-system image may be quite workable.

One can view Intel's Paragon OS as a cluster operating system, although the Paragon is not a cluster in the sense of this paper because it is not made of commodity components. Nevertheless, its design has much in common with todays clusters: independent nodes, stripped down to basic hardware functionality, running their own instances of an OS and communicating via a high-speed interconnect. The integration of the OS on the nodes into one SSI is something many administrators of today's clusters would like to see.

Sun's Solaris MC, on the other hand, is specifically intended for use in clusters as this paper addresses them. It provides some shared state between kernels (although much of the Solaris MC functionality is implemented at the user level). However, a weakness of current distributed operating systems is that state sharing is binary: they use all-or-nothing sharing, which inhibits scalability. The Tornado operating system [8] is designed for 2-level hierarchical clusters of workstations, although it is not yet clear how well this approach will work for more generalized clusters.

### 3.3.3 Heterogeneity

It is arguable whether one should attempt to accommodate heterogeneous hardware at the operating system level within a single cluster. There are definite efficiencies to be gained from homogeneous clusters, and it may well make economic sense to replace an existing cluster rather than doing incremental heterogeneous expansion. Even if one accepts heterogeneity as inevitable, the operating system may not be the best place to address it. What we really want is that, at some layer, we provide a homogeneous set of abstractions to higher layers.

The lowest level on which heterogeneity causes problems is the data representation – big-endian vs. little-endian. If such systems are to be connected, the adaptation of the different representations could also be done on the lowest level possible to gain suitable performance. However, approaches to do endian conversion in hardware have not yet been done (the closest thing we are aware of is the ability of the old MIPS processors to run in either endian-ness, although there was no dynamic conversion).

On the other hand, this can just as easily be at the middleware layer instead of at the operating system layer (the "end-to-end" argument in networking would argue for pushing this to the highest layer possible). Middleware systems have done an excellent job of providing the illusion of homogeneity in heterogeneous systems – consider the success of Java and the Java Virtual Machine. However, creating illusions does cost in terms of performance. Therefore, we consider an operating system that runs on heterogeneous hardware as more of a serendipitous benefit rather than a requirement.

### 3.3.4 User-level communication facilities

A key characteristic of high-performance clusters is the interconnect between the nodes normally not being an Ethernet-based network, but more sophisticated technologies like SCI [8], Myrinet [10], GigaNet [11], or other, mostly proprietary solutions. This kind of hardware offers communication bandwidth in the range of several Gbps, however, to make best use of this performance, it is required that the access to the interconnect adapter involves as little overhead as possible. Therefore, the involvement of the operating system in this kind of communication is not desired; everything is to be done in the user space, preferably by techniques like protected user-level DMA. However, it is expected that the multiplexing of an arbitrary number of processes to a single interconnect adapter is still possible in a secure manner. This imposes a difficult task to the developers of the interconnect adapter and its driver, and also to the designers of the operating system into which the driver is to be embedded. The VIA industry standard [12], as discussed in the communications section of this document, appears to be the future for low-latency, high-bandwidth cluster communication.

Approaches like VIA (and U-Net [13], an earlier design on which many parts of VIA are based) try to minimize the involvement of the OS into inter-process communication by moving as much functionality as possible from the kernel space into user space. This means the buffer and queue management for send and receive operations is done by the user process in user-space. The OS is only involved into the setup of communication channels, but no longer in the communication (depending on the network adapters capabilities). This increases performance by reducing the number of context switches and local buffer copy operations. The most radical form of this low-overhead communication is SCI, which maps memory of remote processes into the address space of the local process, reducing the inter-process, inter-node communication to simple load and store operations.

3.3.5 Optimized parallel I/O

Less work has been done on high-speed file input/output concepts than for high-speed inter-process communication. However, file I/O is crucial for the performance of many types of applications, scientific codes as well as databases. The usual way is to employ a number (one or more) of dedicated I/O nodes in the cluster. However, every shared resource represents a potential bottleneck in a system that has to be scalable. A good example for the problems involved in this is, once again, the Intel Paragon. Its I/O concept was used for traditional file input/output as well as for swapping, is organized in a complex, tree-oriented parallel manner and nevertheless did not deliver optimal performance. Clusters should follow other concepts by doing as much node-local I/O as possible to reduce inter-node communication and I/O contention, while maintaining a consistent global view of the I/O space.

Examples of this approach are PVFS [14] and PPFS [15] in which a flexible number of servers in a cluster provide general I/O services in a distributed manner. For specific I/O needs of scientific applications, specialized middleware libraries like PANDA [16], which supports efficient parallel I/O for huge multidimensional arrays, have been developed.

3.4. Current State-of-the-art

State-of-the-art can be interpreted to mean the most common solution as well as the best solution using today's technology. By far the most common solution current clusters is running a conventional operating system, with little or no special modification. This operating system is usually a Unix derivative, although NT clusters are becoming more common. We do not intend "cluster" to be synonymous with "cluster of PCs," although this is, again, the most common case, for reasons of economics.

The single most popular cluster operating system is Linux [26]. This is primarily for three reasons:
1. It is free,
2. It is an open source operating system, meaning that one is free to customize the kernel to ones liking. To date, this has usually meant specialized drivers for underlying high-speed networks, and other I/O optimizations.
3. For historic reasons: Don Becker selected Linux for the original Beowulf cluster (this was for both technical and social reasons, as the Linux kernel was free of any possible licensing problems, unlike the BSD derivatives at the time), and thus Beowulf-derived systems have also used Linux.

Beowulf [17] is not a single ready-to-run package for Linux clustering, but merely a collection of tools, middleware libraries, network drivers and kernel extensions to enable a cluster for specific HPC purposes. This projects includes valuable components and is another example of how an (open source) OS can be extended and adapted to specific needs.

Sun Microsystems has developed a multi-computer version of Solaris; aptly named Solaris MC [18] Solaris MC consists of a small set of kernel extensions and a middleware library. Solaris MC incorporates some of the research advances from Sun's Spring operating system, including an object-oriented methodology and the use of CORBA IDL in the kernel. Solaris MC provides a SSI to the level of the device, i.e. processes running on one node can access remote devices as if they were local. The SSI also extends to a global file system and a global process space.

The Puma operating system [19], from Sandia National Labs and the University of New Mexico, represents the ideological opposite of Solaris MC. Puma takes a true minimalist approach: there is no sharing between nodes, and there is not even a file system or demand-paged virtual memory. This is because Puma runs on the "compute partition" of the Intel Paragon and Tflops/s machines, while a full-featured OS (e.g. Intel's TflopsOS or Linux) runs on the Service and I/O partitions. The compute partition is focused on high-speed computation, and Puma supplies low-latency, high-bandwidth communication through its Portals mechanism.

MOSIX [20],[21] is a set of kernel extensions for Linux that provides support for seamless process migration. Under MOSIX, a user can launch jobs on their home node, and the system will automatically load balance the cluster and migrate the jobs to lightly-loaded nodes. MOSIX maintains a single process space, so the user can still track the status of their migrated jobs. MOSIX offers a number of different modes in which available nodes form a cluster, submit and migrate jobs, ranging from a closed batch controlled system to a open network-of-workstation like configuration. MOSIX is a mature system, growing out of the MOS project and having been implemented for seven different operating systems/architectures. MOSIX is free and is distributed under the GNU Public License.

Next to the cluster operating systems mainly used in research and for running computationally intensive applications (which do not require a high degree of OS-support), clustering is also in use in the commercial arena. The main commercial applications in this area are data-intensive and often involve database management systems (DBMS). AIX from IBM has a strong position here, running the SP family of cluster-based servers, featuring proven stability, good manageability and a number of clustered databases. From the other commercial Unix variants, Sun's Solaris has a strong focus on clustering, high availability (HA) and is also widely used in research. IRIX from SGI relies a sophisticated NUMA-SMP technology that provides a very specific kind of clustering. Other operating systems with an emphasis on HA (Tandem Non-Stop Unix and Unixware 7) are covered in the related section of this paper.

In the area of high-performance I/O, only a very limited number of proven and usable solutions exist. One of these is GPFS (General Parallel File System) [22] by IBM, specifically designed for the SP server running AIX. It is based on storage arrays (disks) being connected to one or more servers that exchange data with the clients via multithreaded daemons. It fulfils the criterion "usable" in that it features a kernel extension that can be accessed like any standard UNIX file system. This avoids the necessity of recompiling or even modifying applications that are to use GPFS. However, it performs best for large sequential reads and writes due to its technique of striping the data across the disks. Many technically more sophisticated but less general (and thus mostly less usable in a general cluster environment) research projects exist [23],[24] which often deal with the special, but frequently occurring scenario of collective I/O in large scientific applications.

We present a comparison of the most relevant operating systems used for clustering in Appendix A. We are aware that already the selection criteria "most relevant", and also the comparison of the selected systems will lead to discussions. However, covering every aspect of each system is surely beyond the scope of this paper, and a stimulation of the discussion is a desired effect. Apart from our own experiences, we consider the studies that have been performed by D.H. Brown Associates [25],[26].

## 3.5. Future work

As SMP become more common in clusters, we will see a natural hierarchy arise. SMP have tight coupling, and will be joined into clusters via low-latency high-bandwidth interconnection networks. Indeed, we fully expect that heterogeneous clusters of SMP will arise, having single, dual, quad, and octo-processor boxes in the same cluster. These clusters will, in turn, be joined by gigabit-speed wide-area networks, which will differ from SAN primarily in their latency characteristics. This environment will naturally have a three-level hierarchy, with each level having an order of magnitude difference in latency relative to the next layer.

This structure will have its greatest impact on scheduling, both in terms of task placement and in terms of selecting a process to run from a ready queue. Scheduling is traditionally considered an operating system activity, yet it is quite likely that at least some of this work will be carried out in middleware.

For example, as one of our research projects we are investigating thread management for mixed-mode (multi-threaded and message passing) computing using OpenMP and MPI, which we believe is a natural by product of clusters of SMP. Most cluster applications, particularly scientific computations traditionally solved via spatial decomposition, consist of multiple cooperating tasks. During the course of the computation, hot spots will arise, and a self-adapting program might wish to manage the number of active threads it has in each process. Depending on the relationship of new threads to existing threads (and their communication pattern) and the system state, a decision might be made to do one of the following:
- Add a new thread on an idle processor of the SMP where the process is already running.
- Expand the address space via distributed shared memory to include an additional node, and add a thread there.
- Add a thread to a non-idle processor already assigned to the process.
- Migrate the process to a larger SMP (e.g. from 2 nodes to 4 nodes) with an idle processor, and add a new thread there.

The described technique of thread placing and process migration is also related to the high-availability issue that is critical for commercial applications. Automatic and fast migration of running processes, taking benefit from advanced interconnect technology offering e.g. transparent remote memory access, will lift the definition of fail-over times into new dimensions.

We might also examine the issue of configurability. Users might want to alter the personality of the local operating system, e.g. "strip down" to a Puma-like minimalist kernel to maximize the available physical memory and remove undesired functionality. Possible mechanisms to achieve this range from a reload of a new kernel and a reboot to dynamically linking/unlinking code into/out of the kernel. This leads to the question: "How much (and which) functionality does a cluster operating-system need?" The more functionality a system has, the more complicated it gets to maintain and the greater the chance for malfunctions due to bad configurations or errors in the interaction of different parts of the system. Again, Linux is the easiest way for the majority of researchers to study this area.

Another important aspect of cluster computing in which the operating system is strongly involved is distributed file I/O. Current solutions of I/O systems are mostly static, do not adapt very well to the actual workload and thus tend to have bottlenecks, mostly by the limited number of dedicated I/O nodes to which all data has to be routed. The transparent filesystem-level support of distributed and adaptive I/O is an open issue for cluster operating systems. As an example for an attempt, we are currently implementing an MPI-IO implementation that operates on the basis of an existing DSM library on top of an SCI interconnect. This technique may result in a dedicated filesystem for high-bandwidth, low-latency requirements that is totally distributed among the participating clients.

3.6. Conclusions

Cluster operating systems are similar in many ways to conventional workstation operating systems. How different one chooses to make the operating system depends on ones view of clustering. On the one hand, we have those who argue that each node of a cluster *must* contain a full-featured operating system such as Unix, with all the positives and negatives that implies. At the other extreme, we see researchers asking the question, "Just how much can I remove from the OS and have it still be useful?" These systems are typified by the work going on in the Computational Plant project at Sandia National Laboratories. Still others are examining the possibility of on-the-fly adaptation of the OS layer, reconfiguring the available services through dynamic loading of code into the cluster operating system.

It is worth noting that every notable attempt to provide SSI at the OS layer has been regarded as a failure on some level. Sun's Solaris MC has never been offered as a product, and recently sun has approached academic computer scientists to evaluate Linux on Sparc stations as the basis of a cluster product. Intel's Paragon OS is well known for its tendency to lock up the entire machine because of minor problems on one node, as well as its wretched performance on large systems. We are not saying it is impossible to build a scalable SSI at the OS level, we are just saying that no one has done it, and we think there is a good reason. The forthcoming SIO standard will blur the edges between remote and local devices, and perhaps this will lead to more highly scalable SSI.

A final note is that, through a combination of OS improvements and acquisition of relevant technology, Microsoft has become a viable option in the realm of clustering. The HPVM [27] effort has demonstrated that it is possible to build a reasonable cluster from Windows NT boxes, and with the recent installation of clusters such as the one at Cornell University, NT or Windows 2000 is going to be a factor in the cluster OS picture for the foreseeable future.

To sum it up, clusters for technical and scientific computing based on Linux and other standalone Unix platforms like AIX are here, and they work. In the area of commercial cluster computing, Linux still lacks essential functionalities which "conventional" Unix systems and in parts even Windows NT do offer. It is to be observed if the free and distributed development model of Linux will be able to offer proven solutions in this area, too, since these topics are rarely addressed in the Linux developer community. Nevertheless, more exotic OS technology is and will be the current focus of many research efforts, both academic and commercial. There will probably never be "THE" cluster OS, as Linux will adopt research results much more quickly than commercial vendors, particularly Microsoft, if history is a reliable guide.

# 4. Single System Image (SSI)

Rajkumar Buyya, Monash University, Australia, Toni Cortes, Universitat Politecnica de Catalunya, Spain and Hai Jin, University of Southern California, USA

## 4.1 Introduction

A Single System Image (SSI) is the property of a system that hides the heterogeneous and distributed nature of the available resources and presents them to users and applications as a single unified computing resource. SSI can be enabled in numerous ways, these range from those provided by extended hardware through to various software mechanisms. SSI means that users have a globalised view of the resources available to them irrespective of the node to which they are physically associated. Furthermore, SSI can ensure that a system continues to operate after some failure (high availability) as well as ensuring that the system is evenly loaded and providing communal multiprocessing (resource management and scheduling).

SSI design goals for cluster-based systems are mainly focused on complete transparency of resource management, scalable performance, and system availability in supporting user applications [1][2][3][5][7]. A SSI can be defined as the illusion [1][2], created by hardware or software, that presents a collection of resources as one, more powerful unified resource.

## 4.2 Services and Benefits

The key services of a single-system image cluster include the following [1][3][4]:
o *Single entry point:* A user can connect to the cluster as a virtual host (like telnet beowulf.myinstitute.edu), although the cluster may have multiple physical host nodes to serve the login session. The system transparently distributes user's connection requests to different physical hosts to balance load.
o *Single user interface:* The user should be able to use the cluster through a single GUI. The interface must have the same look and feel than the one available for workstations (e.g., Solaris OpenWin or Windows NT GUI).
o *Single process space:* All user processes, no matter on which nodes they reside, have a unique cluster-wide process id. A process on any node can create child processes on the same or different node (through a UNIX fork). A process should also be able to communicate with any other process (through signals and pipes) on a remote node. Clusters should support globalised process management and allow the management and control of processes as if they are running on local machines.
o *Single memory space: Users* have an illusion of a big, centralised main memory, which in reality may be a set of distributed local memories. Software DSM approach has already been used to achieve single memory space on clusters. Another approach is to let the compiler distribute the data structure of an application across multiple nodes. It is still a challenging task to develop a single memory scheme that is efficient, platform-independent, and able to support sequential binary codes.
o *Single I/O space (SIOS):* This allows any node to perform I/O operations on local or remotely located peripheral or disk device. In this SIOS design, disks associated to cluster nodes, network-attached RAIDs, and peripheral devices form a single address space.
o *Single file hierarchy:* On entering into the system, the user sees a single, huge file-system image as a single hierarchy of files and directories under the same root directory that transparently integrates local and global disks and other file devices. Examples of single file hierarchy include NFS, AFS, xFS, and Solaris MC Proxy.

- o *Single virtual networking:* This means that any node can access any network connection throughout the cluster domain even if the network is not physically connect to all nodes in the cluster. Multiple networks support a single virtual network operation.
- o *Single job-management system:* Under a global job scheduler, a user job can be submitted from any node to request any number of host nodes to execute it. Jobs can be scheduled to run in either batch, interactive, or parallel modes. Examples of job management systems for clusters include GLUnix, LSF, and CODINE.
- o *Single control point and management:* The entire cluster and each individual node can be configured, monitored, tested and controlled from a single window using single GUI tools, much like an NT workstation managed by the Task Manger tool.
- o *Checkpointing and Process Migration:* Checkpointing is a software mechanism to periodically save the process state and intermediate computing results in memory or disks. This allows the roll back recovery after a failure. Process migration is needed in dynamic load balancing among the cluster nodes and in supporting checkpointing.

Figure 4.1 shows the functional relationships among various key middleware packages. These middleware packages are used as interfaces between user applications and cluster hardware and OS platforms. They support each other at the management, programming, and implementation levels.

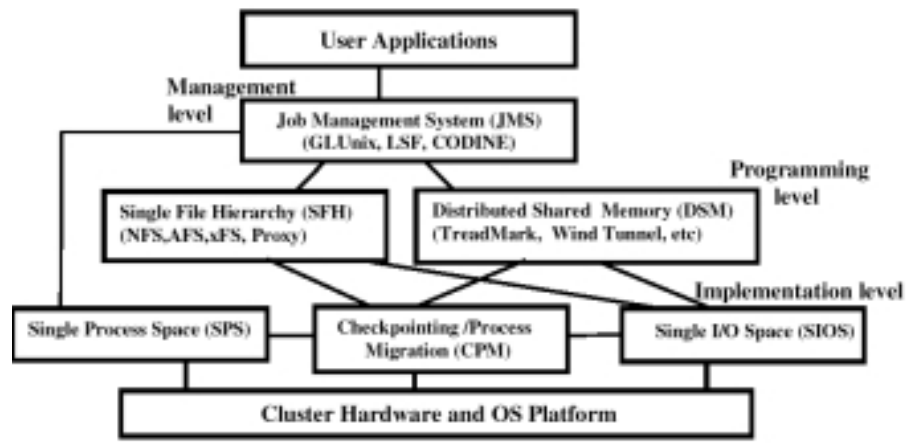

Figure 4.1. The relationship between middleware modules [3].

The most important benefits of SSI include the following [1]:
- o It provides a simple, straightforward view of all system resources and activities, from any node in the cluster.
- o It frees the end-user from having to know where in the cluster an application will run.
- o It allows the use of resources in a transparent way irrespective of their physical location.
- o It lets the user work with familiar interface and commands and allows the administrator to manage the entire cluster as a single entity.
- o It offers the same command syntax as in other systems and thus reduces the risk of operator errors, with the result that end-users see an improved performance, reliability and higher availability of the system.
- o It allows to centralise/decentralise system management and control to avoid the need of skilled administrators for system administration.
- o It greatly simplifies system management and thus reduced cost of ownership.
- o It provides location-independent message communication.

- It benefits the system programmers to reduce the time, effort and knowledge required to perform task, and allows current staff to handle larger or more complex systems.
- It promotes the development of standard tools and utilities.

## 4.3 SSI Layers/Levels

The two important characteristics of SSI [1][2] are:
1. Every SSI has a boundary,
2. SSI support can exist at different levels within a system — one able to be built on another.

SSI can be implemented in one or more of the following levels:
- Hardware,
- Operating System (so called *underware* [5]),
- Middleware (runtime subsystems),
- Application.

A good SSI is usually obtained by a co-operation between all these levels as a lower level can simplify the implementation of a higher one.

### 4.3.1 Hardware Level

Systems such as Digital/Compaq Memory Channel [8] and hardware Distributed Shared Memory (DSM) [8] offer SSI at hardware level and allow the user to view a cluster as a shared-memory system. Digital's Memory Channel is designed to provide a reliable, powerful and efficient clustering interconnect. It provides a portion of global virtual shared memory by mapping portions of remote physical memory as local virtual memory (called reflective memory).

Memory Channel consists of two components: a PCI adapter and a hub. Adapters can also be connected directly to another adapter without using a hub. The host interfaces exchange heartbeat signals and implement flow control timeouts to detect node failure or blocked data transfers. The link layer provides error detection through a 32 bit CRC generated and checked in hardware. Memory Channel uses point-to-point, full-duplex switched 8x8 crossbar implementation.

To enable communication over the Memory Channel network, applications map pages as read- or write-only into their virtual address space. Each host interface contains two page control tables (PCT), one for write and one for read mappings. For read-only pages, a page is pinned down in local physical memory. Several page attributes can be specified: receive enable, interrupt on receive, remote read etc. If a page is mapped as write-only, a page table entry is created for an appropriate page in the interface 128 Mbytes of PCI address space. Page attributes can be used to store a local copy of each packet, request acknowledgement message from receiver side for each packet, define the packets as broadcast or point-to-point packets.

Broadcasts are forwarded to each node attached to the network. If a broadcast packet enters a crossbar hub, the arbitration logic waits until all output ports are available. Nodes, which have mapped the addressed page as a readable area, store the data in their local pinned down memory region. All other nodes simply ignore the data. Therefore once the data regions are mapped and set up, simple store instructions transfer data to remote nodes, without OS intervention.

Besides this basic data transfer mechanism, Memory Channel supports a simple remote read primitive, a hardware-based barrier acknowledge, and a fast lock primitive. To ensure correct behaviour, Memory Channel implements a strict in-order delivery of written data. A write invalidates cache entries on the reader side, thus providing cluster-wide cache coherence.

Digital provides two software layers for Memory Channel: the Memory Channel Services and Universal Message Passing (UMP). The first is responsible for allocating and mapping individual memory page. UMP implements a user-level library of basic message passing mechanisms. It is mainly used as a target for higher software layers, such as MPI, PVM or HPF. Both layers have been implemented for the Digital UNIX and the Windows NT operating systems.

Memory Channel reduces communication to the minimum, just simple store operations. Therefore, latencies for single data transfers are very low. This also enables the Memory Channel to reach the maximal sustained data rate of 88 Mbytes/s with relative small data packets of 32 bytes. The largest possible configuration consists out of 8 12-CPU Alpha server nodes, resulting in a 96-CPU cluster.

4.3.2 Operating System Level

Cluster operating systems support an efficient execution of parallel applications in an environment shared with sequential applications. A goal is to pool resources in a cluster to provide better performance for both sequential and parallel applications. To realise this goal, the operating system must support gang scheduling of parallel programs, identify idle resources in the system (such as processors, memory, and networks), and offer globalised access to them. It should optimally support process migration to provide dynamic load balancing as well as fast inter-process communication for both the system and user-level applications. The OS must make sure these features are available to the user without the need for additional system calls or commands. OS kernel supporting SSI include SCO UnixWare NonStop Clusters [5][6], Sun Solaris-MC [9], GLUnix [11], and MOSIX [12].

SCO UnixWare

UnixWare NonStop Clusters is SCO's high availability software. It significantly broadens hardware support making it easier and less expensive to deploy the most advanced clustering software for Intel systems. It is an extension to the UnixWare operating system where all applications run better and more reliably inside a Single System Image (SSI) environment that removes the management burden. It features standard IP as the interconnect, removing the need for any proprietary hardware.

The UnixWare kernel has been modified via a series of modular extensions and hooks to provide: single cluster-wide file-system view; transparent cluster-wide device access; transparent swap-space sharing; transparent cluster-wide IPC; high performance inter-node communications; transparent cluster-wide process migration; node down cleanup and resource failover; transparent cluster-wide parallel TCP/IP networking; application availability; cluster-wide membership and cluster time-sync; cluster system administration; and load leveling.

UnixWare NonStop Clusters architecture offers built-in support for application failover using an "n + 1" approach. With this approach, the backup copy of the application may be restarted on any of several nodes in the cluster. This allows one node to act as a backup node for all other cluster nodes.

UnixWare NonStop Clusters also supports active process migration, which allows any application process to be moved to another node between instruction steps. This allows continuation without disruption to the application. Active process migration allows dynamic removal and addition of nodes within the cluster.

With the Single System Image (SSI) capability of UnixWare NonStop Clusters, both applications and users view multiple nodes as a single, logical system. SSI also provides automatic process migration and dynamic load balancing. Depending on the workload and available resources in the cluster, the system automatically reassigns processes among available nodes, delivering optimal overall performance. The cluster offers a single UNIX system name space and appears to the application as a very large n-way SMP server. The cluster services maintain the standard service call interface, so upper levels of the operating system do not need to be changed. Applications access clustered services through standard UNIX system libraries, which in turn access clustered services through the service-call interface. Applications do not need to be cluster aware and may run unmodified in the cluster.

The cluster service determines whether a request can be handled locally or must be forwarded to another node. If the request is passed to another node, it uses an internode communication system over ServerNet to communicate to the service peer on another node. The request is then handled by the standard UNIX system service on the targeted node.

Sun Solaris MC

Solaris MC is a prototype extension of the single node Solaris Kernel. It provides single system image and high availability at the kernel level. Solaris MC is implemented through object-oriented techniques. It extensively uses the object-oriented programming language C++, the standard COBRA object model and its Interface Definition Language.

Solaris MC uses a global file system called Proxy File System (PXFS). The main features include single system image, coherent semantics, and high performance. The PXFS makes file accesses transparent to process and file locations. PXFS achieves this single system image by intercepting file-access operations at the vnode/VFS interface. When a client node performs a VFS/vnode operation, Solaris MC proxy layer first converts the VFS/vnode operation into an object invocation, which is forwarded to the node where the file resides (the server node). The invoked object then performs a local VFS/vnode operation on the Solaris file system of the server node. This implementation approach needs no modification of the Solaris kernel or the file system.

PXFS uses extensive caching on the clients to reduce remote object invocations. PXFS uses a token-based coherency protocol to allow a page to be cached read-only by multiple nodes or read-write by a single node.

Solaris MC provides a single process space by adding a global process layer on top of Solaris kernel layer. There is a node manager for each node and a virtual process (vproc) object for each local process. The vproc maintains information of the parent and children of each process. The node manager keeps two lists: the available node list and local process list, including migrated ones. When a process migrates to another node, a shadow vproc is still kept on the home node. Operations received by the shadow vproc are forwarded to the current node where the process resides.

Solaris MC provides a single I/O subsystem image with uniform device naming. A device number consists of the node number of the device, as well as the device type and the unit number. A process can access any device by using this uniform name as if it were attached to the local node, even if it is attached to a remote node.

Solaris MC ensures that existing networking applications do not need to be modified and see the same network connectivity, regardless of which node the application runs on. Network services are accessed through a service access point (SAP) server. All processes go to the SAP server to locate in which node a SAP is on. The SAP server also ensures that the same SAP is not simultaneously allocated to different nodes. Solaris MC allows multiple nodes to act as replicated SAP server for network services.

GLUnix

Another way for the operating system to support a SSI is to build a layer on top of the existing operating system and to perform global resource allocations. This is the approach followed by GLUnix from Berkeley [11]. This strategy makes the system easily portable and reduces development time.

GLUnix is an OS layer designed to provide support for transparent remote execution, interactive parallel and sequential jobs, load balancing, and backward compatibility for existing application binaries. GLUnix is a multi-user system implementation at the user level so that it can be easily ported to a number of different platforms. It is built as a protected, user-level library using the native system services as a building block. GLUnix aims to provide cluster-wide namespace and uses Network PIDs (NPIDs) and Virtual Node Numbers (VNNs). NPIDs are globally unique process identifiers for both sequential and parallel programs throughout the system. VNNs are used to facilitate communications among processes of a parallel program. A suite of user tools for interacting and manipulating NPIDs and VNNs are supported.

GLUnix is implemented completely in the user level and does not need any kernel modification, therefore it is easy to implement. GLUnix relies on a minimal set of standard features from the underlying system, which are present in most commercial operating systems. So it is portable to any operating system that supports inter-process communication, process signalling, and access to load information. The new features needed for clusters are invoked by procedure calls within the application's address space. There is no need to cross hardware protection boundary, no need for kernel trap or context switching. The overhead for making system calls is eliminated in GLUnix. Using shared-memory segments or interprocess communication primitives can do the co-ordination of the multiple copies of GLUnix, which are running on multiple nodes.

The main features provided by GLUnix include: co-scheduling of parallel programs; idle resource detection, process migration, and load balancing; fast user-level communication; remote paging; and availability support.

MOSIX

MOSIX [12] is another software package specifically designed to enhance the Linux kernel with cluster computing capabilities. The core of MOSIX are adaptive (on-line) load-balancing, memory ushering and file I/O optimisation algorithms that respond to variations in the use of the cluster resources, e.g., uneven load distribution or excessive disk swapping due to lack of free memory in one of the nodes. In such cases, MOSIX initiates process migration from one node to another, to balance the load, or to move a process to a node that has sufficient free memory or to reduce the number of remote file I/O operations.

MOSIX operates silently and its operations are transparent to the applications. This means that you can execute sequential and parallel applications just like you would do in an SMP. You need not care about where your process is running, nor be concerned what the other users are doing. Shortly after the creation of a new process, MOSIX attempts to assign it to the best available node at that time. MOSIX then continues to monitor the new process, as well as all the other processes, and will move it among the nodes to maximise the overall performance. All this is done without changing the Linux interface. This means that you continue to see (and control) all your processes as if they run on your node.

The algorithms of MOSIX are decentralised – each node is both a master for processes that were created locally, and a server for (remote) processes, that migrated from other nodes. This means that nodes can be added or removed from the cluster at any time, with minimal disturbances to the running processes. Another useful property of MOSIX is its monitoring algorithms, which detect the speed of each node, its load and free memory, as well as IPC and I/O rates of each process. This information is used to make near optimal decisions where to place the processes.

The system image model of MOSIX is based on the home-node model, in which all the user's processes seem to run at the user's login-node. Each new process is created at the same site(s) as its parent process. Processes that have migrated interact with the user's environment through the user's home-node, but where possible, use local resources. As long as the load of the user's login-node remains below a threshold value, all the user's processes are confined to that node. However, when this load rises above a threshold value, then some processes may be migrated (transparently) to other nodes.

The Direct File System Access (DFSA) provision extends the capability of a migrated process to perform some I/O operations locally, in the current node. This provision reduces the need of I/O-bound processes to communicate with their home node, thus allowing such processes (as well as mixed I/O and CPU processes) to migrate more freely among the cluster's node, e.g., for load balancing and parallel file and I/O operations.

Currently, the MOSIX File System (MFS) meets the DFSA standards. MFS makes all directories and regular files throughout a MOSIX cluster available from all nodes, while providing absolute consistency as files are viewed from different nodes, i.e., the consistency is as if all file accesses were done on the same node.

4.3.3 Middleware Level

Middleware, a layer that resides between OS and applications, is one of the common mechanisms used to implement SSI in clusters. They include cluster file system, programming environments such as PVM [13], job-management and scheduling systems such as CODINE [14] and Condor [15], cluster-enabled Java Virtual Machine (JVM) such as JESSICA [18]. SSI offered by cluster file systems makes disks attached to cluster nodes appear as a single large storage system, and ensure that every node in the cluster has the same view of the data. Global job scheduling systems manage resources, and enable the scheduling of system activities and execution of applications while offering high availability services transparently. Cluster enabled JVM allows execution of Java threads-based applications on clusters without any modifications.

CODINE is a resource-management system targeted to optimise utilisation of all software and hardware resources in a heterogeneous networked environment [14]. It is evolved from the Distributed Queuing System (DQS) created at Florida State University. The easy-to-use graphical user interface provides a single-system image of all enterprise-wide resources for the user and also simplifies administration and configuration tasks.

The CODINE system encompasses four types of daemons. They are the CODINE master daemon, the scheduler daemon, the communication daemons and the execution daemons. The CODINE master daemon acts as a clearinghouse for jobs and maintains the CODINE database. Periodically the master receives information about the host load values in the CODINE cluster by the CODINE execution daemons running on each machine. Jobs, submitted to the CODINE system, are forwarded to the CODINE master daemon and then spooled to disk. The scheduler daemon is responsible for the matching of pending jobs to the available resources. It receives all necessary information from the CODINE master daemon and returns the matching list to the CODINE master daemon which in turn dispatches jobs to the CODINE execution daemons.

CODINE master daemon runs on the main server and manages the entire CODINE cluster. It collects all necessary information, maintains and administers the CODINE database. The database contains information about queues, running and pending jobs and the available resources in the CODINE cluster. Information to this database is periodically updated by the CODINE execution daemons.

The CODINE master daemon has a critical function, as the system will not operate without this daemon running. To eliminate this potential point of failure, CODINE provides a shadow master functionality. Should the CODINE master daemon fail to provide service, a new CODINE master host will be selected and another CODINE master daemon will automatically be started on that new host. All CODINE ancillary programs providing the user or administrator interface to CODINE directly contacts the CODINE master daemon via a standard TCP port to forward their requests and to receive an acknowledgement or the required information.

The CODINE scheduler daemon is responsible for the mapping of jobs to the most suitable queues. Jobs are submitted to the CODINE master daemon together with a list of requested resources. A job that cannot be dispatched immediately waits in a pending queue until the CODINE scheduler daemon decides the requested resources for this job are available. The result of the mapping is communicated back to the CODINE master daemon to update the database. The CODINE master daemon notifies the CODINE execution daemon on the corresponding machine to start the job.

The CODINE execution daemon runs on every machine in the CODINE cluster where jobs can be executed. It reports periodically the status of the resources on the workstation to the CODINE master daemon. The CODINE execution daemon is also responsible for starting and stopping the jobs. For each job, the CODINE execution daemon starts a subordinate shepherd process, which is responsible for running and monitoring its job.

One or more CODINE communication daemons have to run in every CODINE cluster These daemons are responsible for the communication between the other CODINE daemons. This allows asynchronous communication between the various CODINE daemons, speeding up the system and increasing efficiency. The communication daemons control the whole communication via a standard TCP port.

### 4.3.4 Application Level

Finally, applications can also support SSI. The application-level SSI is the highest and, in a sense, most important because this is what the end-user sees. At this level, multiple co-operative components of an application are presented to the user as a single application. For instance, a GUI based tool such as PARMON [15] offers a single window representing all the resources or services available. The Linux Virtual Server [17] is a scalable and high availability server built on a cluster of real servers. The architecture of the cluster is transparent to end-users as all they see a single virtual server. All other cluster-aware scientific and commercial applications (developed using APIs such as MPI) hide the existence of multiple interconnected computers and co-operative software components, and present themselves as if running on a single system.

The Linux Virtual Server (LVS) [17] directs network connections to the different servers according to scheduling algorithms and makes parallel services of the cluster to appear as a virtual service on a single IP address. Linux Virtual Server extends the TCP/IP stack of Linux kernel to supports three IP load balancing techniques: NAT, IP tunneling, and direct routing. It also provides four scheduling algorithms for selecting servers from cluster for new connections: round-robin, weighted round-robin, least-connection and weighted Least-Connection. Client applications interact with the cluster as if it were a single server. The clients are not affected by interaction with the cluster and do not need modification. Scalability is achieved by transparently adding or removing a node in the cluster. High availability is provided by detecting node or daemon failures and reconfiguring the system appropriately.

Linux Virtual Server is a three-tier architecture.
- *Load Balancer* is the front end to the service as seen by the outside world. The load balancer directs network connections from clients who know a single IP address for services, to a set of servers that actually perform the work.
- *Server Pool* consists of a cluster of servers that implement the actual services, such as Web, Ftp, mail, DNS, and so on.
- *Backend Storage* provides the shared storage for the servers, so that it is easy for servers to keep the same content and provide the same services.

The load balancer handles incoming connections using IP load balancing techniques. Load balancer selects servers from the server pool, maintains the state of concurrent connections and forwards packets, and all the work is performed inside the kernel, so that the handling overhead of the load balancer is low. The load balancer can handle much larger number of connections than a general server, therefore a load balancer can schedule a large number of servers and it will not be a bottleneck of the whole system.

Cluster monitor daemons run on the load balancer to monitor the health of server nodes. If a server node cannot be reached by ICMP ping or there is no response of the service in the specified period, the monitor will remove or disable the server in the scheduling table of the load balancer. The load balancer will not schedule new connections to the failed one and the failure of server nodes can be masked.

In order to prevent the load balancer from becoming a single-point-of-failure, a backup of the load balancer is set-up. Two heartbeat daemons run on the primary and the backup, they heartbeat the health message through heartbeat channels such as serial line and UDP periodically. When the heartbeat daemon on the backup cannot hear the health message from the primary in the specified time, it will use ARP spoofing to take over the virtual IP address to provide the load balancing service. When the primary recovers from its failure, then the primary becomes the backup of the functioning load balancer, or the daemon receives the health message from the primary and releases the virtual IP address, and the primary will take over the virtual IP address. The failover or the take-over of the primary will cause the established connection in the state table lost, which will require the clients to send their requests again.

4.3.5 Pros and Cons of Each Level

Each level of SSI has its own pros and cons. The hardware-level SSI can offer the highest level of transparency, but due to its rigid architecture, it does not offer the flexibility required during the extension and enhancement of the system. The kernel-level approach offers full SSI to all users (application developers and end-users). However, kernel-level cluster technology is expensive to develop and maintain, as its market-share is/will be probably limited and it is difficult to keep pace with technological innovations emerging into mass-market operating systems.

An application-level approach helps realise SSI partially and requires that each application be developed as SSI-aware separately. A key advantage of application-level SSI compared to the kernel-level is that it can be realised in stages and the user can benefit from it immediately. Whereas, in the kernel-level approach, unless all components are specifically developed to support SSI, cannot be used or released to the market. Due to this, kernel-level approach appears as a risky and economically non-viable approach. The middleware approach is a compromise between the other two mechanisms used to provide SSI. In some cases, like in PVM, each application needs to be implemented using special APIs on a case-by-case basis. This means, there is a higher cost of implementation and maintenance, otherwise the user cannot get any benefit from the cluster. The arguments on the, so-called, "underware" versus middleware level of SSI are presented in [5].

4.4 Conclusions

SSI can greatly enhance the acceptability and usability of clusters by hiding the physical existence of multiple independent computers by presenting them as a single, unified resource. SSI can be realised either using hardware or software techniques, each of them have their own advantages and disadvantages. The middleware approach appears to offer an economy of scale compared to other approaches although it cannot offer full SSI like the OS approach. In any case, the designers of software (system or application) for clusters must always consider SSI (transparency) as one of their important design goals in addition to scalable performance and enhanced availability.

# 5. Middleware

Mark Baker, University of Portsmouth, UK, and Amy Apon, University of Arkansas, USA.

## 5.1 Introduction

Middleware is generally considered the layer of software sandwiched between the operating system and applications. Middleware provides various services required by an application to function correctly. Middleware has been around since the 1960's. More recently, it has re-emerged as a means of integrating software applications running in a heterogeneous environment. There is large overlap between the infrastructure that it provides a cluster with high-level Single System Image (SSI) services and those provided by the traditional view of middleware. A definition of SSI can be found in Section 4., whereas in this section of the paper we are concerned with middleware, and it can be described as the software that resides above the kernel and the network and provides services to applications.

## 5.2. Background and Overview

Heterogeneity can arise in at least two ways in a cluster environment. First, as clusters are typically built from commodity workstations, the hardware platform can become heterogeneous. As the cluster is incrementally expanded using newer generations of the same computer product line, or even using hardware components that have a very different architecture, problems related to these differences are introduced. For example, a typical problem that must be resolved in heterogeneous hardware environments is the conversion of numeric values to the correct byte ordering. A second way that clusters become heterogeneous is the requirement to support very different applications. Examples of this include applications that integrate software from different sources, or require access to data or software outside the cluster. In addition, a requirement to develop applications rapidly can exacerbate the problems inherent with heterogeneity. Middleware has the ability to help the application developer overcome these problems. In addition, middleware also provides services for the management and administration of a heterogeneous system.

## 5.3. Technological Scope

In this section, we briefly describe a range of technologies that are being used as middleware.

### 5.3.1 Message-based Middleware

Message-based middleware is a technology that uses a common communications protocol to exchange data between applications. The communications protocol hides many of the low-level message passing primitives from the application developer. Message-based middleware software can pass messages directly between applications, send messages via software that queues waiting messages, or use some combination of the two.

Examples of message-based middleware include the three upper layers of the OSI model [1] (i.e., the session, presentation and applications layers), DECmessageQ [2] from Digital, MQSeries [3] from IBM, TopEnd [4] from NCR, and middleware for parallel scientific programming.

Several languages were developed for parallel scientific programming during the late 1980's and early 1990's. Among these, Parallel Virtual Machine (PVM) is a continuing collaborative research effort between Oak Ridge National Laboratory and the University of Tennessee [5]. Due to the efforts of a consortium of about 60 people from universities, government laboratories, and industry, the Message Passing Interface standard [8] interface for message-passing parallel applications was published in May 1994. MPI is the *de facto* standard for cluster parallel programming. Several implementations of MPI are described in Appendix B.

5.3.2 RPC-based Middleware

Although message passing provides a convenient way to structure a distributed application, many developers believe that applications are difficult to develop using message passing alone. Remote Procedure Call (RPC) allows a requesting process to directly execute a procedure on another computer in the cluster and receive a response in the form of a return value. RPC inter-process communication mechanisms serve four important functions [7]:

o   They allow communications between separate processes over a computer network,
o   They offer mechanisms against failure, and provides the means to cross administrative boundaries,
o   They enforce clean and simple interfaces, thus providing a natural aid for the modular structure of large distributed applications
o   They hide the distinction between local and remote communication, thus allowing static or dynamic reconfiguration

The overwhelming numbers of interactions between processes in a distributed system are remote operations. An important characteristic of the implementation of the client-server model in terms of RPC is that the code of the application remains the same if its procedures are distributed or not. This property is because the whole communication system is external to the application.

Client and server do not execute in the same address space, so there are no global variables and pointers cannot be used across the interface. Marshalling is the term used for transferring data structures used in RPC [8] from one address space to another. Marshalling is required, as two goals need to be achieved. The data must be serialized for transport in messages, and the data structures must be converted from the data representation on the sending end and reconstructed on the receiving end.

Middleware tools built over RPC include Network Information Services [9] (NIS) and Network File Services [10] (NFS). Both NIS and NFS were originally developed by Sun Microsystems, but versions of each of them have been released into the public domain. NIS is a network naming and administrative tool for smaller networks that allows users to access files or applications on the network with a single login name and password. NFS allows files and directories to be exported from the server computer to client computers on the network. With NFS, users can have the same view of the file system from any computer in the network. With NFS and NIS, users can login with an ID and password that is the same for any computer in the network, access an application that resides on the file server, and save files into a home directory on the file server from any computer on the network.

5.3.3 CORBA

CORBA describes an architectural framework that specifies the mechanisms for processing distributed objects. The middleware that supports the remote execution of objects is called the *Object Request Broker* (ORB). CORBA is an international standard supported by more than 700 groups and managed by the *Object Management Group* [11] (OMG). The OMG is a non profit-making organization whose objective is to define and promote standards for object orientation in order to integrate applications based on existing technologies.

The *Object Management Architecture* (OMA) is characterized by the following:
o   The Object Request Broker (ORB). The broker forms the controlling element of the architecture because it supports the portability of objects and their interoperability in a network of heterogeneous systems;
o   Object services. These are specific system services for the manipulation of objects. Their goal is to simplify the process of constructing applications;
o   Application services. These offer a set of facilities for allowing applications access databases, to printing services, to synchronize with other application, and so on;
o   Application objects. These allow the rapid development of applications. A new application can be formed from objects in a combined library of application services. Adding objects belonging to the application can extend the library itself.

CORBA Projects

PARDIS [12][13] is based on CORBA in that it allows the programmer to construct meta-applications without concern for component location, heterogeneity of resources, or data translation and marshalling in communication between them. PARDIS supports SPMD objects representing data-parallel computations. These objects are implemented as a collaboration of computing threads capable of directly interacting with the PARDIS ORB – the entity responsible for brokering requests between clients and servers. The PARDIS ORB interacts with parallel applications through a run-time system interface implemented by the underlying the application software package. PARDIS can be used to interface directly parallel packages, based on different run-time system approaches, such as the POOMA library and High Performance C++.

5.3.4 OLE/COM

The term COM has two meanings. It stands for *Component Object Model* [14], which form the object model underpinning Object Linking and Embedding (OLE) version 2.0. It also stands for the *Common Object Model* after an agreement between Microsoft and Digital. The first version of OLE was designed to allow composite documents (text and images) to be handled. The second version of OLE introduced a highly generic object model whose use can be extended well beyond the handling of composite documents. OLE2 offers a set of interfaces (Object Oriented) that allows applications to intercommunicate. The first version of OLE2 required applications to run on the same machine. OLE2 was later extended to run in a distributed fashion on Windows NT and DEC UNIX platforms.

The COM model defines mechanisms for the creation of objects as well as for the communication between clients and objects that are distributed across a distributed environment. These mechanisms are independent of the programming languages used to implement objects. COM defines an inter-operability standard at the binary level in order to make it independent of the operating system and machine.

COM Projects

NCSA Symera [15][16] (Symbiotic Extensible Resource Architecture) is a distributed-object system based on Microsoft's Distributed Component Object Model (DCOM). Symera is designed to support both sequential and parallel applications. The Symera management system consists of an NT service that is installed on each platform in a cluster. The management system hosts objects that allocate resources, schedules jobs, implements fault tolerance as well as object activation and migration. Symera is written in C++ that conforms to the Win32 standard.

5.3.5 Internet Middleware

The TCP/IP suite of network protocols is the enabling technology for the Internet. However, a number of middleware-like technologies are built over TCP/IP. HyperText Transport Protocol (HTTP) allows text, graphics, audio, and video files to be mixed together and accessed through a web browser. The Common Gateway Interface (CGI) standard enables retrieved files to be executed as a program, which allows web pages to be created dynamically. For example, a CGI program can be used to incorporate user information into a database query, and then display the query results on a web page. CGI programs can be written in any programming language.

Since users access some web pages and applications frequently, some web servers will send a small amount of information back to the web client to be saved between sessions. This information is stored as a "cookie", and saved on the local disk of the client machine. The next time that the web page is accessed, the client will send the cookie back to the server. Cookies allow users to avoid retyping identification information for repeated sessions to the same server, and allow sessions to be restarted from a saved state.

Web-based Projects

There are a large number of Web-based software systems that can be considered as middleware for cluster computing. Typically such software consists of a Web-based portal to another middleware system such as CORBA or COM that is used to provide the actual applications services. The Web portal can then be used to, for example, build, submit and visualise the application that is using the backend CORBA or COM services.

5.3.6 Java Technologies

Java Remote Method Invocation [17] (RMI) allows communications between two or more Java entities located in distinct Java Virtual Machines (JVM). Java RMI allows a Java applet or application access another remote object and invokes one of its methods, and so in that sense is similar to RPC, except that it operates on objects. In order to access the remote object the calling application must obtain its address. This is obtained by access to a Registry where the object's name was registered. The Registry acts as a name server for all objects using RMI. The Registry contains a table where each object is associated with a reference, which is the object's interface and unique address of the object.

Jini [28][29] from Sun Microsystems, is an attempt to resolve the inter-operability of different types of computer-based devices, given the rising importance of the network. These devices, which come from many different vendors, may need to interact over a network. The network itself is dynamic - devices and services will need to be added and removed regularly. Jini provides mechanisms to enable the addition, removal, and discovery of devices and services attached to network. Jini also provides a programming model that is meant to make it easier for programmers to enable devices to interact.

Jini objects to move around the network from virtual machine to virtual machine. Built on top of Java, object serialization, and RMI, Jini is a set of APIs and network protocols that can be used to create and deploy distributed systems that are organized as federations of services. A Jini service can be anything that is connected to the network and is ready to perform some useful task. The services can consists of hardware devices, software, communications channels, and even interactive users. A federation of services, then, is a set of services, currently available on the network that a client (meaning some program, service, or user) can bring together to help it accomplish some task.

*Other Java Technologies*

Sun Microsystems has produced a plethora of Java-based technologies that can be considered as middleware [30]. These technologies range from the Java Development Kit (JDK) product family that consists of the essential tools and APIs for all developers writing in the Java programming language to APIs for telephony (JTAPI), database connectivity (JDBC), 2D and 3D graphics, security, and electronic commerce. These technologies enable Java to interoperate with many other devices, technologies, and software standards.

Some Java Projects

BAYANIHAN [18],[19] is a software framework that uses Java and HORB [20] – a distributed object library. The framework allows users to co-operate in solving computational problems by using their Web browser to volunteer their computers' processing power. HORB uses something akin to RMI to pass data between objects. BAYANIHAN also provides a PVM interface for parallel computing.

The Heterogeneous Adaptable Reconfigurable Networked SystemS [21][22] (HARNESS) is an experimental metacomputing framework built around the services of a customisable and reconfigurable Distributed Virtual Machine (DVM). HARNESS defines a flexible kernel and views a DVM as a set of components connected by the use of a shared registry, which can be implemented in a distributed fault tolerant manner. Any particular kernel component thus derives its identity from this distributed registry. The flexibility of service components comes from the way the kernel supplies DVM services by allowing components, which implement services, to be created and installed dynamically. HARNESS uses the micro-kernel approach, where services are added as needed to provide the functionality that the users require.

The Java Network of Workstations [23] framework (JavaNOW) provides a mechanism where large, complex computational tasks can be broken down and executed in a parallel fashion across a network. The concept of JavaNOW is similar to PVM. However, the JavaNOW interface is based on logical distributed associative shared memory instead of message passing. The interface for JavaNOW allows users to store complete objects in shared memory.

JavaParty [24][25] is a system for distributed parallel programming in heterogeneous workstation clusters. JavaParty extends Java by adding the concept of remote objects. The JavaParty consists of a pre-processor and a runtime-system to support the system.

The central idea in the Javelin [26][27] architecture is a computing broker, which collects and monitors resources, and services requests for these resources. A user interacts with the broker in two ways, by registering or requesting resources. The broker matches clients with host machines for running computations. A key design feature of this architecture is that the host software runs completely at the user level, and only requires that registered machines have a Web browser that can run untrusted code, such as Java applets.

5.3.7 Cluster Management Software

Cluster Management Software (CMS) is primarily designed to administer and manage application jobs submitted to workstation clusters. More recently CMS has also be extended to manage other underlying resources as well as software licenses (e.g., CODINE/GRD [31]). The job can be a parallel or sequential application that needs to run interactively or in the background. This software encompasses the traditional batch and queuing systems. CMS can be used to help manage clusters in a variety of ways:

- Optimise the use of the available resources for parallel and sequential jobs;
- Prioritise the usage of the available resources;
- Manage mechanisms to "steal" CPU cycles from cluster machines;
- Enable check-pointing and task migration;
- Provide mechanisms to ensure that tasks complete successfully.

CMS software is widely available in both commercial and public domain offerings. There are several comprehensive reviews of their functionality and usefulness [32][33][34].

## 5.5. Conclusions

The purpose of middleware is to provide services to applications running in distributed heterogeneous environments. The choice of which middleware may best meet an organization's needs is difficult as the technology is still being developed and may be some time before it reaches maturity. The risk of choosing one solution over another can be reduced if [35]:

- The approach is based on the concept of a high-level interface;
- The concept of a service is associated with each interface;
- The product conforms to a standard and supports its evolution.

Object-oriented middleware such as CORBA/COM and Java are widely accepted. The popularity of CORBA was limited until the appearance of Java in the mid-1990's. Current CORBA and COM products are being developed to be inter-operable with Java. Jini, Sun Microsystems' latest distributed environment, has wide possibilities to subsume all the other systems. The key question about what system a particular organization should adopt is no longer as it was a few years ago. Today the key question is, "is the middleware that I chose to adopt interoperable with the other current and emerging standards in middleware?"

# 6. Systems Administration

Anthony Skjellum, MPI Software Technology, Inc. and Mississippi State University, USA
Rossen Dimitrov and Srihari Angulari, MPI Software Technology, Inc., USA
David Lifka and George Coulouris, Cornell Theory Center, USA
*Putchong Uthayopas, Kasetsart University, Bangkok, Thailand*
Stephen L. Scott, Oak Ridge National Laboratory, USA,
Rasit Eskicioglu, University of Manitoba, Canada

## 6.1 Introduction

As industry standard computing platforms such as workstations and small-scale symmetric multiprocessor servers continue to show better and better price performance, more and more significant clustering projects are developing internationally. What separates these systems is how usable they are in terms of actually producing computations of value to their users, and at what "comfort level" for their users. Manageability of a system can be the single most important factor in its practical usability. This is complicated by the current fact that many clusters are in all or in part themselves experimental computing systems for computer science studies of architecture, and/or software, while simultaneously attempting to offer research and/or production platforms for their user base. Even for a single system, the emphasis can vary over time in order to satisfy sponsor requirements and evolving research ideas, whereas a production-oriented user base typically is most interested in application results, system availability, and stability.

The goals of various clustering projects are quite diverse, with extremes evident. First, there is the dedicated cluster for computer science research. The goal of this type of system is typically to undertake performance testing, benchmarking, and software tuning. This work is often done at the networking, application, and operating systems levels. These clusters are not typically used for computational science and quite often are in a state of flux. The opposite extreme is the production-computing environment. The tacit goal of these systems is to provide reliable computing cycles with dependable networking, application software, and operating systems. As noted above, systems may attempt both extremes, either by spatial division of the resources among emphases, or by reallocation policies over time. Realistically, the entire range of work being done on clusters and the quality of the results generated on them are driven strongly by the tools available to the systems administrators. In this section we describe some of the critical issues that arise when planning a cluster computing installation and its ongoing support and maintenance.

No matter what the goals of a particular cluster-computing project may be, good systems manageability will directly equate to better results. Systems like the IBM SP have been successful in providing these types of utilities in the past. However, similar and in many cases superior tools are being developed for clusters out of necessity. Even if the users are just experimenting with a small test cluster, the ability to verify and ensure consistency in hardware, application software, and system settings across the cluster's resources from a single console will improve the efficiency and reproducibility of results. It is equally important for a systems administrator to be able to install system images and layered software in a parallel fashion. All of these issues reduce the cost of ownership, which is an important aspect of the overall value of cluster computing to users, not just their initially low-cost nature.

## 6.2 System Planning

There are several considerations that must be made when planning a cluster installation. Generally they are the types of computers to purchase, the type of networking interconnect, and the system software to use. All of these must be taken into consideration and typically the driving factor for any particular configuration is the requirements of the applications that will be run on it. In this section we describe some of the more popular options and try to focus on the application requirements that warrant their use.

### 6.2.1 Hardware Considerations

The idea of clustering has been around for well over ten years. It originated as a way to leverage the investment of workstations in an organization. Because workstation technology was and continues to progress rapidly, it was found that linking workstations together with Ethernet networks was a good way to get near to traditional supercomputer performance. The performance could be delivered at a low entry price, and furthermore expanded interest in applications that could run in parallel without the huge network performance of traditional parallel supercomputers. Further, it meant the machines, which were commonly left idle, such as desktops at night and weekends, could putatively begin to be used in a productive manner during those periods.

With the ever-increasing performance of the Intel architecture and its broad market base, the cost-performance ratio of clusters of workstations continues to improve almost daily. Because of its origins many people prefer to purchase "beige boxes" or inexpensive generic Intel-based workstations. This provides a low cost/compute cycle ratio. What has changed is how people are using clusters. Today organizations are starting to use clusters as their primary source of computational platform. In doing so the demands for reliability, manageability, and supportability have dramatically increased. Once again, applications dictate what works and what does not. If the purpose of a cluster is hard-core computer science research where drivers and software are being continually evolved, workstations may work just fine. As clusters are placed in production computing environments, where the research that is being undertaken is chemistry, physics, finance, computational fluid dynamics, or other quantitative application areas, and not necessarily only computer science, the issue of uptime becomes critical. In order to meet these needs, computing centers are looking toward high-end industry standard-based servers. The cost-benefit remains but the additional up-front cost typically affords excellent hardware redundancy and maintenance programs that used to only be associated with traditional supercomputers and mainframes. Further, as the size of the cluster begins to increase, many server vendors are providing space efficient rack mountable solutions. These not only provide a smaller footprint but also organized cable, power and cooling management. Figure 6.1 shows a common cluster cable plant. Figure 6.2 illustrates what is possible with a rack-mounted solution.

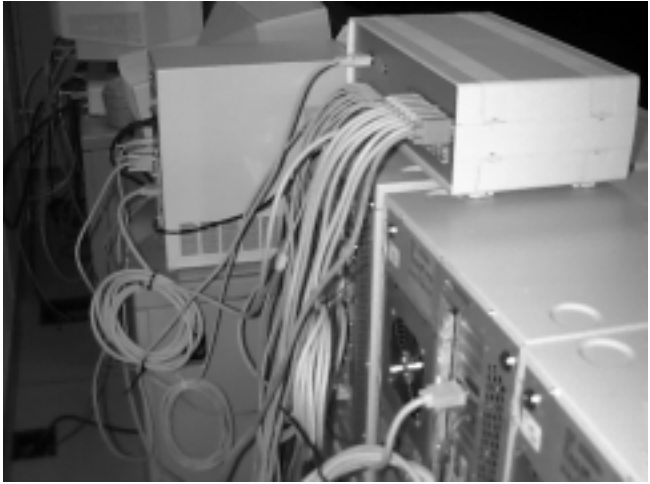

Figure 6.1. A do-it-yourself cluster look and feel.

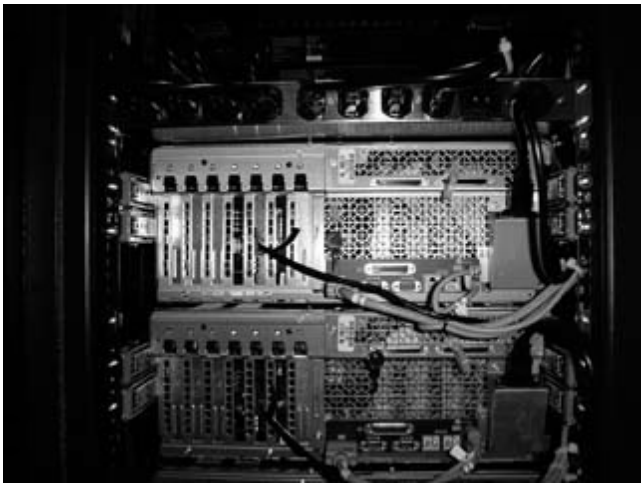

Figure 6.2. A professionally integrated cluster.

6.2.2. Performance Specifications

Information about the type of workload intended to be incorporate in a cluster will help tailor the cluster's hardware to these needs. When building a cluster, it is important to choose hardware that is well suited to the prospective workload. Low-level details such as memory bus speed and PCI bus interfaces are just as important as processor speed or cache sizes. We outline several of these issues in the following sub-sections.

6.2.3 Memory speed and interleave

While faster memory is in general "always better," higher memory density is not always desirable. For example, consider a cluster that requires 256 MBytes per node, and each node has four available slots. While getting two 128 MBytes modules allows for later upgrades, using four 64 MBytes modules may provide increased memory bandwidth if the motherboard chipset supports memory interleaving. Upgradability is a competitive concern, if one preplans to do a mid-life memory upgrade to a system, based on funding or expanding user requirements for in-core problem size support.

### 6.2.4 Processor core speed vs. bus speed

Modern processor cores typically run at a multiple of the speed at which they interface with the rest of the system. For example, while a processor might run at 500MHz, its bus interface might only be 100MHz. On a code, which is memory-intensive, for example, the cluster designer may wish to opt for a system with a lower core frequency but higher bus speed.

### 6.2.5 PCI bus speed and width

PCI interfaces currently come in two widths (32 or 64 bits) and two speeds (33 or 66MHz), with expansions be developed. While most low-end cards will have a 32-bit/33MHz interface, cards such as high-performance network adapters will have wider or faster interfaces. If the motherboard does not support the wider or faster bus, maximum performance will not be obtained. The quality of PCI bus implementations is a hot topic among people considering network card performance, because actual performance varies considerably from chipset to chipset. Investing time to be sure that the chipset works well with the cluster's planned interconnect (described in Section 2) is essential, as is any issues of multiple buses and the planned utilization of local I/O as well as networking. Newer systems do not immediately equate to better PCI performance, though this is the general trend.

### 6.2.6 Multiprocessor issues

Depending on the cluster's workload, one may wish to choose single- or multiprocessor building blocks. For applications whose working set is smaller than the processor cache size, using multiprocessor machines can increase a cluster's aggregate performance significantly. For memory-intensive applications, however, there are tradeoffs. Simple bus-based multiprocessor machines work well in 2-processor configurations, but do not scale well for higher numbers of processors. Crossbar-based or hybrid bus/crossbar systems deliver higher memory bandwidth and better scalability but are usually priced higher, and memory consistency management heightens the context switch times of such systems. Context switch time impacts performance of thread-oriented applications, and may somewhat impact applications that do a lot of I/O.

### 6.2.7 Cluster Interconnect Considerations

Network interconnects are essential components of high-performance computational clusters. Four major factors determine the importance of cluster interconnects:
1. Communication-intensive parallel applications require efficient data transfers even at low processor scales. Applications that spend a large portion of their execution time in communication benefit most from improvements in network speeds. Fine-grain algorithms that are sensitive to short message latency and coarse grain algorithms that are sensitive to peak bandwidth of bulk transfers require balanced and efficient communication systems.
2. Efficiency of the network system includes the effective drain on processor cycles associated with transfers resulting both from hardware and associated software stack. Furthermore, as technical applications often possess the ability to overlap communication and computation, network architectures with software stacks that support user-to-user overlap of communication and computation can reduce application runtimes significantly, even for large-scale systems (since problems are often scaled as cluster processor counts as scaled).
3. Large-scale clusters with number of processors exceeding 1,000 are currently being deployed. Maintaining sufficient parallel efficiency at these scales requires highly optimized network interconnects.
4. The rapid growth of microprocessor speed and memory throughput allows for a cost-efficient increase of parallel performance. However, the effective benefit of this increase may be significantly diminished if the cluster is interconnected with an inadequate network that may become a performance bottleneck, in latency, bandwidth, and/or consumption of CPU cycles to deliver needed performance.

These issues, and others, are discussed thoroughly in Section 2.

## 6.3 Software Considerations

Hardware is not the only set of concerns that play a major role in the decision process when building a commodity cluster. The choice of the operating system and the other necessary software also affect the usability and manageability of the cluster. In most cases it is the proposed use of the cluster that influences the decision for the software that goes into the system, including the operating system. Clusters present a logical extension and/or an equal environment to traditional multiprocessor and multi-computer (massively parallel) systems. Typically, clusters are chosen with price-performance in mind, and the problems that are solved on clusters are those that are ported from multi-computer environments, though applications targeted especially for clustered architectures are slowly taking off. This issues surrounding the choice, features and functionality of Cluster operating systems are discussed in Section 3.

### 6.3.1 Remote Access

How users and administrators access cluster resources remotely is an important consideration. Both UNIX and Windows offer a variety of options. Windows 2000 has made great strides to provide the same types of tools that UNIX provides.

Windows Capabilities

Windows 2000 has a built in Telnet server. Under Server and Advanced Server it limits the number of telnet connections to two. Under the services for UNIX tool kit from Microsoft there is an alternative Telnet Service that does not limit the number of connections. One important factor that is currently lacking in the Windows Telnet environment is a secure shell capability. However, one can use `SSH` to encrypt terminal server sessions. Terminal Server from Microsoft only allows Windows based clients to connect, while Citrix [1] has an add on product, which provides Terminal Server clients for many UNIX flavors as well as Macintosh and a web client that permits connections from a standard web browser.

Microsoft Internet Information Services (IIS) provides FTP, Web, News and Email services for Windows 2000. IIS can be configured with various levels of security for each of the services it provides. One features of IIS is its integration with the Windows security model. This makes the development of secure web sites easier.

Finally, Microsoft allows users to run new and existing X Windows applications. Products such as Hummingbird Exceed [2] and Cygnus [3] are designed for Windows platforms.

UNIX Capabilities

Since UNIX has been used in cluster environments longer than Windows most of the access tools have been available for a while. Most UNIX versions provide `SSH`, Telnet, X Windows and FTP. Integrating secure UNIX Web sites can be accomplished, using for example the Apache Web server [4].

### 6.3.2. System Installation

When managing a cluster of any size it is extremely important to be able to ensure a consistent system image across all cluster nodes. This helps eliminate variables when trying to debug system configuration problems and user program errors. System tools like DHCP go along way toward making it possible to generate a system image and propagate it across all cluster nodes by minimizing the number of node specific customizations, like IP addresses.

Windows Capabilities

Windows 2000 includes the capability to remotely install all cluster nodes and to distribute layered software. Remote Installation Service (RIS) [5] allows an administrator to generate a system installation image and then push it to new cluster nodes. There are third party tools also, such as Norton's Ghost [6] and Imagecast [7] that perform the same function.

Windows 2000 also has a built in capability to "package" a software installation and then distribute it to all cluster nodes. The way it is implemented is particularly convenient. Windows keeps track of where in the installation of a particular package is on a node. If the network fails or the node crashes Windows will resume the installation where it left off.

A real strength of the Windows system and software distribution is the way it has been integrated into the Windows 2000 operating system making it easy for anyone to take advantage of.

UNIX Capabilities

The Linux installation basically consists of – describe the hardware; configure the software, and load – load – load… It can be tedious for one machine, yet completely unbearable and unnecessary for the multiple, mostly identical, cluster nodes. There are presently three approaches to solving this problem:
1. The machine vendor supplied disk duplication;
2. The use of a network disk installation or image replication package;
3. The use one of a self-install pre-packaged Beowulf2 cluster-in-a-box solution.

While having a vendor duplicate a cluster node image sounds rather simple, there are still some problems to avoid. The first hurdle to clear is to obtain a machine identical to those to be used in the cluster. The best way to do this is to convince a vendor that an initial loan machine is part of the purchase pre-qualification process. When the machine arrives, install the operating system and the entire necessary programming environment that will reside on a cluster node. Then, backup that disk and either give the disk or the entire system to the vendor so that they may use their disk duplication system to copy the disk contents to all purchased cluster nodes. Another advantage of doing this, in addition to the apparent time saved, is that a vendor may keep the base disk image and provide all future machines, of the same hardware configuration, with the same software installation. There will be nothing further to do when the machines arrive if the cluster is setup to do dynamic IP via an DHCP server. However, if using statically allocated IPs, each machine will require its network configuration to be reset after initial boot (and then be rebooted before use).

Another mechanism is to require the vendor to write the MAC address on the outside of each system box. This makes setting up static IP with a specific node numbering scheme much simpler. The alternative is to find it written on the NIC or to boot each machine one at a time and capture the MAC address as the machine starts.

A final note here that cannot be over emphasized – find a knowledgeable Linux vendor. Although, it may save money going to the lowest bidder, it may not be such a good deal if the vendor simply hits the power switch on a Linux box after the burn-in period without going through the proper shutdown procedure. You may find out if they did this during the initial cluster power-on as a number of nodes will fail to boot without direct user interaction.

The second method and one of the options if the cluster is not a new purchase is to use one of the tools to either do a replicated network installation or network disk image duplication. Linux Utility for cluster Install (LUI) [8] is an open source tool developed by IBM to perform a replicated network install across cluster nodes. VA SystemImager [9] is a network disk image duplication tool available as open source from VA Linux.

LUI installation consists of unpacking a tar file, soon to be available via RPM, onto the machine that is to become the LUI Server. The LUI Server will be the machine from which cluster builds' will be orchestrated and will maintain a database of cluster configurations and resources. To use LUI the server, the cluster (client) nodes, and finally the resources to load onto the client cluster nodes must be configured. LUI provides a GUI for this purpose. The server information required is:
- Machine name,
- IP address,
- Net mask.

The cluster node information required is:
- Machine name,
- IP address,
- net mask,
- MAC address,
- Fully qualified hostname.

The optional node information includes:
- The default route,
- Installation gateway,
- Number of processors,
- PBS string.

The last two data fields are included as a result of collaboration with the Open Source Cluster Application Resources (OSCAR) [11] project, (see sub-section 6.4), and are used for the automated OSCAR install of OpenPBS [9]. Various resources may be specified, including:
- Disk partition table,
- User specified exit code,
- File system,
- Kernel, and kernel map,
- Ramdisk,
- RPM,

- Source (copied as a file.)

The final GUI setup step requires that resources be associated with each cluster node. A future release of LUI will include a grouping feature so that nodes may be grouped together and then the group associated with a set of resources. To effect node installation, one need to only network boot the cluster client nodes LUI automatically installs the specified resources. This of course requires a PXE [12] enabled network card and BIOS capable of its use. LUI has a number of significant advantages over disk image solutions – one is that it stores the configuration, very small in comparison to a disk image, and rebuilds each time for installation. Since the LUI Server need only be online during the installation process, a computer outside of the normal cluster resources may be used to "service" the installation of the actual cluster nodes. This machine could be the cluster administrator's desktop or notebook machine. One negative aspect of LUI is that although it has the ability to move any software package across the cluster during installation, it does not have the ability to automatically install that package on each cluster node. This is a major drawback with respect to passing an entire computing environment across the cluster.

VA Systemimager is a network disk image duplication tool developed by VA Linux. Systemimager must be installed on the image server and the Systemimager-client must be installed on the master client. The image server will be the machine to initiate the image duplication process as well as holding the storage of all disk images. The master client is the cluster node that is configured and maintained as the example node to be distributed to all other cluster nodes. In addition to the installation of the appropriate Systemimager packages on the image server and master client – all the final software needed on the cluster must be installed on the master node before it is loaded across the cluster – this process is similar to that used with the vendor disk duplication discussed earlier. Once the master client is fully configured, its image may be pulled from the master client to the image server. All cluster nodes may now be booted via a diskette, CD, or the network, each will pull the specified image from the image server. As with LUI, a PXE enabled network card and BIOS is required for the network boot option.

While neither LUI nor Systemimager provide all the desired capabilities, they complement one another in numerous ways and when combined with other tools provide the ability to extract the best of each. This is the approach taken with the OSCAR project.

The third method of building a complete cluster-computing environment is that of using one of the self-install pre-packaged Beowulf cluster-in-a-box solutions. A number of companies including Scyld [13], and Turbolinux [15] have started to market commercial cluster solutions that include some type of auto-install feature. However, these options tend to contain proprietary solutions for building a Beowulf cluster and are rather expensive (as compared to a no-cost solution). OSCAR is an open source consortium solution that provides an auto-install cluster-on-a-CD that contains the current "best of practice" for clusters consisting of less than 64-nodes. OSCAR consists of a LUI initiated install and subsequent configuration of cluster nodes with the following cluster computing environment, M3C [14], $C^3$ [16], OpenPBS, SSH/SSL, DHCP, NFS, TFTP, RSYNC [76], Systemimager, PVM, and MPICH. The configuration consists of providing LUIs' required and optional information via the LUI GUI. Installation and configuration then takes place in a similar fashion to a standard LUI installation. The difference from the LUI only installation is that when finished the entire cluster computing environment is configured.

### 6.3.3. System Monitoring & Remote Control of Nodes

System performance monitoring is the act of collecting cluster's system performance parameters such as node's CPU utilization, memory usage, I/O and interrupts rate and present them in a from that can be easily retrieved or displayed by the system administrator. This feature is important for the stable operation of large cluster since it allows the system administrator to spot potential problems earlier. Also, other parts of the systems software, such as task scheduling, can benefit from the information provided, e.g. using it to perform better load balancing.

Structure of Monitoring Systems

The basic system-monitoring tool consists of the components shown in Figure 6.3.

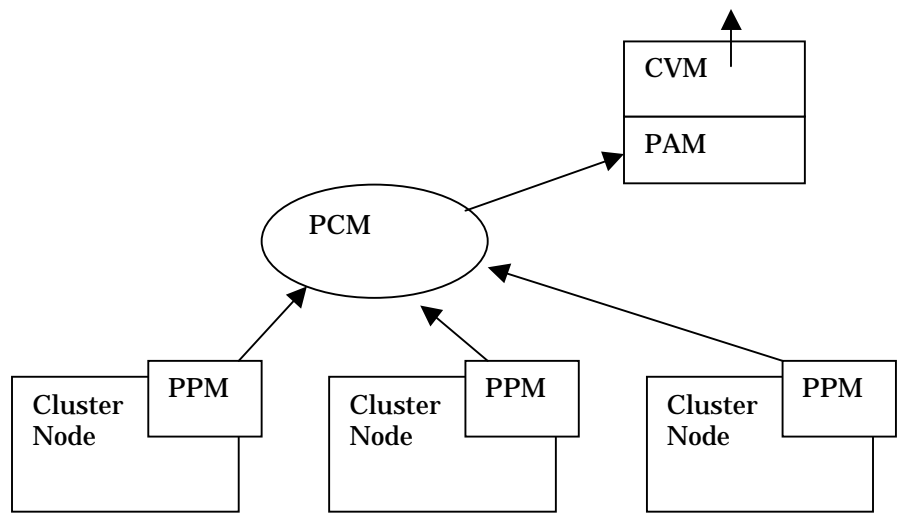

Figure 6.3. The Organization of Typical System Performance Monitoring Tools

- The Performance Probing Module (PPM): This is part of the monitoring system that interfaces with local operating system to obtain the performance information. The information obtained is later sent or collected by other monitoring sub-systems. In many implementations [17], PPM also performs certain checking on node condition and generates event notification to indicate the occurrence of certain condition.

- The Performance Collection Module (PCM): This part of the monitoring system acts as information collectors that collect information from every node and store it for later retrieval. One of its main is to cache performance information that reduces the overall impact of monitoring on the system being monitored.

- A Performance API Library Module (PAM): This is a library that allows a programmer to write a applications to access the performance information.

- The Control and Visualization Module (CVM): This is an application that presents the performance information to user in a meaningful graphical or text format. Another function that the CVM performs allows a user to control the set of nodes and information that user wants to see. One example of this module from software called SCMS [21] is illustrated in Figure 6.4.

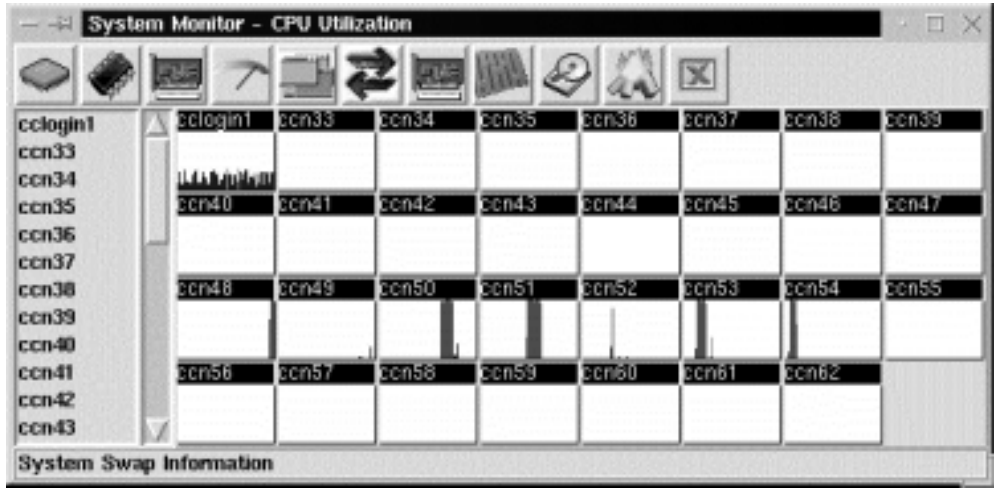

Figure 6.4. The CVM part of SCMS [21](showing CPU load on 31 of 256 nodes on Chiba City Linux Cluster, Argonne National Laboratory)

Obtaining the Performance Information

There are many techniques being used by the developer to obtained performance information from the systems. These techniques can be summarized as follows:

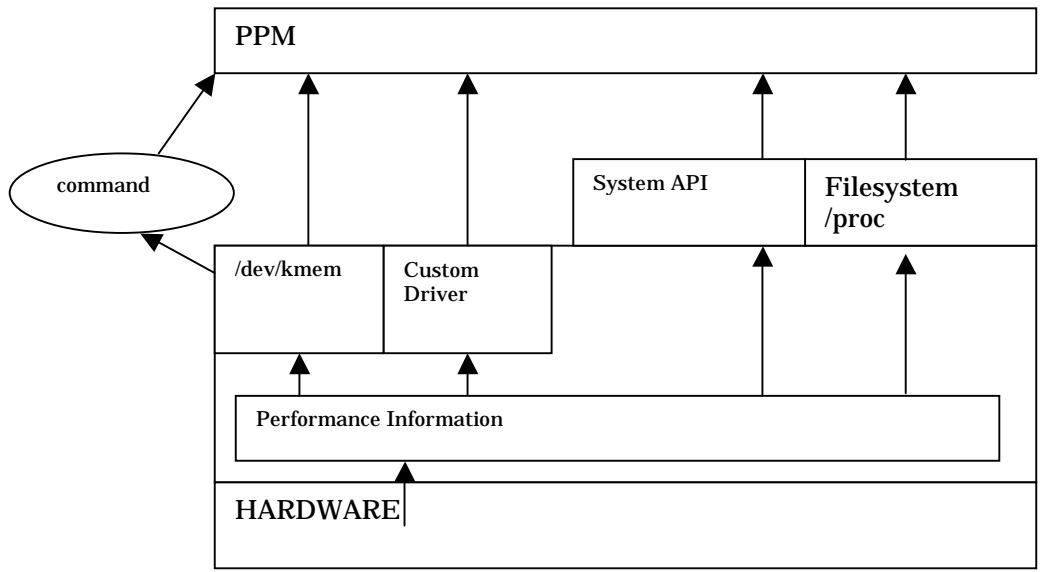

Figure 6.5. An example of how to read system information

Probing by direct access to kernel memory

In this approach, the data is obtained by reading it directly from a known address (using an interface such as `/dev/kmem`). This approach makes probing depend on the kernel version.

Even a slight change in the kernel can cause a failure of probing the system. To avoid this problem, most modern kernels usually provide a system performance interface API. The advantage of this approach is the speed of information collection.

Some recent tools [22] try to make fast access to kernel information by utilizing a custom device driver that directly accesses kernel information but keeps the application interface API fixed. The benefit of this is fast access and kernel independence. The device driver, however, must be kept updated with kernel changes. This is rather difficult considering the fast pace of kernel development in Linux environment.

Probing by File System Interface

Most UNIX systems, including Linux, currently have a interface through a virtual file system named `/proc`. Information from inside kernel such as CPU, process, file, networking statistics will appear in `/proc` for users to access. The advantages of this approach are better portability and kernel independence. The location of information found in `/proc` tends to be the same for all kernel versions. Moreover, the speed of access is comparable to direct kernel access since `/proc` is a virtual interface through the Virtual File System (VFS) that goes directly into kernel memory. Some projects such as SMILE [17] also developed a library layer (Hardware Abstraction Layer – HAL) on top of `/proc` that allows better portability and ease of programming.

Probing by System APIs

Certain operating systems, such as Microsoft Windows, provide APIs like the Windows Management and Instrumentation (WMI) [23] to gather system information. The advantage to using such an API is the ease of programming and kernel independence. WMI is implemented as a distributed COM object that can easily called from almost any programming and scripting language under Windows. There are similar APIs available in the UNIX environment but none quite as comprehensive or consistently implemented as WMI.

Probing by System Command

Some UNIX based tools use the result generated from the execution of standard UNIX command such as `uptime` or `ps` to probe system performance. This can provides portability between families of UNIX systems. However, this is an inefficient approach since its place a heavily load on the system every time the commands are executed. This approach is by and large used in primitive tools or used in early development prototypes before more efficient methods are implemented.

Collecting the Performance Information

Once the system information has been read on each node, this information must be delivered to the application that request it.

A choice as to whether to have the application perform the collection of information itself or having some sub-system collect and maintain the information for the application needs to be made. Most monitoring systems opt for the later case, since there are many benefits to be gained, such as reducing the network traffic, ease of management for multiple sessions. But its choice may limit scalability as will be discussed later.

The act of collecting the information itself is a collective operation (like gather/scatter). So far, most monitoring systems rely on point-to point communication from nodes to information collector. High-level communication systems, such as MPI have well defines and efficient collective operations. The use MPI might not be possible yet since MPI is not dynamics enough to cope with some situations. For instance, if the monitoring system used MPI, when one of the nodes dies, every MPI processes must be terminated and the monitoring system must be restarted. Currently MPI has no clear interface for the interaction with processes outside the MPI scope. So, it is not possible at this time to write a hybrid and flexible MPI-based system management tool.

Scalability

Many works on monitoring systems [22][21][24] are based on a centralized performance collection paradigm. This simple implementation works efficiently for a small to medium size cluster (16-64 nodes). However, the trend is to build ever-larger clusters. Currently, 64 to 256 nodes systems are not uncommon. Some clusters have more than 1000 nodes [25][26]. With these large clusters, the centralized approach will be inadequate. One technique suggested in [27] is to use the cascading hierarchy of monitoring domains. In this approach, the monitoring is divided into domains. Each domain consists of a set of nodes and a *monitor proxy* that receives monitoring requests from higher domains, merges and sends information upward, and handles fault events and forwards them to the upper level proxy.

Optimizing the Network Traffic

There are many techniques being used by various performance monitoring systems to optimize network traffic. Examples of these techniques are:

- Adjusting the monitoring frequency: The function of the monitoring system is to presents the current system state of the cluster. This tends to be a non-trivial task since the state of the cluster consists of hundreds of continuously changing variables. Hence, the precision of monitoring depends on the frequency that the monitoring system collects the information. However, more up-to-date information usually means a higher frequency of information collection. This will generate a high traffic load for the system. Some system such as use a PCM to separate the frequency of requesting the information by the application program from the real collection frequency of information cluster nodes. This approach allows the tuning of the level of accuracy required against the network traffic.
- Data set selection: In a practical system, a system administrator usually wants to observe only a certain number performance parameters. Therefore, allowing the user select only information they are interested can reduce the amount of data movement. This support is typically built into the API and user interface to permit users to mask/filter the information they need. This mask or filter must be then send to each node to notify PBM so it sends back only information needed. Certain Boolean operations such as AND or OR might be needed to combine the masks from multiple applications running concurrently and requesting the information from the monitoring system at the same time. This feature appears in [19].
- Data merging: In [27], a hierarchical structure of monitoring has been used. A benefit of this, apart from higher scalability, is the ability to merge redundant information collected in each level of hierarchy. By merging this information, the traffic is reduced.

Reducing The Intrusiveness

Performance monitoring must be minimally intrusive to the system it is monitoring. The sources of intrusiveness are reading of the system information and the execution of event notification. When reading the system information using /proc usually causes a low impact on the system since this requires only the memory access. However, in [22] the level of intrusiveness has been reduced further by using a custom driver that collects system information and sends all of it directly to probing module. The trade off between low intrusiveness and portability is an issue here. For the execution of event notification, the intrusiveness usually depends on the complexity of rule, number of events, frequency of checking, and efficiency of implementation. The only optimization possible is to select an appropriate number of events and best interval of checking. In general, one possible technique is to distribute the execution of certain parts of the monitoring system to dedicated hosts. For example, assigning one node to be a management node, then executing PCM, PAM, and CVM on that node.

Web Technology and Cluster Management/Monitoring Systems

Web-based technologies provide many benefits to users such as:
- Internet access to system management and monitoring capabilities.
- Browser-based user interface,
- The capability to interact with the monitoring system through low speed communication channel.

A typical web based monitoring system is illustrated in Figure 6.6. (This design is used by KCAP [28].

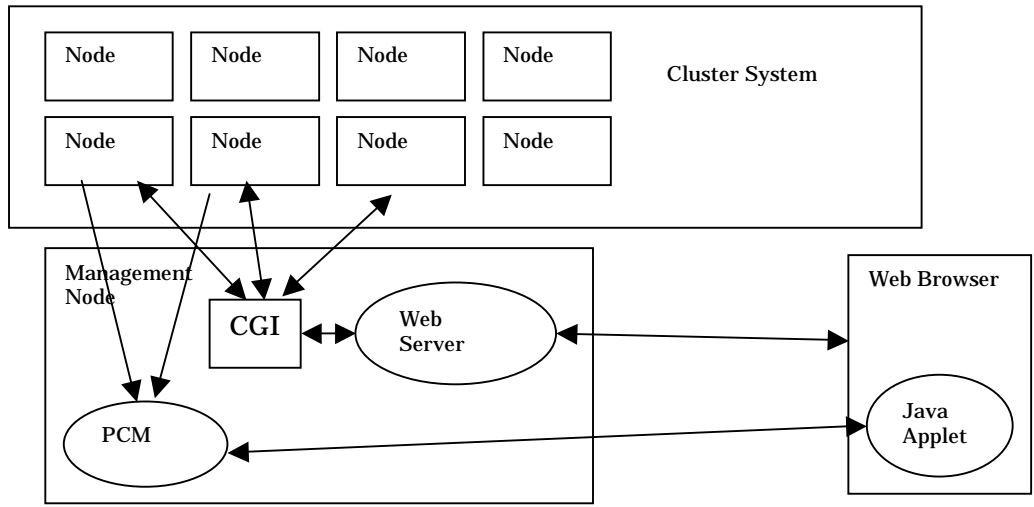

Figure 6.6. The structure of a Web-based management and monitoring system

In this configuration, a dedicated node called management node runs the PCM and Web server. The Java applet is the interface to connect to the real-time monitoring system. As the user accesses a certain Web page, the Java applet is loaded. This applet makes the connection back to the PCM, gets the system information, and presents it to the user graphically. For some other services, such as checking the number of users, number of running processes, CGI-bin can be used instead. Nevertheless, the system administrator must set up security to allow the remote execution on the cluster nodes. This may not be ideal where high security is required of the system.

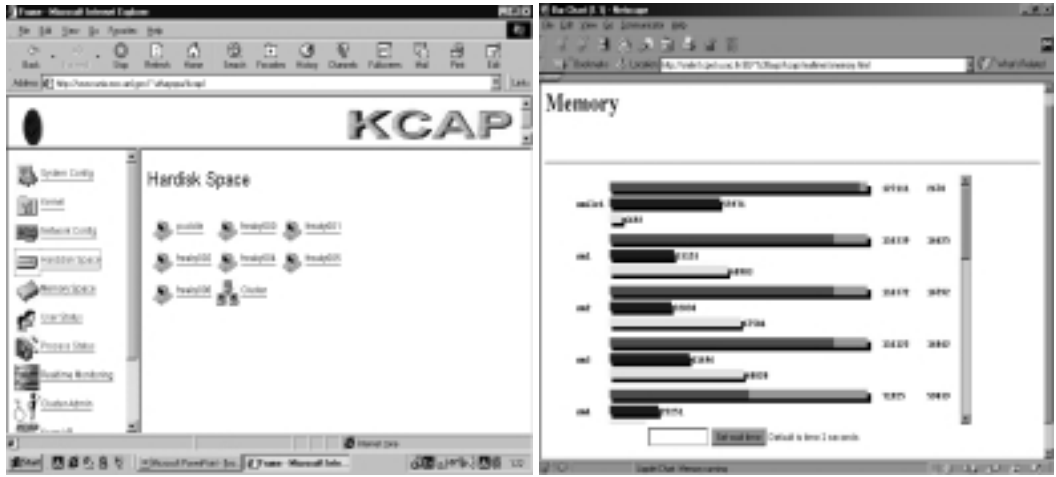

Figure 6.7. An example of a Web-based system monitoring and management [28]

Emerging Web standards such as VRML97 [29] can be used to increase the information delivered to user by allowing a user to navigate around the system be monitored. By interfacing Java using the External Authoring Interface (EAI) [30], dynamics behavior of the cluster can be displayed in real time. This method has been used in the KCAP3D [31] package to visualize a cluster. In this system the cluster configuration is converted into navigable VRML using a Perl based parser. This approach allows users to select and view components that they want to interact with in 3D.

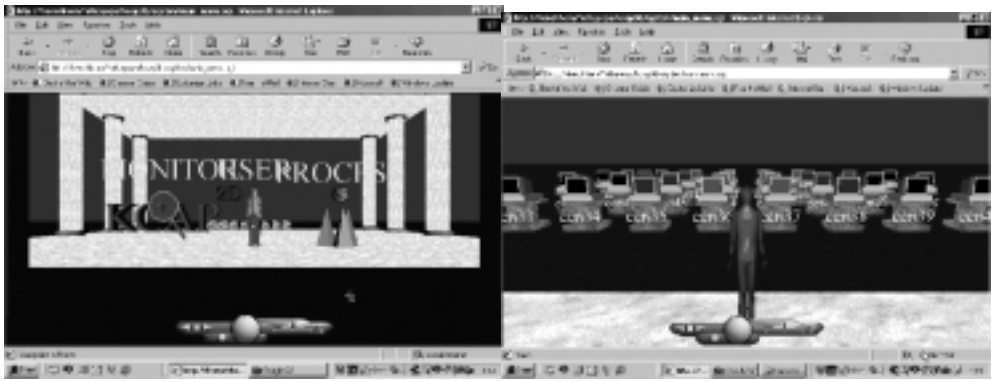

(a) 3-D Icon showing global function                                         (b) A Cluster farm

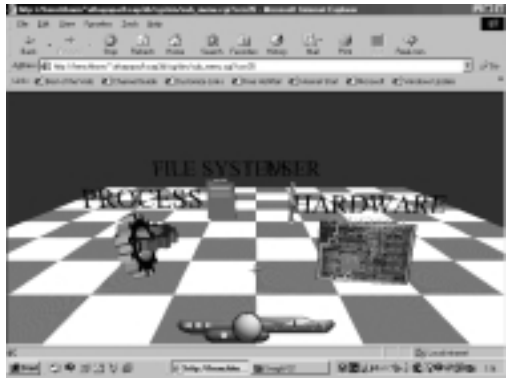

(c) Inside a node

Figure 6.8. Walk-in visualization of Chiba City generated by KCAP3D

6.4 Remote Management: Tools and Technology

On a UNIX based system any software that resided on each node can be easily accessed using X windows. However, the situation is somewhat different for UNIX commands. In the past remote execution have generally been done using traditional commands such as `rcp` and `rsh`. The system administrator, however, needs to relax security to allow cluster wide execution of these commands. This is fine for clusters that are behind firewalls but it might present a security risk for open clusters. It should be noted that the speed of `rsh` execution depends on how users are validated. In the systems that use a centralized server, such as NIS, executing hundreds of `rsh` sessions on remote nodes will slow due to the interaction with NIS. So, many systems solve this problem by replicating control files, such as `/etc/passwd`, `/etc/group`, `/etc/shows` instead. The advantage of this is two fold. First, since authorization is done locally, remote access for short-lived tasks is speeded up substantially. Second, by removing the centralized NIS bottleneck, a user can login to any node the need to interact with the central NIS service.

Traditional UNIX commands do not support the model of parallel computing. There is an effort to extends and standardize UNIX command over parallel computing platforms Parallel Tools Consortium project called Scalable UNIX Tool [32]. The current implementation of this tool is based mostly on shell scripts, which are slow and do not scale well. There are two major requirements for fast implementation of scalable UNIX tools:
- Fast process startup;
- Efficient collective communication.

For the fast process startup, this seems to be the duty of cluster middleware and operating system. Collective operations seem to be at the heart of cluster tools. There is a need for tools to start tasks on multiple nodes and collect the results back quickly to a central point. Therefore, fast collective communications are extremely important. MPI [33][34] and PVM implementations seems to be a good choice for tools to implement fast and scalable UNIX commands.

Naming is also another important issue. In order to operate efficiently in cluster environment, the tools must have:
- The ability to name one object e.g. node, file;
- The ability to name set of related object e.g. `ps` to a set of nodes.

Traditional UNIX commands rely on regular expression to address single and multiple objects. The Scalable UNIX Tools project also proposes the use of enhanced regular expression to handle the selection of nodes for command execution. However, this is not enough for large clusters, which might not use uniform naming. In this case, packages such as SCMS handles this by using GUI instead. The advantage of this is that users can freely select unrelated nodes for command execution. This tool indicates growing importance of GUI component to system administration tool.

6.4.1 Cluster Command & Control ($C^3$)

The Cluster Command & Control ($C^3$) tool suite from Oak Ridge National Lab (ORNL) [16] provides capabilities that may be used to perform various system administration and user tasks across single and multiple clusters for tasks such as:
- Remote monitoring and control of nodes,
- Copy/move/remove files,
- Remote shutdown/reboot of nodes,
- Security maintenance, and
- Parallel execution and invocation of any cluster wide task.

The true value of the $C^3$ tools was shown recently when applying a security patch across the 64-node ORNL cluster HighTorc [35]. The patch was applied in parallel via the `cl_exect` command in under one minute.

Eight command line general user tools have been developed toward this effort with both a serial and parallel version of each tool. The serial version aids in debugging user tasks as it executes in a deterministic fashion while the parallel version provides better performance. Briefly, these tools are as follows:
- `cl_pushimage` is the single machine *push* in contrast to the Systemimager *pull* image solution.
- `cl_shutdown` will shutdown or reboot nodes specified in command arguments.
- The `cl_pushimage` and `cl_shutdown` are both root user system administrator tools.

The other six tools, `cl_push`, `cl_rm`, `cl_get`, `cl_ps`, `cl_kill`, and `cl_exec` are tools that may be employed by any cluster user both at the system and application level.

- `cl_push` enables a user to push individual files or directories across the cluster.
- `cl_rm` will permit the deletion of files or directories on the cluster.
- `cl_get` copies cluster based files to a user specified location.
- `cl_ps` returns the aggregate result of the ps command run on each cluster node.
- `cl_kill` is used to terminate a given task across the cluster.
- `cl_exec` is the $C^3$ general utility in that it enables the execution of any command across the cluster.

The parallel version of each tool is invoked by appending a "t" to the end of the command. For example the parallel version of the `cl_push` command is `cl_pusht`. From the command line each $C^3$ tool may specify cluster nodes to operate on as individual cluster nodes, a file node list, or use the $C^3$ default file `/etc/c3.conf` node list.

`cl_pushimage` enables a system administrator logged in as root to *push* a cluster node image across a specified set of cluster nodes and optionally reboot those systems. This tool is built upon and leverages the capabilities of Systemimager. While Systemimager provides much of the functionality in this area it fell short in that it did not enable a single point *push* for image transfer. `cl_pushimage` essentially *pushes* a request to each participating cluster node to *pull* an image from the image server. Each node then invokes the *pull* of the image from the outside cluster image server.

The `cl_shutdown` avoids the problem of manually talking to each of the cluster nodes during a shutdown process. As an added benefit, many motherboards now support an automatic power down after a halt – resulting in an "issue one command and walk away" administration for cluster shutdown. `cl_shutdown` may also be used to reboot the cluster after pushing a new operating system across with `cl_pushimage`.

`cl_push` provides the ability for any user to *push* files and entire directories across cluster nodes. `cl_push` uses `rsync` to push files from server to cluster node.

The converse of `cl_push` is the `cl_get` command. This command will retrieve the given files from each node and deposits them in a specified directory location. Since all files will originally have the same name, only from different nodes, each file name has an underscore and IP or domain name appended to its tail. IP or domain name depends on which is specified in the cluster specification file. Note that `cl_get` operates only on files and ignores subdirectories and links.

cl_rm is the cluster version of the rm delete file/directory command. This command will go out across the cluster and attempt to delete the file(s) or directory target in a given location across all specified cluster nodes. By default, no error is returned in the case of not finding the target. The interactive mode of rm is not supplied in cl_rm due to the potential problems associated with numerous nodes asking for delete confirmation.

The `cl_ps` utility executes the `ps` command on each node of the cluster with the options specified by the user. For each node the output is stored in /$HOME/ps_output. A `cl_get` is then issued for the `ps_output` file returning each of these to the caller with the node ID appended per the `cl_get` command. The `cl_rm` is then issued to purge the `ps_output` files from each of the cluster nodes.

The `cl_kill` utility runs the kill command on each of the cluster nodes for a specified process name. Unlike the kill command, the `cl_kill` must use the process name as the process ID (PID) will most likely be different on the various cluster nodes. Root user has the ability to further indicate a specific user in addition to process name. This enables root to kill a user's process by name and not affect other processes with the same name but run by other users. The root user may also use signals to effectively do a broad based kill command.

The `cl_exec` is the general utility tool of the C$^3$ suite in that it enables the execution of any command on each cluster node. As such, `cl_exec` may be considered the cluster version of `rsh`. A string passed to `cl_exec` is executed "as is" on each node. This provides flexibility in both the format of command output and arguments passed in to each instruction. The `cl_exect` provides a parallel execution of the command provided.

## 6.5 Scheduling Systems

It is not an unusual practice that users manually reserve the nodes they need for running their application in the absence of any specialized tools, by announcing their intent to use these nodes through e-mail or a message board to users, prior to starting their job. Such practice is acceptable as long as everyone abides by the protocol and resolves any usage conflicts in a friendly manner. However, as organizations grow larger and the complexities in hardware, software and user base increase, these kinds of arrangements may not be too efficient or effective at all. What is needed is an automated tool that will accept user requests to reserve cluster resources and execute their jobs automatically, without much human intervention. This collection of tools is part of any well-designed scheduling system.

Apart from scheduling user jobs, a scheduler is also expected to arbitrate access to the cluster to optimize the resource usage, provide security, and to perform accounting. Based on the particular job's resource requirements, the scheduler allocates an optimized set of resources dynamically, from the available set of resources. This optimization is deemed to be important especially when the cluster consists of a large set of heterogeneous resources, and administration wants to impose a certain group policy on the resource usage that the scheduler should take into consideration before allocating the resources to the job.

More often, the scheduling software is integrated into the cluster management [36] middleware itself, so that the scheduler can make its resource allocation decisions based on feedback received from the management modules – mainly about resource availability and their load characteristics. The feedback may be used by the scheduler to statically load balance the cluster nodes, apart from deciding the feasibility of running the job on them. In this respect, the cluster scheduler undergoes the same decision process the typical operating system scheduler undergoes:
- Which job it should execute next from the ready queue(s)?
- Are there enough resources to meet the job requirements?
- How to allocate the most optimal resources to the job?
- Does the owner of the job have enough privileges to use the resources?
- Does the job meet its accounting restrictions?
- If preemption is permitted, which running job to preempt?

Note that some or all of these decisions are strongly influenced by one or more of the following factors:
- The underlying cluster architecture;
- Operating system support;
- Administrative policy on resource usage;
- Support for multiple queues and/or multiple jobs per node;
- The degree of heterogeneity in the cluster;
- Support for priorities on jobs;

A variety of scheduling technologies are available, ranging from basic batch job schedulers to dynamic load balancers. The majority of the schedulers are designed to work in a clustered as well as in a multiprocessor environment.

- CODINE [37], for instance, can operate in a heterogeneous environment to schedule the nodes in a clustered SMP as well as the processors on a Vector supercomputer. It also provides tools for managing the cluster policy, and dynamically load-balancing the jobs.
- CONDOR [38], on the other hand, works to improve the throughput of the resources by scheduling jobs on idle workstations. It can also checkpoint and dynamically migrate jobs across workstations.
- The Portable Batch System (PBS) [39] has a flexible Scheduler module that can be replaced by the site-specific Scheduler to incorporate the sites own scheduling policy. The Basic Scheduling Language (BaSL) is a scripting language provided in PBS to be used to write the custom Scheduler, apart from using other more familiar languages like C and Tcl. PBS also supports the creation of multiple job queues, so that access control lists can be used on these queues to control which users have access to submit jobs to these queues. The scheduler can be redesigned to process the queues in a prioritized fashion.

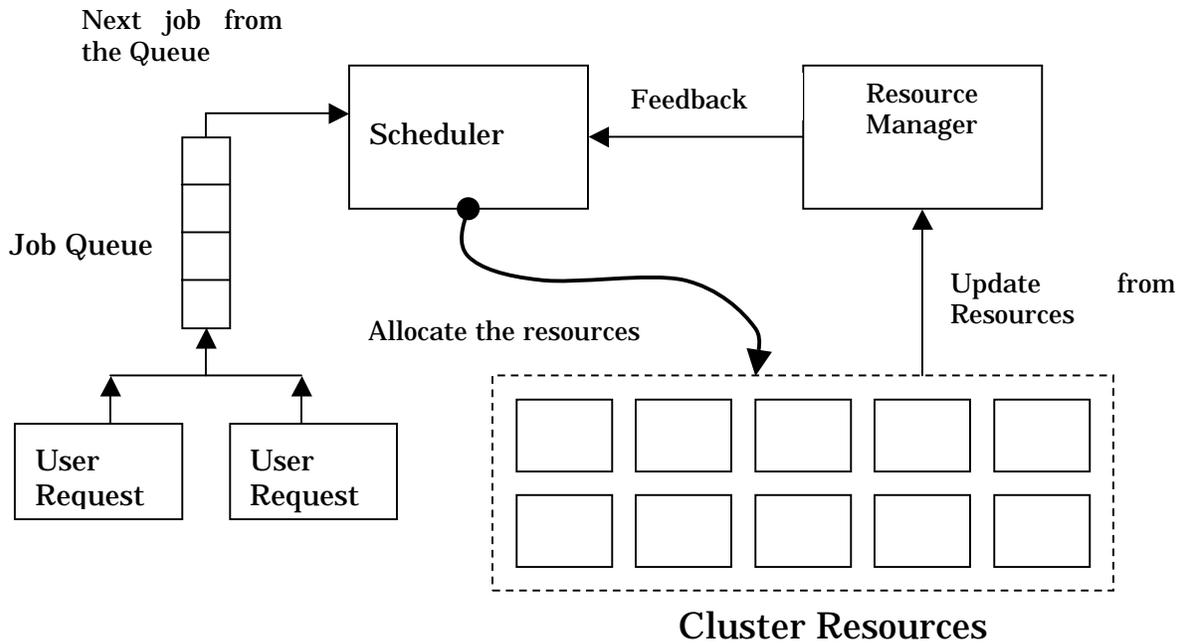

**Figure 6.9.** The conceptual architecture of a cluster scheduler and its managed resources

6.6 Conclusions

Many interesting and complex issues are involved in the creation and utilization of commodity clusters and system administration drives much of these issues. Clusters of small scale (e.g., under 32 interconnected nodes) are widespread, and larger systems (e.g., up to 1,000 or more interconnected node) are becoming prevalent today.  All the major building blocks needed for successful clustering exist somewhere or another:
- Cost-effective workstations and small-scale SMP servers,
- High speed networks with low-latency, high-bandwidth and acceptable CPU overhead,
- Operating systems (with choice of Linux, other UNIX variants or Windows),
- Middleware for message passing based on the MPI standard,
- Cluster scheduler batch and interactive systems,
- Software and system understanding tools,
- System maintenance tools for clusters, and appropriate device drivers.

Successful projects are evidently those that take a balance of cost, component-oriented risk, capability per system, and scale in order to deliver their users the best production environment possible for the investment made, or which clearly emphasize the "research" nature of clusters for computer-science-centric work. System administration's role in selecting, operating, and presenting the cluster resource to users cannot be underestimated to the success of a given system.

The state-of-the-art today involves a lot of picking and choosing in each area of concern, whether one does a "do it yourself cluster," or one works with an integrator/cluster OEM. The ability to break out the capital and maintenance cost of each item (some of which are free under the Linux strategy), allows users to control their investments according to their expected uses.  However, the self-integration approach can also lead to problems and higher cost of ownership, and perhaps more expenses in terms of human resources for administration.  The quality of the underlying software and hardware components, and the quality of their integrated behavior, all drive the percentage of available cycles that will actually be available to be used by the audience of a specific cluster installation.  Quick upgrade, recovery, and other factors enabled by diagnostic and support tools help this area.  Well-understood procedures for debugging and repairing clusters, enabled by maintenance tools, are clearly essential to system administration.

Typical cluster sites dedicated as production platforms definitely receive benefit from packaged solutions and support, while many academic and research clusters thrive on the mix-and-match combination of freeware, low-cost networks, and inexpensive workstation components ("beige boxes"). Software tools, middleware, and maintenance tools, both freeware and commercial software, are emerging to address both the Linux, other UNIX, and Windows markets.  Significant expansion of options and of packages among software tools, systems, and networks are likely to emerge in future, offering packaged cluster solutions that include system maintenance tools.  It is possible to offer the familiar "UNIX command line" and tool chain both in Linux, and under Windows; Windows offers other possibilities that may prove attractive as cluster users move into enterprises and need to integrate with desktop systems, or utilize desktop systems as so-called "supercomputers at night."

At present, the freeware world still relies heavily on software ported from earlier generations of multi-computers and multiprocessors; there are not enough new freeware items being created, nor do existing freeware products necessarily track hardware technology changes fast enough for cluster users.  This causes a benefit-gap that is evidently only going to be filled by software created based on demand and a market willing to pay for support. Training and general knowledge of system administrators is likely to increase over time, driven strongly by the growing user base and popularity of clusters.  Interest among integrators and other OEMs to provide complete solutions are obviously growing.  The willingness of users to run their applications on clusters will only widen if the system maintenance tools, middleware, and runtime environments grow in their user friendliness, allowing users with less and less cluster-specific knowledge reap significant benefits, while remaining relatively novice about the cluster's complexity.  Clearly, many more killer applications and avid users will emerge as clusters become easier to use and well integrated computational environments with common features, capabilities, and well-defined maintainability.  The options, choices, and state-of-the-art described in this section point system administration in directions that hopefully will lead to such easier-to-use cluster environments in the not-too-distant future.

# 7. Parallel I/O

Erich Schikuta, University of Vienna, Austria and Helmut Wanek, University of Vienna, Austria

## 7.1 Introduction

In the last few years many applications in high performance computing shifted from being CPU-bound to be I/O bound, as in the area of high energy physics, multimedia, geophysics and so on. Thus the performance cannot be simply scaled up by increasing the number of CPUs, but it is often necessary to increase the bandwidth of the I/O subsystem. This situation is commonly known as the I/O bottleneck in high performance computing. The result was a strong research stimulus on parallel I/O topics. However, the focus was on massive parallel processor (MPP) systems, mainly neglecting clusters of processing units, as workstations or personal computers. Besides the cumulative processing power, a cluster system provides a large data storage capacity. Usually each unit has at least one attached disk, which is accessible from the system. Using the network interconnect of the cluster, these disks can build a huge common storage medium.

I/O in parallel computing traditionally uses a small number of I/O nodes, which are connected to some storage media (for the sake of brevity these media will be referred to as "disks" in the following though other types of media like tapes or CD's may be used too). I/O nodes can either be dedicated or non-dedicated. Dedicated nodes process I/O operations only, while none-dedicated nodes perform some parts of the computation as well. For a disk read the I/O nodes read the requested data from disk and scatter it to the computing nodes, which issued the request. Similarly data is gathered from the computing nodes and written to disk for write operations. Typically an I/O node uses several disks in parallel (e.g. a RAID system) to increase I/O bandwidth and to ensure data integrity.

In order to improve I/O throughput for these systems some classical strategies (like two phase I/O [1], data sieving [2] and collective I/O [3]) have been worked out. But the fact that many processors have to share a relatively small number of disks still leads to a severe I/O bottleneck.

The advent of cluster computing, which is rapidly becoming more and more popular amongst scientists (as can be easily seen in the ever increasing number of cluster systems among the top 500 computers [4]), has induced some interesting possibilities to reduce the I/O bottleneck. Since each processing unit[2] normally has a local disk, only the data that really has to be shared by different processing units needs to be sent across the network interconnect. Moreover increasing the capacity of overall storage can be done in a cost efficient manner without hurting overall I/O throughput too much by just adding more disk space on an arbitrary number of processing units or by adding additional processing units to the cluster.

However to find the "correct" distribution of data among the available disks (i.e. to achieve maximum I/O throughput) turns out to be an extremely complex task. Many parameters that are highly interdependent have to be taken into consideration. The most important of these parameters are: data transfer rates and sizes of disks, network bandwidth, disk contention, network contention, capacity and speed of memory and cache modules.

---

[2] In the following we differentiate between nodes (on a classical supercomputer) and units (of a cluster).

So there is a need for sophisticated software, which can automatically determine an optimal distribution of data among the available disks. Though there exist a number of publications and solutions for different sub-aspects of this optimization problem (e.g. [5], [6]) no general resolution is available yet.

7.2 Parallel I/O on Clusters

Compared to classical supercomputers cluster systems typically have a slower interconnect between processors though this difference is currently diminishing because of new and easy to use networking techniques such as 1Gbps Ethernet, Myrinet and Beowulf. The processors of a cluster system tend to be more state-of-the-art due to the long design phase needed for classical supercomputers. So on the one hand we have a somewhat less elaborate interconnect between units, on the other hand it is rewarded with cheaper and more powerful system components. Most important for I/O on a cluster system is the fact that each processor has access to one or more local disks and that the system scales better and cheaper.

As an example the Intel ASCI Red TOPS Supercomputer consists of 4640 computing nodes and 74 I/O nodes with an 800 Mbytes/s interconnect. The maximum I/O throughput available is bounded by the interconnect bandwidth (since each node has to get the data somehow from the I/O nodes) and thus is 59,2 Gbytes/s. To raise this limit the interconnect bandwidth has to be increased, which is very expensive or in most cases impossible.

For a cluster with 4640 units a 100Mbps network and disk transfer rates of 20MBytes/s (typical average priced disk), the theoretical maximum I/O throughput (if each unit only accesses data on its local disk) is 92.8 Gbytes/s. This is also highly scalable because each additional processing unit will increase this value, which is not dependent on network bandwidth.

For most real world applications however local disk accesses alone will not be sufficient. Some data has to be shared by different processors and data also has to be read/written from/to external sources. If a processor needs some remote data (data, which is not stored on its local disk) this data has to be transferred via the network, which is relatively slow. The key factor to high I/O performance thus is to reduce these remote data accesses to a minimum by using a distribution of data among the computing units appropriate to the problem at hand. The distribution also has to reflect the remote accesses in order to minimize the total I/O time. Suppose that a program running on the cluster described above needs 1MByte of data to be accessible by all the processors concurrently. Several strategies can be considered like:
o Put a copy of that data on each disk (replication);
o Read the data from a single disk, broadcast it to all the units and keep it in the memory of each processor (caching);
o Distribute the data among the available disks. Whenever a part of the data is needed during program run the processor "owning" the data reads that part and broadcasts it to the other units (distribution);
o Various combinations of the above.

It is by no way easy to determine which solution is the best. Everything depends on what the program actually is doing. Are there some other disk accesses too, is there much communication over the network? To calculate the optimal I/O throughput for a particular I/O strategy a cost model is required. The following parameters are most crucial for such a model:

- *Distribution of data on disks:* Which parts of which specific file are stored on which disk. For each processor data can be stored on local disk(s) or on any disk accessible via the cluster's communication network;
- *Data layout on disks:* Which blocks are stored contiguously; are blocks stored on outer tracks (faster) or on inner tracks (slower);
- *Latencies and transfer rates of disks:* The average disk-seek time and the speed of physical read/write operations;
- *Communication network latency and transfer rate:* Has to be considered for remote accesses;
- *Sizes of disk blocks, memory and communication buffers:* Larger buffers can help to avoid overhead incurred by repetitive calls to system functions (e. g. one call to send/receive 1MBytes instead of 10 calls sending/receiving 100KBytes each). Buffers that are too large though can also decrease performance (if the message size is only 1KBytes on average, sending a 1MBytes buffer for each message will clearly degrade network throughput);
- *Size of memory and cache:* Determines how extensively caching and pre-fetching techniques may be used;
- *Transfer rates for memory to memory copy:* Unfortunately it is not generally possible to avoid all copying of buffers from the application's memory space to the I/O runtime system and finally to the network card using techniques like zero copy protocols and DMA, especially for a system which should be applicable universally;
- *I/O scheduling:* The order in which blocks are read/written. Affects not only if and how much contention occurs, but also has a big impact on disk seek times;
- *Caching, prefetching and write behind policies:* Clever techniques can greatly enhance performance by essentially overlapping disk accesses, network communication and the actual calculation routines;
- *Contention:* May occur on different levels (disk contention, network contention, memory contention) and it can have different causes (intra-application versus inter-application).

One of the biggest advantages of cluster systems over conventional supercomputers is the fact that clusters can be extended in a cheap and easy way. Just add some additional processing units to the network to increase the computing power of the cluster. In many cases this solution will scale well, but with respect to I/O some new problems arise. Often it's not possible to use exactly the same type of processing unit that has been used in the original cluster (new and faster processors are available and maybe the older type can not be bought any more). Therefore quite a number of clusters will become heterogeneous over time.

Heterogeneity can be found at different levels:

- Hardware can contain different;
    - Processors,
    - Memories (capacity and speed),
    - Disks (capacity and speed),
    - Network interfaces,
    - Network types (1Gbps, 100Mbps, 10Mbps)
- Software can be different;
    - Operating system,
    - File system,
    - Messaging system.

With respect to I/O optimization heterogeneity basically means that there is not one set of parameters as declared above, but that each processor has its own set of parameters and that these parameters also may differ greatly from one processor to another.

For multi-user systems at least a basic functionality for user authentication, access control, priority handling and data security is needed too. In these systems the various programs running concurrently will influence each other's I/O operations to a big extent. So the I/O system has to be able to dynamically adapt its I/O strategies to a changing environment. If for instance a program is started that uses the same files as another program already running but uses a completely different data distribution, it may be advantageous to change the data layout on disk. Such a redistribution of data among the disks however has to be done transparent to all the active programs.

A recent trend in supercomputing is also to connect different clusters via Internet connections and thus sharing the resources of multiple clusters for scientific calculations. From a standpoint of parallel I/O this so-called "meta computing" or "Grid computing" [21] techniques require some additional effort. The level of heterogeneity is once again increased (especially in network connections, which consist of a mixture of relatively fast local interconnects and very slow internet connections). Data may no more be distributed freely over all the computing units since data distribution over the internet connections is very slow and also the data may belong to a specific person or institution and thus is not allowed to be moved from a specific cluster. Furthermore Internet connections are far less reliable than local networks, so the I/O system has to be more fault tolerant. Finally since data is sent over the Internet certain security issues (like for instance data encryption) also have to be considered.

## 7.3 Techniques

All developed parallel I/O techniques try to achieve one or a number of the following goals:
- Maximization of the I/O bandwidth by the usage of disk devices in parallel;
- Minimization of the number of disk accesses to avoid multiple latency costs of physical disk read or write operations;
- Minimization of total application run time by overlapping computation, network communication and I/O operations;
- Minimization of re-organization and administration costs.

Generally the I/O techniques can be grouped into three different classes:
- Application level;
- Device level,
- Access anticipation methods.

*Application level methods* organize the mapping of the application's main memory objects (e.g., buffers) to respective disk objects to make disk accesses more efficient by exploiting the data locality (i.e. access data where it is stored). Examples for these methods, which are also known as buffering algorithms, are the Two-Phase method [1], where in the first phase the data is collectively accessed where locally stored and in the second phase it is transferred to the requesting nodes over the network (see PASSION [14]), and the Extended Two-Phase method [15].

A further commonly used application level technique is data sieving, which reads large sequential blocks of the disk at once into main memory and then extracts the smaller requested data blocks to avoid disk latency [2].

*Device level methods* reorganize the disk access operations according to the application's request to optimize the performance. Independent I/O node servers, located at the physical I/O devices, collect the requests, reorganize them and perform the accesses accordingly, distinguishing between the application's logical requests and the physical disk accesses. Representatives of this approach are disk-directed I/O [16] or server-directed I/O [17].

*Access anticipation methods* try to foresee the I/O access characteristics of the application based on programmer's hints (can be placed by the programmer directly into the application's code or delivered automatically by appropriate tools (e.g. a compiler)), anticipated knowledge (e.g. data distribution learned from previous runs or from persistent data layout) or reaction to identified behavior (e.g. runtime data flow analysis).
Examples for this group are informed pre-fetching [18] or Two-Phase Data Administration [12].

7.4 Systems

Three different parallel I/O software architectures can be distinguished:
- Application libraries;
- Parallel file systems;
- (Intelligent) I/O systems/parallel databases.

*Application libraries* aim for the application programmer and consist basically of a set of highly specialized I/O functions. These functions allow an expert with specific knowledge of the problem to model his solution approach. However this situation shows both advantages and disadvantages. On the one hand it provides a powerful development environment and leads to high-performance solutions, but on the other hand it often lacks generality and dynamic adaptability by creating only tailor-made, static software for specific problems. Another problem of these libraries is the fact that they are mostly too difficult to be used by non-expert users resulting in a limited distribution and acceptance of the community. Typical representatives are MPI-IO [7], an I/O extension of the standardized message passing interface library or ADIO [8], a standard API yielding an abstract device interface for portable I/O.

Every major supercomputer manufacturer supports high performance disk accesses via a proprietary parallel file system (PFS). These file systems operate independently from the application thus allowing a certain degree of flexibility. However due to their proprietary design they often lack generality allowing only access techniques specifically supported by the hardware in focus. This situation limits the possibilities of an experienced application developer, but provides even to the casual user an environment to reach at least a certain performance of the disk accesses. Well-known proprietary file systems are Intel's CFS and PFS or IBM's PIOFS (Vesta). In the last few years also some non-proprietary systems got attention, as Galley [9] or PVFS [10], which also focused cluster architectures.

*Intelligent I/O systems* hide, similar to database systems, the physical disk accesses from the application developer by providing a transparent logical I/O environment. The typical approach followed is that the user describes what she wants and the systems tries to optimize the I/O requests and accesses invisibly by applying powerful optimization techniques. These systems try to reach maximum generality and flexibility as well as the highest possible overall performance of client applications. This approach is followed in Armada [11], Panda [17], ViPIOS [12] and PPFS [13].

7.5 Case Study: The Data-intensive Computational Grid at CERN

In the past disk I/O intensive problems were very often omitted, bypassed or neglected in practical high performance computing due to their obvious complexity. However, with the rise of cluster systems and the usage of high performance methods and techniques for conventional problems parallel I/O techniques and methods yield an immense importance. An overview of parallel I/O applications can be found in [22]. To show the importance of cluster I/O techniques we will present the Data-intensive Computational Grid, which is also an eminent example for the use of clusters to cope with enormous data volumes.

In 2005 a new particle accelerator and detector, the Large Hadron Collider (LHC), will be constructed at CERN, the European Organization for Nuclear Research, which will generate information on particle collision in huge amounts of data. During a lifespan of about 15 to 20 years a single detector like the Compact Muon Solenoid (CMS) will produce about 1 PetaByte ($10^{15}$ Bytes) of data per year. This gigantic cumulated information will be analyzed, queried and processed simultaneously by thousands of High Energy Physics (HEP) researchers in a mostly on-line mode. The computing capacity necessary to fulfill this task will be several magnitudes greater than that actually used by the current experiments at CERN. Thus it will be necessary to access and integrate additional computational facilities at several HEP sites distributed across Europe, America and Asia. This situation led to the Computational Grid Project, which focuses on three issues: management of very large amounts of distributed data, high throughput computing (giving preference to capacity over performance) and automatic management of computing resources. In the course of this project tools and facilities will be developed to support an efficient, distributed access to the stored information. In 2006 the estimated computing resources to reach the defined goals are 10500 computing nodes, 2.1 PetaBytes disk capacity and an aggregate disk I/O rate of 340 GBytes/s; even the development test bed in use at the end of 2000 will comprise 400 computing nodes, 20 Terabyte disk capacity and an disk I/O rate of 5 GBytes/s.

Two aspects of this project are of specific interest for parallel I/O, the data administration environment and data replication. The data of the experiment will be distributed at CERN among huge cluster farms where each node stores its information in an object oriented database system. The administration on top of the database systems is done via multi-dimensional index structures. Research is on the way to analyze and develop appropriate index structures, as bit map indices, to allow highly parallel access to the data [19].

The experiment's computing model is not only restricted to the central computing facilities at CERN but also consists of a network of world-wide distributed regional centers, where the data is replicated in order to allow distributed and fast user analysis for some 2000 researchers. Thus it is necessary to develop appropriate replication methods and cost models [20].

It is planned to develop the necessary tools and techniques on the foundation of the Globus environment [21], which is a meta-computing infrastructure toolkit, providing basic capabilities and interfaces in areas such as communication, information, resource location, resource scheduling, authentication, and data access.

## 7.6 Future trends

The main goal of future research has to be to make massively parallel I/O as simple and usable as conventional I/O today. Besides many promising directions of research in parallel I/O, we think that two areas will have eminent importance in the near future, optimization techniques and meta information.

### 7.6.1 Optimization

As pointed out already the tuning of parallel applications to achieve maximum I/O performance is a very demanding undertaking even for a specialist in parallel I/O. In order to provide the average application programmer, who normally is oblivious to all the underlying details, with a reasonable I/O throughput software tools are strongly needed that can automatically optimize parallel I/O operations. Current systems just optimize certain details (like for example checkpoint operations) and often consider subsets of the parameters only. They rely to a great deal on various cost models used to estimate the performance of I/O operations. Based on these estimations different I/O strategies are evaluated and the best strategy is chosen and finally executed. AI algorithms (like simulated annealing and genetic algorithms) and neural networks have proven valuable to search the vast space of possible strategies for a near optimum solution. However these solutions do not only lack generality but also are of a static nature to a big extent. The series of I/O operations has to be known in advance and once optimized no changes can be made without having to rerun the optimization process. This is a serious restriction especially for multi-user systems where the sequence of I/O operations cannot possibly be predicted. In the future we hope to see more general parallel I/O systems that can easily adapt to changing system environments too.

7.6.2 Meta Information

Reading and writing raw data may be sufficient for scratch files and input and output of number crunching applications. But as soon as data is stored persistently a need for additional information arises. The date and time of file creation and the latest update or maybe even a list of updates are of importance in almost all cases. Furthermore it often is of interest what kind of data (field names, data types) is stored in a specific file. Storing this semantic information along with the file allows different and independently developed applications to easily share data by operating on common files, provided the semantic information can be interpreted correctly by all of the participating programs. XML obviously is a promising standard for storing semantic information, which can be shared by various applications. OLAP and data warehousing applications can also use the semantic information to integrate the data stored distributed across the cluster into some sort of a cluster wide distributed and parallel database.

7.7 Conclusions

The need for the storage of huge data sets is apparent, which is shown by the doubling in sales of storage systems each year. The need for support of fast and efficient access of this data is even more urgent, due to the ascendance of totally new application domains, as in the area of multimedia, knowledge engineering, and large-scale scientific computing.

Cluster systems can provide a viable platform for these I/O intensive tasks. However the lessons learned from parallel I/O on MPPs are helpful but too specific and not comprehensive enough to be mapped directly to clusters. Thus cluster I/O will need new approaches and models to cope with the special characteristics of cluster systems.

Further new application domains arise heavily stimulated from the database community, which forces the research to focus new directions (e. g. OLAP and data warehousing). Thus we believe that the cluster I/O topic will be a promising area for research in the coming years.

# 8. High Availability

Ira Pramanick, Sun Microsystems, USA

## 8.1 Introduction

System availability has traditionally been a requirement for mission critical applications. Nowadays, it is also becoming increasingly important in most commercial and many academic arenas where the goal of providing minimal downtime to users is receiving significant attention. System downtimes can be either for scheduled maintenance, or due to the occurrence of a failure of the components in the system.

Broadly speaking, there are three flavors of system availability: continuous availability, fault tolerance, and high availability. Continuous availability implies non-stop service, representing an ideal state. To achieve continuous availability, the system would need to be made up of perfect components that could never fail, either in hardware or in software. Hence, by definition, there are no actual systems that can provide this type of availability.

Fault tolerance [1] masks the presence of faults in a system by employing redundancy in hardware, software and/or time. Hardware redundancy consists of adding replicated custom hardware components to the system to mask hardware faults. Software redundancy includes the management of redundant hardware components and ensuring their correct use in the event of failures in the system. Time redundancy refers to the repetition of the execution of a set of instructions to ensure correct behavior even if a failure occurs.

High availability (HA) [2][3] provides a cheaper alternative to fault tolerance. Here, the system uses off-the-shelf hardware components in a redundant fashion, together with software solutions that mask system failures to provide uninterrupted services to a user. Typically, highly available solutions guarantee survival from single points of failure in the system. An HA system is usually a cluster [4] of machines or computer nodes.

This section of the white paper focuses on high availability in cluster computing [5][6]. It starts with an overview of the main features of a typical HA solution. This is followed by a brief description of the salient features of several HA research projects and then by a listing of several current commercial HA solutions. The next subsection discusses the main issues that come up when designing and using an HA system. The section ends with some concluding remarks.

## 8.2 Background and Overview

A typical HA solution consists of software modules which run on a cluster consisting of two or more off-the-shelf computer nodes, that together form the HA system. Cluster applications or services run on a subset of the cluster nodes, where configuration parameters of the application as well as that of the cluster determine the nodes in this subset. At one end of the spectrum are active-standby configurations where the applications run on some of the nodes (the active nodes) with the remaining nodes acting as redundant backups for those services. At the other end of the spectrum are active-active configurations where each node acts as an active server for one or more applications and potentially serves as a backup for applications that are running on the other nodes. When a node fails in the system, applications running on that node are migrated to one of their configured backup nodes. Under such circumstances, these backup nodes may get overloaded with applications leading to sub-optimal performance. This is generally acceptable since degradation in performance is preferable to the total unavailability of the service in question.

An HA system can also be a single-node system. HA in a single-node scenario refers to providing uninterrupted services in the wake of intra-node component failures. Just as in the general cluster HA case; software techniques are used in conjunction with hardware redundancy of intra-node components, to mask these component failures. Masking the failure of a network adapter is an instance of providing intra-node HA.

Thus, HA and clustering are closely related. Except for intra-node availability that applies to a single computer node as well as to nodes of a cluster, high availability of services directly implies a clustering solution. The basic feature of an HA system is that the failure of a node results in all HA applications that were running on that node being migrated or failed over to another node in the system. Similarly, failure of other components in the system results in appropriate failovers for that resource, provided the system is configured to have redundancy in that resource and there is support in the HA software for masking the failure of that resource. In addition to this automatic failover in the event of a component failure, HA systems also typically provide the user with commands to manually migrate services to another component. This is commonly referred to as a switchover.

Current HA solutions range from clusters consisting of two to eight nodes, although there are a few commercial products that are advertised to work on much larger clusters. Since the target market for most HA applications involves nodes that are servers, the need for much larger than 32 node clusters for HA is not deemed very important currently. This has an implicit impact on the choice of underlying algorithms used in an HA solution where scalability is not the main driving force. This is in contrast to algorithms used in high performance computing (HPC), which assumes that a cluster will consist of a large number of nodes. Recent market trends indicate that even for the high availability arena, there is now a demand for application availability on large clusters consisting of small nodes. This is leading to a greater emphasis on scalability during the design of HA solutions.

Broadly speaking, an HA solution consists of two parts, an HA infrastructure and HA services. The HA infrastructure consists of software components that cooperate with each other and enable the cluster to appear as a single system to a user. Their functions include monitoring cluster nodes, monitoring cluster-related processes on the cluster nodes, controlling access to shared cluster resources, and enforcing resource constraints both to ensure data integrity and to satisfy user requirements. The infrastructure needs to provide these functions when the cluster is in a steady state and more importantly, while nodes in the cluster are going up and down.

The HA services are clients of the HA infrastructure, and use the facilities exported by the latter to provide semi-seamless service availability to users. There is usually a performance degradation when a system resource such as a cluster node fails and a service is failed over to another resource in the system, but there is no service interruption per se. There is a plethora of HA applications available on various HA products in the marketplace. Common examples include the various databases, file systems, mail servers and web servers. Many of these applications are prepackaged with the HA product itself, while the others are supplied by various application vendors.

An application that executes on a single node needs to be made cluster-aware so that it can be run as an HA service. This typically involves adding wrapper scripts around the application in question such that it fits the basic HA framework; i.e., it can be started and stopped on any cluster node and can be monitored for liveness on any cluster node. This conversion of an application into an HA version is done via the HA Application Programming Interfaces (HA APIs). Currently, there is no industry-wide standard for HA APIs, and each HA product comes with its own version. All the APIs contain methods that will enable an application to switchover or failover as the case may be, and these include methods to start an application, stop an application, to monitor an application-specific resource set, to mention a few.

8.3 Research Projects

There have been several notable research projects on fault tolerance and availability. This sub-section discusses a representative sample of these research efforts. These are:
1. The ISIS Project [7];
2. The Horus Project [8];
3. (The Solaris MC Project [9];
4. The High Availability Linux Project [10];
5. The Linux Virtual Server Project [11].

ISIS and its successor, Horus, were a result of research work done at Cornell University on fault tolerance in distributed systems. Both systems implemented a collection of techniques for building software for distributed systems that performed well, was robust despite both hardware and software crashes, and exploited parallelism. Both systems provided a toolkit mechanism for distributed programming, whereby a distributed system was built by interconnecting fairly conventional non-distributed programs, using tools drawn from the kit. They included tools for managing replicated data, synchronizing distributed computations, automating recovery, and dynamically reconfiguring a system to accommodate changing workloads.

ISIS became very successful with several companies and Universities employing the toolkit in settings ranging from financial trading floors to telecommunications switching systems. It was moved from Cornell University to Isis Distributed Systems, a subsidiary of Stratus Computer, Inc. in 1993 where it was sold as a product for five years.

The Horus project was originally launched as an effort to redesign the Isis group communication system. It, however, evolved into a general purpose communication architecture with advanced support for the development of robust distributed systems in settings for which Isis was unsuitable, such as applications that have special security or real-time requirements. Besides the practical uses of this software, the project contributed towards the theory of virtual synchrony, a runtime model used for some implementations of data replication and fault-tolerance. Horus was also much faster and lighter weight than the Isis system.

Solaris MC was a prototype distributed operating system designed for a cluster of computer nodes that provided a single-system image such that users and applications could treat the cluster as a single computer running the Solaris operating system. It was implemented as a set of extensions to the base Solaris UNIX system and provided the same ABI/API as Solaris, thereby running unmodified applications. The components of Solaris MC were implemented in C++ through a CORBA-compliant object-oriented system with all new services defined by the Interface Definition Language (IDL). Objects communicated through a runtime system that borrowed from Solaris doors and Spring sub-contracts. Solaris MC was designed for high availability: if a node failed, the remaining nodes remained operational. It had a distributed caching file system with UNIX consistency semantics, based on the Spring virtual memory and file system architecture. Process operations were extended across the cluster, including remote process execution and a global `/proc` file system. The external network was transparently accessible from any node in the cluster. The Solaris MC project has been the basis of Sun Microsystems' next generation Sun Cluster product line.

The High Availability Linux Project and Linux Virtual Server are both HA solutions for the Linux operating system, built via a community development effort. Both these groups are beginning to collaborate their efforts and although these HA solutions are currently not at par with the commercial products as far as features go, they are rapidly moving towards that goal. The HA Linux Project also includes efforts to port some commercial HA products to Linux, a notable example of this being SGI's FailSafe product. The Linux Virtual Server is a highly scalable and a highly available server built on a cluster of real servers. The architecture of the cluster is transparent to end users, and the users see only a single virtual server. It uses various forms of IP-level load balancing, and is advertised to work for up to 100 servers. It is released under a GNU license.

### 8.4 Commercial Products

There are several commercial HA solutions available in the marketplace today. Each HA product comes with its set of features, details of which can be found at the corresponding vendor's website. These points of reference will also have the latest publicly available information regarding the products. The reader is encouraged to visit these websites for most up to date information on these products. The following is a list of some of these products. This is not a comprehensive list.

(i)     Compaq's TruClusters [12].
(ii)    Data General's DG UX Clusters [13].
(iii)   HP's MC/ServiceGuard [14].
(iv)    IBM's HA Products [15].
(v)     Microsoft Cluster Services [16].
(vi)    NCR's LifeKeeper [17].
(vii)   Novell's HA solutions [18].
(viii)  RSi's Resident Server Facility [19].
(ix)    Sequent's ptx/CLUSTERS [20].
(x)     SGI's IRIS FailSafe [21].
(xi)    Siemens Reliant Monitor Software [22].
(xii)   Stratus's CA Solution [23].
(xiii)  Sun Clusters [24].
(xiv)   TurboCluster Server [25].
(xv)    VERITAS Cluster Server [26].

### 8.5 Technological Scope

The design of an HA solution presents many technical challenges, both from the perspectives of infrastructure design as well as of its use by an application. This sub-section will touch upon some of these issues.

An HA infrastructure can be part of the base operating system (OS) or it can be a separate layer that sits on top of the OS. There are examples of both kinds of design in the commercial products available to the cluster community today. Integrating an HA solution into the underlying OS has several advantages including better performance and greater ease of use. The main advantages of an HA solution that sits on top of an OS include independence of the solution with respect to the underlying OS leading to portability across various hardware platforms, and often a smaller development cycle. Both from a user's perspective and that of

HA applications that sit on top of an HA framework, the single system image (SSI) capability of an HA solution is critical to its ease of use. SSI is necessary for providing a single-entity view of the cluster and for easing the task of administering a cluster. This is a rapidly growing area of research for cluster computing in general. For an HA solution, the most important elements of SSI include global file systems, global devices and global networking in the cluster.

As with any multi-resource system, load balancing becomes important in a clustering solution and assumes additional dimensions of complexity when its integration with HA is taken into account. The main facet of load balancing in an HA product includes decisions about the node to which services should be failed over if the node on which these were running goes down. The decision for this can be either resource-driven, where only a subset of the cluster nodes have the appropriate resources, or it can be user-driven, where the user defines the failover order a priori. The issue of load balancing on a quiescent cluster (where the nodes are in steady state, some up and others down) is similar to that on a non-HA cluster with resource constraints.

Scalability of HA solutions has not received its due attention since typical HA cluster nodes tend to be fat server nodes. Hence, there does not seem to be a compelling reason to support very large (greater than 32 nodes) HA clusters. That is, algorithms used in the design and implementation of the HA infrastructure do not necessarily scale. As HA becomes increasingly popular and begins to be used by many different kinds of applications, a need for supporting larger clusters will become important. This has already started happening in the marketplace, where customers are demanding highly available applications on nodes that are thin nodes, which are clustered together with other thin or fat nodes. This may also happen if highly scalable HPC solutions need to be integrated with HA solutions.

As HA becomes increasingly important, there is a significant growth in the number of applications that will need to be made highly available. The traditional "how difficult is it to program" issue that has been a great source of extensive research in parallel and distributed computing and clustering in the HPC arena, begins to loom up in the HA world as well. There is ongoing work in the HA area to establish easier ways for users to write custom HA applications.

Ease of cluster management is extremely important in general and for HA in particular. HA products are often difficult to understand from a user's perspective, who often prefer to treat their HA cluster as a "system" that acts as a server or a server group, and provides seamless access to services in the wake of different types of failures. Cluster administrators do not want to be burdened with details of how the HA software works. The success of an HA solution is largely dependent on how easy it is to administer and maintain. A graphical user interface (GUI) for cluster administration, management and monitoring (and sometimes even controlling) is a good starting point. However, the real enabler here is SSI, which allows administration tools, graphical or not, to present a unified picture of the cluster to an administrator. All this becomes particularly critical in an HA environment, where incorrect cluster management and/or user error should minimally affect system downtime.

One of the advantages of cluster computing and a reason for its growing success is the fact that a cluster can consist of heterogeneous nodes. This is in contrast to multi-computers where the nodes are usually homogeneous nodes. Support for heterogeneous nodes in a cluster is, in fact, almost always a necessity for a user. Users often test out applications on small test/development nodes, before deploying these on bigger nodes in the cluster. Heterogeneity support and ways to effectively utilize a heterogeneous cluster have been actively researched in the HPC area for the past ten years, and these topics are now receiving attention in the HA area as well. For an HA system, these issues are further complicated by considerations such as the degree of integration with the underlying operating system, vendor support in terms of the APIs used by HA applications, portability of the APIs, and support for rolling upgrades.

## 8.6 Conclusions

High availability is becoming increasingly important since there is an increasing demand for minimal downtime of systems. Clustering provides the basic infrastructure, both in hardware and software, to support high availability. Typical HA solutions span clusters of sizes ranging from two to eight server nodes today.

There are many different HA products available to the cluster community, each with several HA applications enabled for them in a pre-packaged form, and others being supported by various vendors. HA is in its infancy stage currently, and presents exciting opportunities for research and development. This is especially true with respect to its seamless integration with other facets of cluster computing such as high performance computing, scalability, and standardization of application programming interfaces.

# 9. Numerical Libraries and Tools for Scalable Parallel Cluster Computing

Jack Dongarra, University of Tennessee and ORNL, USA, Shirley Moore (formerly Shirley Browne), University of Tennessee, USA, and Anne Trefethen, Numerical Algorithms Group Ltd, UK

## 9.1 Introduction

For cluster computing to be effective within the scientific community it is essential that there are numerical libraries and programming tools available to application developers. Cluster computing may mean a cluster of heterogeneous components or hybrid architecture with some SMP nodes. This is clear at the high end, for example the latest IBM SP architecture, as well as in clusters of PC-based workstations. However, these systems may present very different software environments on which to build libraries and applications, and indeed require a new level of flexibility in the algorithms if they are to achieve an adequate level of performance. We will consider here the libraries and software tools that are already available and offer directions that might be taken in the light of cluster computing.

## 9.2 Background and overview

There have been many advances and developments in the creation of parallel code and tools for distributed memory machines and likewise for SMP-based parallelism. In most cases the parallel, MPI-based libraries and tools will operate on cluster systems, but they may not achieve an acceptable level of efficiency or effectiveness on clusters that comprise SMP nodes. Little software exists that offers the mixed-mode parallelism of distributed SMPs. It is worth considering the wealth of software available for distributed memory machines; as for many cases this may be entirely suitable. Beyond that we need to consider how to create more effective libraries and tools for the hybrid clusters.

The underlying technology on which the distributed memory machines are programmed is that of MPI [1]. MPI provides the communication layer of the library or package, which may, or may not be revealed, to the user. The large number of implementations of MPI ensures portability of codes across platforms and in general, the use of MPI-based software on clusters. The emerging standard of OpenMP [2] is providing a portable base for the development of libraries for shared memory machines. Although most cluster environments do not support this paradigm globally across the cluster, it is still an essential tool for clusters that may have SMP nodes.

## 9.3 Numerical Libraries

The last few years have seen continued architectural change in high performance computing. The upside of this development is the continued steep growth in peak performance, but the downside is the difficulty in producing software that is easy to use, efficient, or even correct. We will review a list of these challenges.

A major recent architectural innovation is clusters of shared-memory multiprocessors referred to as a Constellation. This is to avoid confusion with the term Cluster, which usually is used in the context of a group of PCs connected through a switched network. A cluster can include as its nodes small multiprocessor SMPs. A constellation integrates nodes of large

SMPs. The distinction is where the majority of parallelism exists. If it is at the top clustering level (the message passing domain), then the system is a cluster. If it is at the bottom SMP level (the shared memory domain), then the system is a constellation. A constellation incorporates more processors per SMP node than the number of nodes integrated at the system level. Constellations are the architectures of the ASCI machines, and promise to be the fastest general-purpose machines available for the next few years, in accordance with the industrial trend to invest most heavily in large market sectors and use the same building blocks to service the smaller high-end market.

It is the depth of the memory hierarchy with its different access primitives and costs at each level that makes Constellations more challenging to design and use effectively than their SMP and MPP predecessors. Ideally, a programmer should be able to produce high performance parallel code on a Constellation using an abstract programming model that is not dependent on implementation details of the underlying machine. Since users may have different notions of what is acceptable high performance, we expect the layered programming model to allow the underlying machine to be exposed at varying levels of detail. This requires good communication libraries, data structure libraries, and numerical algorithms, which we propose to build.

Although Constellations will be the dominant architecture for high end computing, most programs will be from either SMP or MPP versions, and even new code developments will likely be done on smaller machines. Therefore, users need a uniform programming environment that can be used across uniprocessors, SMPs, MPPs, and Constellations. Currently, each type of machine has a completely different programming model: SMPs have dynamic thread libraries with communication through shared memory; MPPs have Single Program Multiple Data (SPMD) parallelism with message passing communication (e.g., MPI); Constellations typically have the union of these two models, requiring that the user write two different parallel programs for a single application, or else treat the machine as "flat", with a corresponding performance loss. Well-designed libraries can hide some of these details, easing the user's transition from desktop to SMP to Constellations.

In addition to Constellations, architectures that are likely to be important are distributed networks and IRAM. IRAM stand for Intelligent-RAM, and is an example of PIM, or Processor-in-Memory. IRAM consists of a chip containing both processors and a substantial memory, and is a foreseeable architectural change, as devices get smaller and more integrated. Distributed Networks, which can be and are used now, have very slow and even unreliable access to remote memories, whereas IRAM promises to significantly flatten the memory hierarchy, with the on-chip memory effectively acting like an enormous cache. This profusion of architectural targets makes software development for scientific computing challenging.

One common feature is to build libraries that are *parameterized* for the architectural features most influencing performance. We are used to dealing with cache sizes, latency and bandwidth in our use of performance modelling in previous work, but the complexity of Constellations and other architectures presents new challenges in this regard.

Another architecturally driven algorithmic opportunity arises on the Pentium family of processors, which are not only widely used on desktops, but comprise the ASCI Red machine, and will be widely used in smaller clusters as inexpensive computing platforms. This ubiquity of Intel platforms leads us to ask how to exploit special features of the Intel architecture to do better high performance computing. In addition to excellent support for IEEE standard 754 floating arithmetic, the basic arithmetic is done to 80-bit precision rather than 64-bit. There are a variety of algorithms that perform more quickly and/or more accurately by using these features. These algorithms can be encapsulated within standard libraries, and so do not require user sophistication or even awareness for their use.

Another challenge from the proliferation of computing platforms is how to get high performance from computational kernels like matrix-matrix multiplication, matrix-vector multiplication, FFTs, etc. There are systems such as ATLAS, PhiPac, and FFTW that use a sophisticated search algorithm to automatically find very good matrix-multiply kernels for RISC workstations with C compilers that do good register allocation and basic instruction scheduling; using this approach one can produced matrix multiply and FFT routines that are usually faster than the hand tuned codes from IBM, SGI and some other manufacturers[3].

9.4 Current State-of-the-art

Of the hundred or more parallel numerical packages available some provide a conventional library interface to routines written in C, Fortran or C++. Others provide more of a parallel environment for the application developer.  At the moment few, if any, mix distributed and shared-memory parallelism.

Recent surveys on parallel numerical analysis software [3][19][20] include approximately 50 different libraries and packages for parallel architectures.  All of these would be suitable for cluster computing although they may not be fully efficient for particular configurations.  The software discussed in the report forms a subset of all parallel packages that are available either commercially or distributed on Netlib[4] or on the National HPCC Software Exchange, NHSE[5].

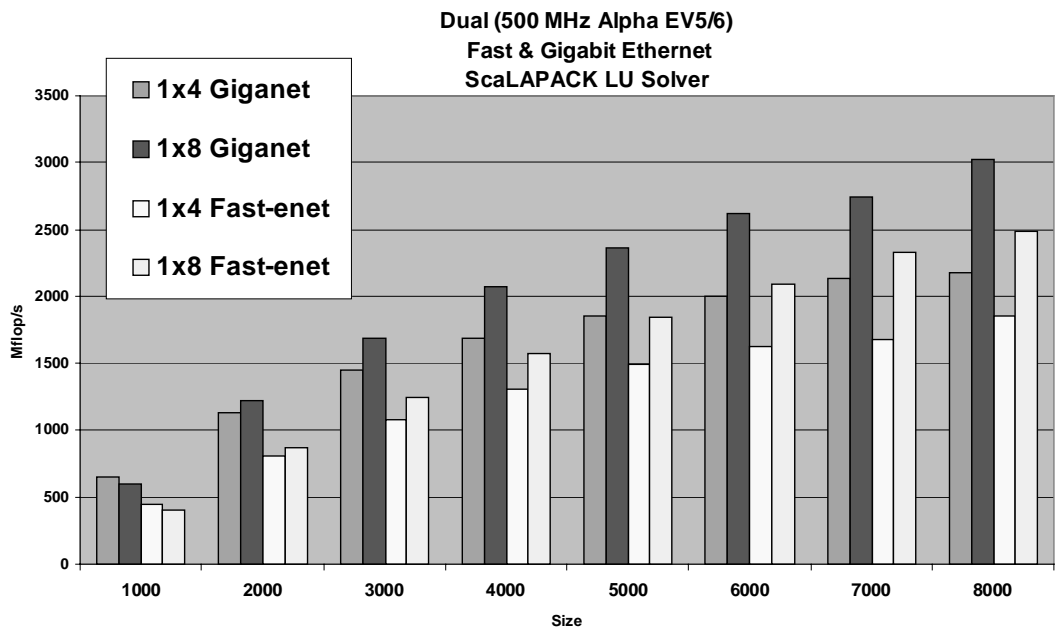

---

[3]  See: http://www.netlib.org/atlas/ and http://www.fftw.org/
[4] Netlib – http://www.netlib.org
[5] NHSE – http://www.nhse.org

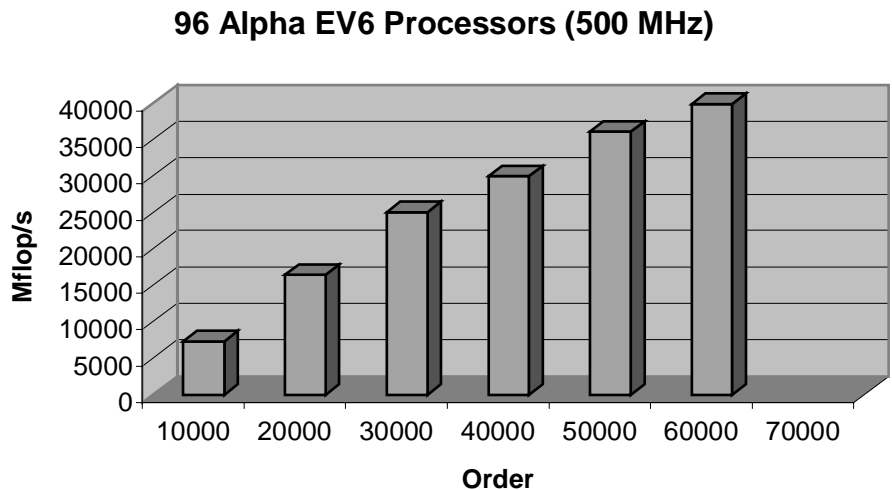

**Figure 8.1.** Performance of ScaLAPACK LU Solver on Alpha Clusters

As one might expect, the majority of available packages in this area are in linear algebra. Both direct solvers and iterative are well represented. Direct solver packages include ScaLAPACK [4] and PLAPACK [5] both of which are based on the parallelization of LAPACK [5] but by different approaches. The graphs above show good scalability of the ScaLAPACK LU solver on Alpha clusters with fast interconnection networks.

Iterative solver packages include Aztec [7], Blocksolve [8] and also PSPARSLIB [9]. Each of these provides a selection of iterative solvers and preconditioners with MPI providing the underlying communications layer. Similarly there are packages and libraries for eigenvalue problems including PARPACK [10] and PeIGS [11]; the first based on Arnoldi iterations and the second on Cholesky decomposition; both using MPI.

Other areas covered by existing parallel libraries include optimization and PDE solvers. All of the libraries mentioned above are available from Netlib. Commercial products are provided by many of the machine vendors and NAG provides a commercial, supported, general Parallel Library [12] based on MPI and also an SMP library [13][14] based on OpenMP.

The libraries that have been mentioned so far all have a traditional library interface. One of the packages that offer parallel functionality in the setting of an environment is PETSc [16]. PETSc provides an object-based interface to C-coded algorithms. In this case the application developer does not need to be aware of the message-passing or underlying mechanisms. PETSc provides a set of tools for solving PDE problems including iterative methods and other underlying functionality. This requires that the users develop their application in the PETSc style rather than calling a particular routine from PETSc. This provides an object-oriented interface that of course is also the natural interface for C++ -based packages such as ISIS++ and ELLPACK [17].

It is clear that there is a wealth of software available for clusters; however, as noted above this is in general either developed for heterogeneous distributed memory architectures or SMP machines. We still have some distance to travel before we can provide effective and efficient numerical software for the general cluster.

## 9.5 Future work

To achieve transparent cluster computing requires algorithms that can adapt to the appropriate configuration to match the cluster requirements. Recent studies show that there are benefits to be gained by rethinking the parallelization issues combining distributed and shared-memory models [18]. It is unlikely that libraries that provide this effective mixed-mode parallelism will be available for some time. It is more likely that libraries that are inherently designed for distributed memory but will be adapted have appropriate SMP-based kernels. This is likely to provide the natural progression to general numerical libraries for clustered computing.

## 9.6 Program Development and Analysis Tools

Rapid improvements in processor performance and multiprocessing capabilities of PC-based systems have led to widespread interest in the use of PC clusters for parallel computing. The two major operating systems that support multiprocessing on such systems are Linux and Windows NT. Scalable parallel computing on PC clusters requires the use of a message passing system such as MPI, although OpenMP and other forms of thread-based parallelism may also be used on SMP nodes. Programming languages of interest for scientific computing on PC clusters include Fortran, C, and C++. Appendix B presents a survey of available program development and analysis tools for PC cluster environments. These tools include compilers and pre-processors, MPI implementations, and debugging and performance analysis tools, as well as integrated development environments (IDEs) that combine these tools. Some IDEs are developed entirely by one vendor or research group, while others are designed to work together with third party tools, for example with different compilers and MPI implementations.

## 9.6 Conclusions

A good base of software is available to developers now, both publicly available packages and commercially supported packages. These may not in general provide the most effective software, however they do provide a solid base from which to work. It will be some time in the future before truly transparent, complete efficient numerical software is available for cluster computing. Likewise, effective programming development and analysis tools for cluster computing are becoming available but are still in early stages of development.

# 10. Applications

David A. Bader, New Mexico, USA and Robert Pennington, NCSA, USA

## 10.1 Introduction

Cluster computing for applications scientists is changing dramatically with the advent of commodity high performance processors, low-latency/high bandwidth networks and software infrastructure and development tools to facilitate the use of the cluster. The performance of an individual processor used in a high-end personal workstation rivals that of a processor in a high-end supercomputer, such as an SGI Origin and the performance of the commodity processors is improving rapidly. For example, MIMD Lattice Computation (MILC) [1] executes at:

- 55 Mflop/s on an Intel 300 MHz Pentium II,
- 105 Mflop/s on an Intel 550 MHz Pentium III Xeon,
- 165 Mflop/s on a single 250 MHz R10000 processor in an SGI O2000.

In the aggregate, the performance is even more impressive. The excellent performance of the interconnect network and related software is shown by comparing the same application running on a large cluster and a large SGI Origin. On 128 processors using MPI, MILC executes at:

- 4.1 Gflop/s on 128 Intel 300 MHz Pentium IIs,
- 8.4 Gflop/s on 128 Intel 550 MHz Pentium III Xeons,
- 9.6 Gflop/s on a 128 processor SGI O2000 with 250 MHz R10000 processors.

The successful standardization of system-level interfaces and supporting libraries has removed many of the issues associated with moving to a new computational platform. High-performance clusters are one of the new platforms that are coming to the forefront in the computational arena. There are still a number of outstanding problems to be resolved for these systems to be as highly effective on applications as a supercomputer. For the current generation of clusters, the most significant areas for the applications scientists are performance of the I/O systems on a cluster system and compilation, debugging and performance monitoring tools for parallel applications.

## 10.2 Application Development Environment

It is essential that the applications scientist be able to move the code to new systems with a minimum amount of effort, and have the same or similar tools and environment available to use on different systems. As recently as a few years ago, the lack of a canonical high-performance architectural paradigm meant that migrating to a new computer system, even from the same vendor, typically required redesigning applications to efficiently use the high-performance system. As a result, the application had to be redesigned with parallel algorithms and libraries that were optimized for each high-performance system. This process of porting, redesigning, optimizing, debugging, and analyzing, the application is generally prohibitively expensive, and certainly frustrating when the process is completed just in time to greet the next generation system.

The importance of supporting portable code that uses accepted standards, such as MPI [[2]], on clusters cannot be overstressed. The current generation of clusters are becoming useful to researchers and engineers precisely because of their direct support of these standards and libraries. Research software messaging systems and interfaces from vendors such as the Virtual Interface Architecture [3] may provide the underlying protocol for a higher level message passing interface such as MPI but are not directly useful to the vast majority of applications scientists.

The development environment for clusters must be able to provide the tools that are currently available on supercomputing systems. The current sets of applications running on supercomputing clusters are generally developed, debugged and tested on supercomputing systems. For first generation applications that may be developed for large scale clusters, tools such as debuggers, profilers and trace utilities that work identically or very similarly to those available on supercomputers need to be available to the applications developers. Common tools such as shell, make, tar, and scripting languages also need to be included in the development environment.

## 10.3 Application Performance

To illustrate the competitive performance of clusters, we show an application, MILC, which has been run on a large-scale cluster in the Alliance, the NT Supercluster [4] at the NCSA. This application is also being used on the Roadrunner Supercluster [5] at the University of New Mexico. In addition, one of the major performance issues confronting clusters is exemplified in the data for the ARPI-3D weather mode code, which shows timings for computations, message passing, initialization I/O and I/O at the completion of the application.

### 10.3.1 MILC

Kostas Orginos and Doug Toussaint from the University of Arizona undertook these benchmarks. They are for the conjugate gradient calculation of quark propagators in quantum chromodynamics with Kogut-Susskind quarks. The conjugate gradient typically takes 90% of the time in full QCD calculations, so it is the reasonable thing to measure. These are with the simplest version of the action, which is fairly standard. In each case, they have chosen the lattice size so that there are 4096 lattice sites per processor. This is in the range typically used in production runs. For the simplest variants of the code, this works out to about 4Mbytes of data per process.

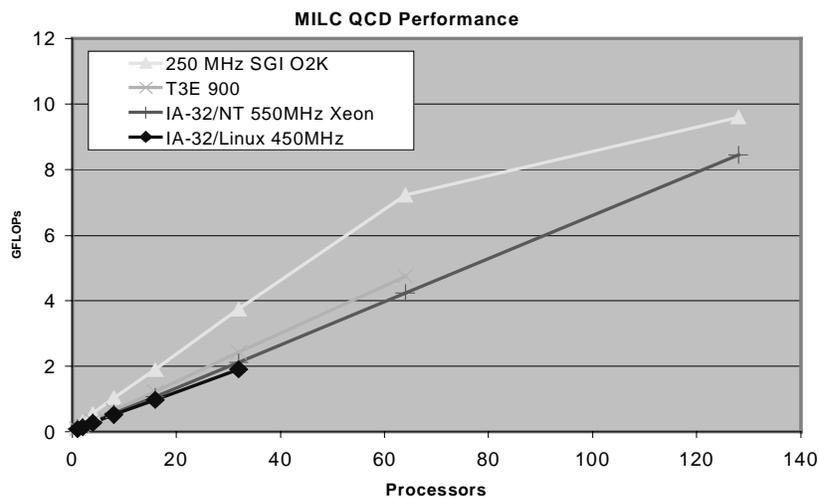

10.3.2 ARPI 3D

These graphs show the performance of the ARPI 3D weather research model on two large-scale clusters, the NCSA NT Supercluster and the UNM Roadrunner Linux Supercluster. These particular tests use a 3D numerical weather prediction model to simulate the rise of a moist warm bubble in a standard atmosphere. The simulations were run to 50 seconds with a 35x35x39 grid, in which time each processor wrote three 2.107 Mbytes/s files to a centralized file system. The data was provided by Dan Weber at the University of Oklahoma from his runs on the two systems.

The NT Supercluster data was taken in two different modes. In the first mode, only one processor in the dual processor systems was being used by the application and the second processor was idle. In the second mode, both processors in the dual processor systems were being used by the application. The dual processor runs show approximately a 20% performance degradation compared to runs using the same number of processors in separate systems. The Roadrunner data is available for only the one processor per dual processor system case.

This is a very well instrumented code and it shows where the time is spent on the system to determine where the performance bottlenecks are located in the systems. The instrumentation provides times for the:

o   Overall runtime of the application,
o   Amount of time spent performing initialization with an input data read from a common file system,
o   Amount of time spent doing the computations
o   Amount of time spent performing message passing
o   Amount of time spent writing out the resultant data files to a common file system.

The first graph shows the overall performance of the application on the two systems. The amount of time necessary for the runs increases roughly linearly as the number of processors increases.

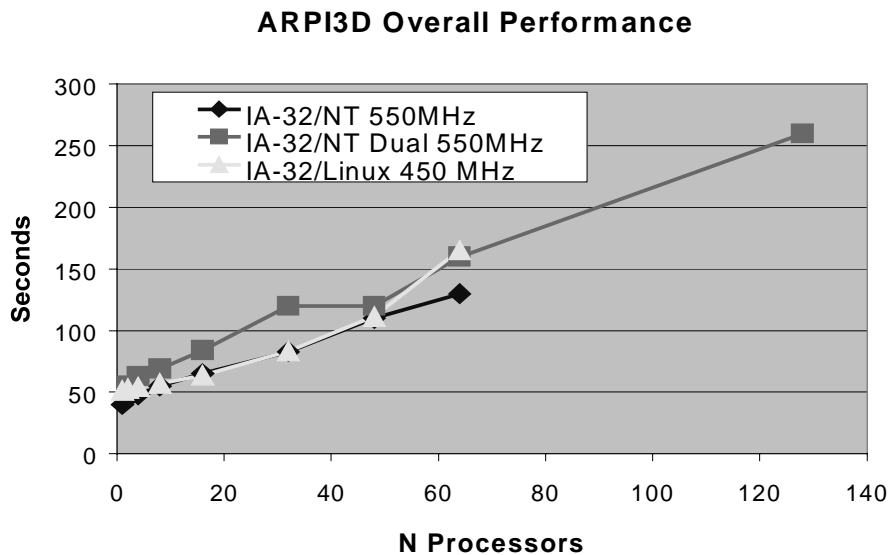

**ARPI3D Overall Performance**

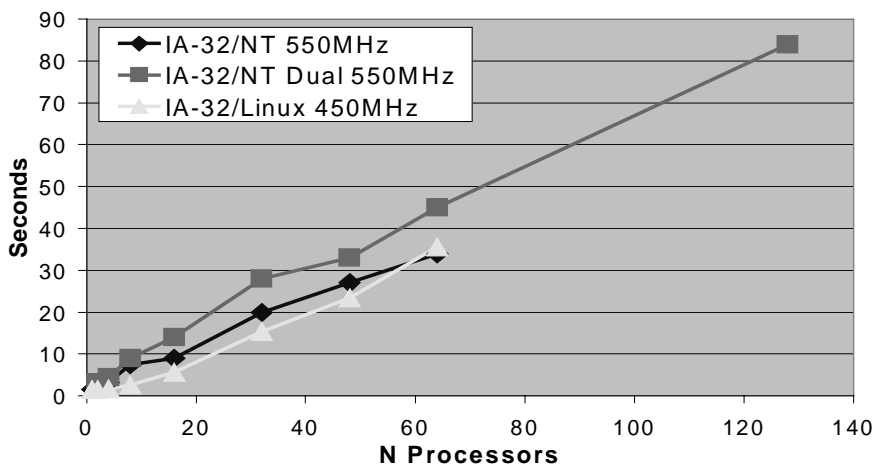

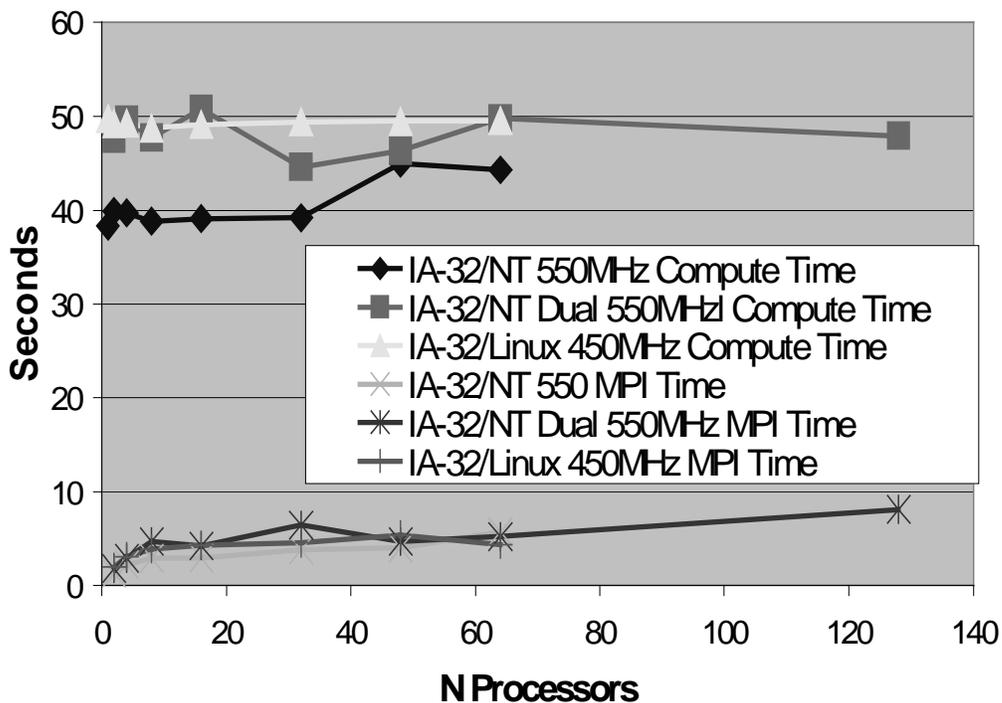

The size of the computational task per processor remains the same and the time taken for the computations is clearly shown to be nearly constant as a function of number of processors in the second graph, roughly 40 to 50 seconds, depending on the processor speed. Also shown in the second graph is the time spent by the application performing MPI calls and this is insensitive to the number of processors at 5 to 10 seconds for all of the runs, illustrating excellent scalability.

## ARPI3D Output I/O Timings

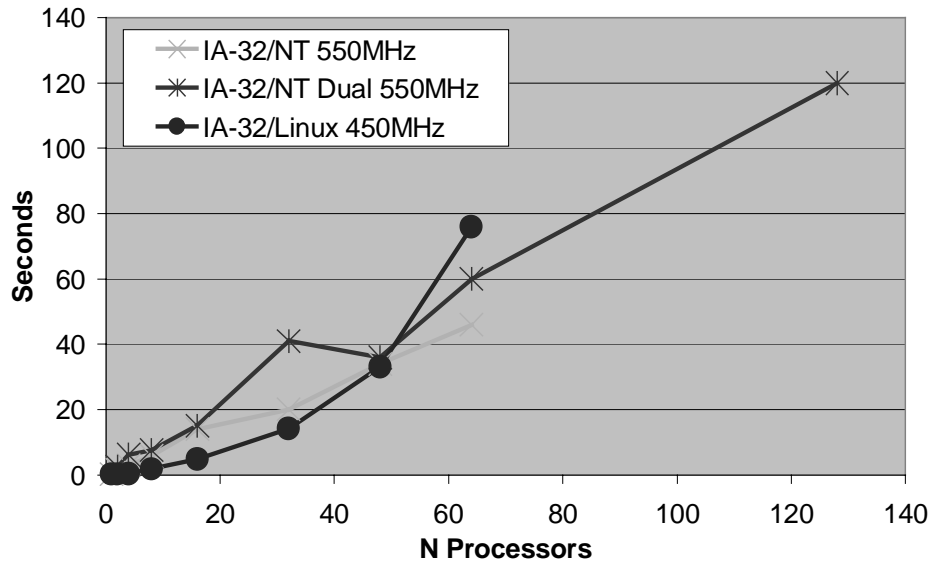

The third and fourth graphs respectively show the time spent reading in an initialization file and writing out an output file. The input data file that is read by all of the processors from the common file system is approximately 8 Kbytes and each processor writes three output data files of 2.107 Mbytes/s. The total volume of data moved is proportional to the number of processors in use, up to 800 Mbytes/s of output for the 128 processor NT run. Both Superclusters have a similar type of configuration with a common filesystem available from a fileserver to the compute nodes across the commodity Ethernet, differing primarily in the speed of the link from the file server to the switch and the file sharing software.

The IA-32/NT Supercluster file serving configuration:

o   Fast Ethernet from the file server to the switch
o   Compute nodes are all on individual Fast Ethernet nodes on the switch.
o   SMB to all nodes

The IA-32/Linux Roadrunner file-serving configuration:

o   Gigabit Ethernet from the fileserver to the switch compute nodes are all on individual Fast Ethernet ports on the switch
o   NFS to all nodes.

At smaller numbers of processors, the increased bandwidth of the server to the switch for the Roadrunner Linux system is apparent. At larger numbers of processors, the performance of the NT file server is clearly better. The NT Supercluster I/O times for input and output both increase linearly with the number of processors. In neither case is the I/O performance of the system sufficient for what is actually needed by the application. The overall increase in execution time as a function of number of processors is due almost entirely to the increased time necessary for file I/O as the number of machines increases.

## 10.4 Conclusions

Applications can perform very well on current generation clusters with the hardware and software that is now available.  There are a number of areas where major improvements can be made such as the programming tools to generate new applications on these systems and the I/O systems that are available for clustered systems.

# 11. Embedded/Real-Time Systems

Daniel S. Katz, Jet Propulsion Laboratory, California Institute of Technology, Pasadena, CA, USA and Jeremy Kepner, MIT Lincoln Laboratory, Lexington, MA, USA

## 11.1 Introduction

Embedded and real-time systems, like other computing systems, seek to maximize computing power for a given price, and thus can significantly benefit from the advancing capabilities of cluster computing. In addition to $/Mflops/s, embedded and real-time systems often have severe constraints on size, weight and power as well as latency and reliability. Satisfying these constraints are usually achieved by reducing components such as disk drives, memory and access ports. These additional constraints have traditionally led too more customized solutions within a limited marketplace. More recently, embedded computer vendors have adopted the practice of using clusters of mainstream RISC processors configured in rack-mounted cabinets. In addition, many cluster vendors are adopting their rack-mounted systems to more compact environments. Reliability constraints in these markets have traditionally been handled by traditional fault avoidance (e.g. applying higher quality and more expensive fabrication procedures on older designs) and fault tolerance (e.g. replication) techniques. While effective, these techniques tend to work against the goals (cost, power, size and weight) of embedded real time systems.

This section of the paper seeks to provide an overview of the cluster computing technology found in the embedded computing world. We begin with a brief description of some of the major applications that use embedded computers. The next two sections present hardware issues (processors, interconnects and form factors) and software issues (operating systems and middleware). Finally, we conclude with a discussion of the current trend towards embedded systems looking more like clusters and clusters looking more like embedded systems.

## 11.2 Applications

In this section, we briefly discuss three application areas that are driving the development of embedded clusters.

### 11.2.1 Space Applications

One area where embedded clusters are being examined is in space. Space-borne instruments are providing data at ever increasing rates, far faster than is feasible to send the data to Earth, and the rate of growth of the data from the instruments is faster than the rate of growth of bandwidth to Earth, so this problem will only get worse in the future. One obvious answer to this is to process the data where it is collected, and to only return the results of the analysis to Earth, rather than the raw data. (A similar argument can be made regarding the signal delay between Earth and space when considering autonomous missions, leading to the same answer: placing the computing where the decisions must be made.)

Traditionally, very little data analysis has been done in space, and what has been done has relied on radiation-hardened processors. These processors are quite old by the time they complete the radiation-hardening process, and do not solve the problem of bandwidth limitation. A proposed solution is an embedded cluster of COTS (Commercial-Off-The-Shelf) processors, where the processors can be selected and placed in the system shortly before mission launch. In as much as COTS processors are not radiation hardened, this requires software that can detect and correct errors caused by the cosmic ray environment found in space.  Such a system (intended to deliver 99% reliability through using multiple layers of fault detection and mitigation, starting with application-based fault-tolerance for parallel applications [[1]]) is being developed by the Jet Propulsion Laboratory under the Remote Exploration and Experimentation Project [2].  Initial tests appear to demonstrate that many common linear routines such as fast Fourier Transforms (FFTs) and linear algebra, which consume a large fraction of CPU time in many data processing applications, can be wrapped to enable detection of greater than 99% of errors of significant size at a cost on the order of 10% overhead in fault-free conditions [3].  The Project's current models predict that common hardware such as a PowerPC 750 would see about 2.5 single event upsets (SEUs; random transient bit flips) per hour, and off-chip cache would see about 1 SEU per Mbyte per hour in either low-Earth orbit or Deep Space [4].  These are low enough numbers to make software-implemented fault tolerance (SIFT) appear viable, although this cannot be proven without significant testing.

The constraints of the space environment (mass, power, size, resistance to vibration, shock, thermal cycling and natural space radiation) demand work in packaging that is very different than for most ground-based clusters. Additionally, replacement of faulty components is either impossible or extremely expensive, and uploading new or changing existing software is very difficult from a system design as well as operational perspective. Finally, reliability of the software as well as the hardware/system is a significant concern due to the difficulty in validating computational results, and the potential impact of erroneous behavior with respect to decision-making and scientific data analysis. Thus, a different set of requirements exists for space-borne computers.

### 11.2.2 Signal Processing Applications

The signal processing applications that embedded systems serve are often highly data parallel and naturally lend themselves to the kind of coarse-grained parallelism ideal to clusters. However, the Quality of Service (QoS) and latency requirements of real-time systems usually dictate a need for interconnects with both deterministic performance and higher performance than are typically found in clusters [5]. Thus the solutions developed to serve these needs are similar to conventional rack mounted clusters with high performance interconnects, denser packaging and lighter operating systems.

### 11.2.3 Telecommunications Applications

In telecommunications, reliability and real-time response have always been critical factors. Early electronic switching systems used specially designed fault-tolerant computers such as the AT&T 3b2. With the current explosion in network servers, in order to achieve acceptable cost it is necessary to use the types of standard commercial machines and operating systems discussed in this section. This industry is currently experimenting with commodity cluster systems to solve the problem of meeting their ultra-high reliability and real-time response constraints while reducing costs.

## 11.3 Hardware

Form factor is a critical factor in embedded computing. Thus compute density (Gflops/s per cubic foot and Gflops/s per watt) is often as, or more important, than aggregate processing power. Typically, embedded system vendors are able to achieve roughly a factor of 10 increase in compute density over conventional systems. These gains are achieved by constructing boards containing multiple nodes (typically 2 or 4). Each node consists of a low power processor (e.g., Motorola PowerPC) with a limited amount of memory (e.g. 128

MBytes). In addition, there are no local disks and access to the node is limited to the interconnect, which may be custom or commodity (e.g. Myrinet), and will have been packaged to minimize size and power consumption. These various design tradeoffs allow embedded vendors to fit nearly 100 processing nodes in a volume that can fit underneath a typical office desk.

As mentioned previously, many embedded systems need to withstand much more severe conditions than standard clusters. These systems may be used in the aerospace or military industries, leading to requirements on tolerance to shock, vibration, radiation, thermal conditions, etc. While many of today's commercial components can handle these conditions, they are not packaged to do so, as this increases cost and is not needed by most ordinary users. Thus, for this niche market, different vendors have sprung up to package standard commercial parts with more consideration of these concerns.

There are a variety of vendors that manufacture systems along the above lines. Mercury, CSPI, and Sky are three of the more popular systems. Some of the general capabilities are shown in Table 11.1. For a comparison, a cluster vendor (AltaTech) is also shown.

| Vendor  | CPU         | Interconnect     | OS      | CPU/ft³ |
|---------|-------------|------------------|---------|---------|
| Mercury | PowerPC     | Raceway          | MCOS    | ~10     |
| CSPI    | PowerPC     | Myrinet          | VxWorks | ~10     |
| Sky     | PowerPC     | Sky Channel      | SKYmpx  | ~10     |
| Alta    | Intel/Alpha | Ethernet/Myrinet | Linux   | ~1      |

Table 11.1 – General Capabilities

In addition to the vendors that specialize in embedded systems, a number of other companies build embedded systems, both parallel and distributed for their customers. These vendors may take systems from the standard vendors listed above and ruggedize and/or repackage them, and they include many US defense contractors (Lockheed, Honeywell, General Dynamics, etc.)

Many embedded systems are also targeted for real-time applications with extremely low latency requirements (e.g., radar signal processing). To achieve these requirements it is often necessary to adopt a pipeline architecture with different processing occurring at each stage of the pipeline. Typically, each stage exploits coarse grain parallelism but the 'direction' of this parallelism is along different dimensions of the data at different steps. To fully exploit a parallel computer in such circumstances requires transposing (or "corner turning") the data between steps. Thus, the interconnects provided by embedded systems often have higher bandwidth and lower latencies than those of shared memory supercomputers, let alone clusters of workstations.

11.4 Software

The real-time requirements of embedded systems necessitate special operating systems and middleware to reduce latency and to fully exploit the interconnects and processors. This has resulted in a wide variety of Unix flavored operating systems: VxWorks, Lynx, MCOS, SKYmpx, LinuxRT, IrixRT. Typically, these operating systems trade off memory protection, multi-user and multi-threaded capabilities to get higher performance. Traditionally, VxWorks has been one of the most common in many industries. However, the need for portable software has led many to examine alternative operating systems for embedded clusters. These include LinuxRT and IrixRT. The advantage of these choices is that software can be developed (at least to some point) on common desktop machines, and easily transferred to the embedded clusters. Lynx and other POSIX-compliant systems are used similarly, under the assumption that software developed on one POSIX-compliant operating system can be easily ported to another. The primary distinction between these operating systems and VxWorks is that VxWorks does not provide process-based memory protection, which may be important in prevention of fault propagation from one process to another.

One of the most significant positive trends in embedded computing has been the adoption of common middleware libraries to ease portability between systems. The two major areas where this has occurred are in communications and math libraries. In the past, special vendor libraries were required to fully exploit their custom networks. Recently vendors have adopted MPI. With careful optimization, they are able to achieve performance similar to that of their proprietary libraries. MPI has been a large step forward for the embedded community, but it does not address all of the communication needs of these systems. This has led to the development of a message passing standard that provide such critical features as Quality of Service (QoS), MPI/RT [6]. In addition to MPI/RT, the Data Reorganization Interface (DRI [7]) has been another standardization effort to provide a common interface to large data movements.

The mathematics libraries developed by embedded vendors are similar to other optimized math libraries in that they provide a variety of standard mathematical operations that have been tuned to a particular processor. The functional focus of the embedded libraries has primarily been on basic signal processing operations (e.g. FFT, FIR filters, linear algebra) for complex floating-point data. Because Fortran compilers for these systems are hard to find, these libraries usually only have a C implementation. One additional feature of these libraries has been the ability to pre-allocate memory. For most operations, this eliminates potential latency. Optimized math libraries are critical to achieving real-time performance and thus these libraries are heavily used in embedded real-time software. This heavy reliance can lead to a significant portability bottleneck. Fortunately, one of the most successful efforts of this community has been the adoption of a standardized Vector, Signal, and Image Processing Library (VSIPL [8]). In addition to being valuable to embedded systems, this library has a significant potential benefit to the general high performance computing community.

Creating an optimized implementation of a standard can be a significant undertaking for embedded systems vendors. Increasingly, these libraries are being implemented by third party software developers (e.g. MPI Software Technologies, Inc). However, as the embedded community is a niche market, software vendors generally do not have enough demand to optimize their products for the variety of embedded systems, and thus products that can optimize themselves, such as ATLAS and FFTW (as mentioned in section 7 of this white paper) become increasingly important.

11.5 Conclusions

While there are certainly a number of differences between embedded clusters and standard clusters that have been brought out in this section, there are also a number of similarities, and in many ways, the two types of clusters are converging. Mass-market forces and the need for software portability are driving embedded clusters to use similar operating systems, tools, and interconnects as standard clusters. As traditional clusters grow in size and complexity, there is a growing need to use denser packaging techniques and higher bandwidth, lower latency interconnects. Real-time capability is also becoming more common in traditional clusters in industrial applications, particularly as clusters become more interactive, both with people and with other hardware. Additionally, fault-tolerance is becoming more important for standard clusters: first, as they are increasingly accepted into machine rooms and subject to reliability and up-time requirements; and second, as feature sizes and operating voltages are reduced, cosmic-ray upsets will occur more frequently. The only area that promise to continue to separate the two cluster worlds is the general need for ruggedized packaging for many embedded clusters.

# 12. Education

Daniel C. Hyde, Bucknell University, USA and Barry Wilkinson, University of North Carolina at Charlotte, USA

12.1 Introduction

Clusters offer an exciting opportunity for *all* institutions of higher education to teach and use high performance computing for the first time without access to expensive equipment. The Task Force on Cluster Computing[6] (TFCC) can help in many ways.

Educators need help in answering such practical questions as: Is a cluster the best computing platform for undergraduate parallel programming courses? How can a cluster be set-up? What is the best software to use and where can it be obtained? How do I to manage a cluster? TFCC can facilitate the sharing of the experiences of others.

Teaching about clusters is new. Educators in both university and industrial settings (industrial seminars and short courses) need direction in selecting course contents and the fundamental principles to be taught. The TFCC can help by providing a forum and leadership in developing and sharing curricular materials.

Academics at universities can provide industry with conceptual frameworks for clusters, explore the fundamental characteristics of clusters and devise models on how to model cluster behaviour and performance. The TFCC can facilitate positive interactions between the cluster computing industrial research and the academic research communities.

12.2 Background

A cluster is a type of parallel or distributed system that consists of a collection of interconnected whole computers used as a single, unified computing resource. Clusters come in at least the following major flavours depending on their purpose, briefly:

1. *High Performance Computing Flavour.* An example is a Beowulf [1]. The purpose is to aggregate computing power across nodes to solve a problem faster. For example, high performance scientific computing typically spreads portions of the computation across the nodes of the cluster and uses message passing to communicate between the portions.
2. *High Throughput Computing Flavour.* These clusters harness the ever-growing power of desktop computing resources while protecting the rights and needs of their interactive users. These systems are a logical extension of the batch job environments on old mainframes. Condor is a good example [2].
3. *High Availability Computing Flavour.* These clusters are designed to provide high availability of service. Many vendors (e., g., IBM, DEC, SUN) provide high availability (HA) systems for the commercial world. Most of these systems use redundancy of hardware and software components to eliminate single points of failure. A typical system consists of two nodes, with mirrored disks, duplicate switches, duplicate I/O devices and multiple network paths. See Pfister's book [3].
4. *High Performance Service Flavour.* A cluster of nodes is used to handle a high demand on a web service, mail service, data mining service or other service. Typically, a request spawns a thread or process on another node.

---

[6] TFCC – http://www.ieeetfcc.org

## 12.3 What is Now Happening in Universities?

### 12.3.1 Graduate Level

At the graduate level, universities often offer a variety of parallel computing courses on architectures, algorithms, and programming. These courses have much in common with flavour 1 above, but ignore characteristics of the other three. These offerings are aimed at a small (less than 1%) and shrinking scientific computing market. Courses are also offered in distributed *something*, e.g., distributed operating systems, distributed computing, distributed systems, and distributed databases. Historically, graduate-level courses have used supercomputers available remotely, or local multiprocessors systems provided for research activities. Nowadays, high performance clusters are replacing such systems, for both research and teaching.

### 12.3.2 Undergraduate Level

There has been parallel computing and distributed computing courses at the undergraduate level for a few years, although distributed computing was barely mentioned in the ACM Curriculum '91 document. The newly formed IEEE/ACM Curriculum 2001 Committee has a Netcentric Task Force as part of their effort [4]. Members of TFCC should provide valuable input to the dialog that will surround the creation of this new curriculum.

Undergraduate parallel computing courses have typically used inexpensive transputer systems or multiprocessor simulators. A computing cluster offers a particularly attractive solution, and courses are appearing based upon clusters [5].

In addition to specific courses, the concepts to cover all four of the above flavours of clusters are scattered in bits and pieces in many courses. For example, topics needed for cluster computing such as scheduling, performance measurement, communication protocols, client-server paradigm, reliability theory, security, monitoring, and concurrency control are either not taught or scattered throughout a university's curriculum.

## 12.4 What is Needed to Improve the Situation?

The following need to be addressed:

1. A consensus should be established on the concepts to be taught and on the fundamental principles pertaining to clusters that will serve our students.
2. Examples of high quality courses are needed, including teaching strategies, lecture notes and laboratory exercises.
3. Material is needed to prepare our students to be able to design, analyze, implement, test and evaluate clusters effectively. A model of the behaviour of clusters is needed that can be used to model performance and availability.
4. More good textbooks are needed, especially for undergraduates. A recent textbook specifically for undergraduate teaching parallel programming on clusters is by Wilkinson and Allen [6].
5. More good reference books are needed for practitioners in the field and for advanced graduate students to acquire background. A recent excellent example is the two-volume set edited by Buyya [7][8].

## 12.5 How can the TFCC help?

The TFCC will play a leadership role in education of clusters. To support educators, the TFCC will:

1. Collect educational materials and freely disseminate them
2. Serve as a forum for discussion.
3. Sponsor workshops and educational sessions at conferences.
4. Encourage the creation of resources such as books, bibliographies and web pages.
5. Encourage the writing of reviews of books, especially textbooks on cluster computing.
6. Be an active participant of the research culture on clusters.
7. Facilitate positive interactions between the industrial research community and the academic research community on cluster computing.

## 12.6 Conclusions

In months of its existence, the IEEE Computer Society's Task Force on Cluster Computing (TFCC) has made significant contributions to supporting cluster computing. It has brought together a large number of researchers and educators who would not otherwise had the opportunity to interact (including the two authors of this section). The work of the task force encompasses all aspects of cluster computing research as well as education.

To support educators, a web site has been created [9] which contains links to educational sites, related books and journals, freely available software, projects from academia and industry, white papers, and descriptions of hardware components. A book donation program has been established with the generous cooperation of major publishers for the donation of some current books on clusters. Also, the TFCC has established an electronic list server and an on-line web-based newsletter for all its members.

The TFCC is committed to its efforts within the education arena of cluster computing. The success of these efforts requires active participation by the TFCC members from the education community. We strongly encourage you to become involved.

# Appendix A

Linux

In the public research community, Linux is the most frequently used operating system simply for two reasons: it is released as true open source (in differentiation on the conditions under which the source code for Sun Solaris and Microsoft Windows NT is available), and it is cheap in terms of both software and required hardware platforms. It offers all the functionality that is expected from standard Unix, and it is developing fast as missing functionality can be implement by anyone who needs it. However, these solutions are usually not as thoroughly tested as releases for commercial Unix variants. This requires frequent updates, which do not ease the job of the administrators to create a stable system, as it is required in commercial environments. The large-scale commercial use as a cluster OS is also severely limited by the missing HA functionality and unverified SMP scalability on CPU numbers greater than 2 or 4.

For scientific and research applications, these functional gaps are not really problematic, and the Linux concept is a great success, as can be seen from the well-known Beowulf project. This leads to a big and growing user community developing a great number of tools and environments to control and manage clusters, which are mostly available for free. However, it's often hard to configure and integrate these solutions to suit ones own, special set up as not all Linuxes are created equal: the different distributions are getting more and more different in the way the complete system is designed. Even the common base, the kernel, is in danger, as solution providers like TurboLinux [1] start to integrate advanced features into the Kernel.

In general, in has to be noted that the support for Linux by commercial hardware and software vendors is continuously improving, so it can no longer be ignored as a relevant computing platform.

Windows NT

Although NT contains consistent solutions for *local* system administration in a workstation or small-server configuration (consistent API, intuitive GUI-based management, registry database), it lacks several characteristics that are required for efficient use in large or clustered servers. The most prominent are standardized remote access and administration, SMP scaling in terms of resource limits and performance, dynamic reconfiguration, high-availability and clustering with more than a few nodes.

Through Microsoft's omnipresence and market power, the support for NT by the vendors of interconnect hardware and tools developers is good, and therefore a number of research projects in clustering for scientific computing is based on Windows NT (HPVM [2] and MPI [3][4][5]. Microsoft itself is working hard to extend the required functionality and will surely improve. The so-called "Datacenter Edition" of its forthcoming Windows2000 (which was delayed and postponed several times) has to prove how far NT has caught up with the established UNIX solutions.

AIX

IBM's AIX operating system, running on the SP series of clusters, is surely one of the most advanced solutions for commercial and scientific cluster computing. It has proven scalability and stability for several years and a broad number of applications in both areas. However, it is a closed system, and as such the development is nearly totally in the hand of IBM, which however has enough resources to create solutions in hardware and software that commercial customers demand, such as HA and DFS. Research from outside the IBM laboratories is very limited, though.

Solaris

In a way, Solaris can be seen as a compromise or merge of the three systems described above: it is not an open system, which ensures stability on a commercial level and a truly identical interface to administrators and programmers on every installation of the supported platforms. Of course, easy kernel extensions or modifications such as with Linux are not possible, making it less interesting for research in this area. It should be noted that the source code for Solaris and also Windows NT is available upon request, but not for free extension and publication of modifications.

It offers a lot of functionality required for commercial, enterprise-scale cluster-based computing like excellent dynamic reconfiguration and fail-over and also offers leading inter- and intra-nodes scalability for both, scientific and commercial clustering. Its support by commercial vendors is better for high-end equipment, and the available software solutions are also directed towards a commercial clientele. However, Solaris can be run on the same low-cost off-the-shelf hardware as Linux as well as on the original Sun Sparc-based equipment, and the support for relevant clustering hardware like inter-connects is given (Myrinet, SCI). Software solutions generated by the Linux community are generally portable to the Solaris platform with little or no effort.

The advantage of this situation is that solutions developed on simple, low-cost systems can easily be adapted for use in a computing environment of commercial-quality level. Of course, such an approach requires support and engagement of the manufacturer, Sun in this case.

References

[1] TurboLinux TurboCluster Server 4.0, TurboLinux, Inc., http://www.turbolinux.com/products/tltcs.html
[2] HPVM, http://www-csag.ucsd.edu/projects/hpvm.html
[3] PATENT, PaTENT MPI 4.0, Genias Software, http://www.genias.de
[4] Multi-Platform MPICH, Lehrstuhl für Betriebssysteme, RWTH Aachen, http://www.lfbs.rwth-aachen.de/~joachim/MP-MPICH.html
[5] MPICH for Windows NT, Argonne National Laboratories, http://www.mcs.anl.gov/mpi/mpich/mpich-nt)

# Appendix B

## Compilers and Preprocessors

The Absoft Pro Fortran toolset [1] for Windows 95/98/NT/2000 includes globally optimizing Fortran 77, Fortran 90, and C/C++ compilers integrated with a development environment designed for Fortran users. Pro Fortran 6.2 is a Fortran toolset optimized for single processor Windows95/98 and Windows NT/2000 systems. Pro FortranMP 6.2 includes in addition a thread-safe runtime library and the VAST-F/Parallel preprocessor for automatically restructuring Fortran code for execution on dual processor systems. The Fx source-level debugger supporting Fortran and C/C++ is included with both toolsets, as is a performance profiler. All compilers and tools can be accessed via an integrated development environment (IDE) that automatically builds make files and generates header dependencies. The Absoft compilers are also available for Linux.

Compaq Visual Fortran [2] (formerly Digital Visual Fortran) is available for Windows 95/98 and Windows NT along with a development environment.

Visual KAP [3] is a preprocessor that automatically parallelizes Fortran 77/90/95 and ANSI C source code. Visual KAP runs on PentiumII/Pentium Pro/Pentium based machines under Windows NT 4.0 or Windows 95, and targets the Compaq Visual Fortran compiler. Since Windows 95 does not support parallelism, parallelism optimizations are available under Windows NT. Visual KAP works with optimizing compilers to provide additional speedups beyond what the compiler's built-in optimizer provides. Visual KAP has both command-line and graphical user interfaces.

Microsoft Visual C++ [4] is part of the Microsoft integrated development environment (IDE) called Visual Studio 6.0. The Visual C++ compiler can process both C source code and C++ source code. The compiler is compliant with all ANSI standards and has additional Microsoft extensions. The C++ Standard Template Library (STL) is included. The Visual C++ debugger has a graphical user interface for setting breakpoints, viewing classes and variables, etc. The debugger has an edit and continue feature that allows the user to change an application and continue debugging without manually exiting the debugger and recompiling.

VAST/Parallel [5] from Pacific-Sierra Research Corporation includes the VAST-F/Parallel and VAST-C/Parallel automatic parallelizing preprocessors for Fortran and C, respectively. VAST/Parallel transforms and restructures the input source to use parallel threads so that the program can automatically make use of parallel processors. They also support the OpenMP standard for user-directed parallelization. VAST/Parallel can optionally produce a diagnostic listing that indicates areas for potential improvement to assist the developer in tuning performance further. VAST/Parallel works with the DEEP development environment by gathering compile-time data for DEEP and inserting instrumentation code for run-time data gathering. DEEP uses this information to display compile-time optimization notes (e.g., which loop nests have been parallelized, which data dependencies are preventing parallelization) and run-time performance data (e.g., which loop nests use the most wallclock time, which procedures are called the most.

## MPI Implementations

MPICH [6] is a freely available reference implementation of MPI developed by Argonne National Laboratory and Mississippi State University. Version 1.1.2 of MPICH is available for Linux on PCs and is in use on the Linux RoadRunner Supercluster at the University of

New Mexico. Problems with MPICH programs on LINUX suddenly failing with lost TCP connections are on the list of things to fix in future MPICH releases. The MPICH development team at Argonne is involved in a research effort with LBNL to develop MPICH on top of VIA, the Virtual Interface Architecture, a specification of an industry-standard architecture for scalable communication within clusters. MPICH for Windows NT is now a beta released.

MPI-FM [7], developed by the Concurrent Systems Architecture Group at the University of Illinois at Urbana-Champaign and the University of California, San Diego, is a version of MPICH built on top of Fast Messages. MPI-FM requires the High Performance Virtual Machine (HPVM) runtime environment that is available for both Linux and Windows NT. MPI-FM is in use on the NCSA NT Supercluster [8].

MPI/Pro [9] is a commercial MPI implementation from MPI Software Technology, Inc. The current release of MPI/Pro is for Windows NT on Intel and Alpha processors, but MPI/Pro for Linux will be released soon. MPI/Pro is based on a version of MPICH for Win32 platforms that was developed at Mississippi State. MPI Software Technology is working on a new version of MPI/Pro that does not include any MPICH code and that supports VIA.

WMPI (Windows Message Passing Interface) [10] was the one of the first implementations for Win32 systems. WMPI is an ongoing effort by The University of Coimbra, Portugal. The first version was released in 1996 and was a port of the MPICH/P4 to Win32 systems. More recently a new internal architecture has been developed that provides greater functionality to the runtime library and enables WMPI to move towards a dynamic environment based on the MPI-2 standard. The latest released version, with the new internal design, is able to use multiple devices simultaneously and is also thread-safe.

PaTENT MPI 4.0 [11] is a commercial MPI implementation for Windows NT available from Genias GmbH. PaTENT stands for Parallel Tools Environment for NT. Genias plans to release a suite of development tools for MPI programs on NT, including debugging and performance analysis support.

FT-MPI [12] developed as part of the HARNESS [13] project at the University of Tennessee, Knoxville is a subset of the MPI 1.2 and MPI 2 standards that provides additional fault tolerance to allow applications to handle process failures, and subsequent communicator corruption. Failure recovery consists of rebuilding communicators, by either shrinking, padding, re-spawning or user run-time, communicator specific methods. FT-MPI uses the SNIPE_Lite [14] multi-protocol communications library and supports, TCP/UDP sockets, shared memory, BIP [15], GM and ELWMCP.

Development Environments

Some vendors of Linux and Windows products are coming out with integrated development environments (IDEs) that integrate tools such as compilers, preprocessors, source code editors, debuggers, and performance analysis tools. Some IDEs are a complete product from a single vendor. Others provide a framework and some tools but integrate other tools, such as compilers, from other vendors.

DEEP [16] from Pacific-Sierra Research stands for Development Environment for Parallel Programs. DEEP provides an integrated interactive GUI interface that binds performance, analysis, and debugging tools back to the original parallel source code. DEEP supports Fortran 77/90/95, C, and mixed Fortran and C in Unix and Windows 95/98/NT environments. DEEP supports both shared memory (automatic parallelization, OpenMP) and distributed memory (MPI, HPF, Data Parallel C) parallel program development. A special version of DEEP called DEEP/MPI [17] is aimed at support of MPI programs.

Debuggers

DEEP/MPI [17] from Pacific Sierra Research [18] is a development environment for MPI parallel programs. DEEP/MPI debugging support is available for Linux on PCs. DEEP/MPI provides a graphical interface for parallel debugging of Fortran or C MPI programs. Capabilities include setting breakpoints, watching variables, array visualization, and monitoring process status. DEEP/MPI for Linux has been tested with MPICH and LAM MPI 6.1.

The PGDBG Debugger is part of the Portland Group, Inc. (PGI) PGI Workstation 3.0 development environment for Intel processor-based Linux, NT, and Solaris86 clusters. PGDBG supports threaded shared-memory parallelism for auto-parallelized and OpenMP programs, but does not provide explicit support for MPI parallel programming.

GDB, the GNU Debugger, is available for both Linux and Windows NT on PCs from Cygnus [19]. GDB supports debugging of multithreaded code but does not provide explicit support for MPI parallel programming.

Performance Analyzers

DEEP/MPI [17] from Pacific Sierra Research is a development environment for MPI parallel programs. DEEP/MPI performance analysis support is available for both Windows NT and Linux on PCs. DEEP/MPI provides a graphical user interface for program structure browsing, profiling analysis, and relating profiling results and MPI events to source code. DEEP/MPI for Linux has been tested with and supplies MPI profiling libraries for MPICH and LAM MPI 6.1. A driver called *mpiprof* is used to instrument, compile, and build MPI programs. Running *mpiprof* results in the creation of a subdirectory called *deep* in which the static information files created for each file you instrument are saved. When you execute your MPI program as you normally do, runtime information files are also stored in this subdirectory. To analyze following execution, start DEEP/MPI by executing the command *deep* which will bring up a File Dialog box asking you the location of your program. Currently DEEP requires that all source files be located in the same directory and that all static and runtime created files be located in the *deep* subdirectory, but this restriction is expected to be removed in future releases.

VAMPIR [20] is a performance analysis tool for MPI parallel programs. The VAMPIRtrace MPI profiling library is available for Linux on PCs. VAMPIR provides a graphical user interface for analyzing tracefiles generated by VAMPIRtrace.

The Jumpshot graphical performance analysis tool is provided with the MPICH [6] distribution. Jumpshot analyzes tracefiles produced by the MPE logging library, which is an MPI profiling library also provided with MPICH for Linux. Jumpshot is written in Java and interprets tracefiles in the binary clog format by displaying them onto a canvas object. Jumpshot itself is available for Windows (a JVM for Windows is provided with the Jumpshot distribution), and MPICH for Windows NT is supposed to be available soon. By default, Jumpshot shows a timeline view of the state changes and message passing behaviour of the MPI processes. Clicking any mouse button on a specific state instance will display more information about that state instance. Clicking any mouse button on the circle at the origin of a message will display m ore information about that message. That Jumpshot's performance decreases as the size of the logfile increases is a known bug, and can ultimately result in Jumpshot hanging while it is reading in the logfile. There is a research effort underway to make Jumpshot significantly more scalable. Jumpshot can be run as an application or as an applet using a web browser or applet viewer.

PGPROF [21] is part of the Portland Group, Inc. (PGI) PGI Workstation 3.0 development environment for Intel processor-based Linux, NT, and Solaris86 clusters. PGPROF supports threaded shared-memory parallelism for auto-parallelized and OpenMP programs, but does not provide explicit support for MPI parallel programming. The PGPROF profiler supports function-level and line-level profiling of C, C++, Fortran 77, Fortran 90, and HPF programs. A graphical user interface displays and supports analysis of profiling results.

Paradyn [22] is a tool for measuring and analyzing performance of large-scale long-running parallel and distributed programs. Paradyn operates on executable images by dynamically inserting instrumentation code while the program is running. Paradyn can instrument serial programs running on WindowsNT/x86 but does not yet support any MPI implementation on that platform.

VTune [23] is a commercial performance analyzer for high-performance software developers on Intel processors, including Pentium III, under Windows NT. VTune collects, analyzes, and provides architecture-specific performance data from a system-wide view down to a specific module, function, or instruction in your code. VTune periodically interrupts the processor and collects samples of instruction addresses and matches these with application or operating system routines, and graphically displays the percentage of CPU time spent in each active module, process, and processor. VTune provides access to the Pentium hardware performance counters under Windows NT.

PAPI [24] is a platform-independent interface to hardware performance counters. A PAPI implementation is available for Linux/x86.

Proceedings of EuroPar'99, LNCS 1685, pp633-642, August 31-September 3, 1999, Toulouse, France.
[16] DEEP, http://www.psrv.com/deep.ht
[17] DEEP/MPI, http://www.psrv.com/deep_mpi_top.html
[18] PGI Workstation, http://www.pgroup.com/prod_description.html
[19] Cygnus, http://www.cygnus.com/
[20] Vampir, http://www.pallas.de/pages/vampir.htm
[21] PGPROF, http://www.pgroup.com/ppro_prof_desc.html
[22] Paradyn, http://www.cs.wisc.edu/paradyn/
[23] Vtune, http://www.pts.com/static/Vtune.html
[24] PAPI, http://icl.cs.utk.edu/projects/papi/